\documentclass[prl, twocolumn, superscriptaddress]{revtex4-2}

\usepackage[usenames]{color}
\usepackage{xcolor}
\usepackage{upgreek}
\usepackage{amsmath}
\usepackage{amssymb} 
\usepackage{graphicx}
\usepackage[utf8]{inputenc}
\usepackage{physics}
\usepackage{empheq}
\usepackage{siunitx}
\usepackage{placeins}
\usepackage{bbm}
\usepackage{float}

\newcommand{\defeq}{\vcentcolon=}
\newcommand{\eqdef}{=\vcentcolon}
\usepackage[colorlinks=true,bookmarks=false]{hyperref}
\hypersetup{linkcolor=blue,citecolor=blue,urlcolor=blue}  

\definecolor{green1}{rgb}{0.0, 0.5, 0.0}
\definecolor{green2}{rgb}{0.55, 0.71, 0.0}

\newcommand{\psiD}{\hat{\psi}_1}
\newcommand{\psiB}{\hat{\psi}_2}
\newcommand{\psiM}{\hat{\psi}_{0}}
\newcommand{\psiF}{\hat{\psi}_3}

\newcommand{\psiDNH}{\psi_1}
\newcommand{\psiBNH}{\psi_2}
\newcommand{\psiMNH}{\psi_{0}}
\newcommand{\psiFNH}{\psi_3}

\newcommand{\rhoMM}{{\rho}_{00}}
\newcommand{\rhoDD}{{\rho}_{11}}
\newcommand{\rhoBB}{{\rho}_{22}}
\newcommand{\rhoFF}{{\rho}_{33}}
\newcommand{\rhoMD}{{\rho}_{01}}
\newcommand{\rhoMB}{{\rho}_{02}}
\newcommand{\rhoMF}{{\rho}_{03}}
\newcommand{\rhoDB}{{\rho}_{12}}
\newcommand{\rhoDF}{{\rho}_{13}}
\newcommand{\rhoBF}{{\rho}_{23}}

\definecolor{green1}{rgb}{0.0, 0.5, 0.0}
\definecolor{green2}{rgb}{0.55, 0.71, 0.0}

\newcommand{\sgn}[1]{\text{sgn}\qty[{#1}]}

\newcommand{\beginsupplement}{%
	\setcounter{table}{0}
	\renewcommand{\thetable}{S\arabic{table}}%
	\setcounter{figure}{0}
	\setcounter{section}{0}
	\setcounter{equation}{0}
	\renewcommand{\thefigure}{S\arabic{figure}}
	\renewcommand{\theHfigure}{S\arabic{figure}}
	\renewcommand{\thesection}{S\Roman{section}}
	\renewcommand{\theHsection}{S\Roman{section}}
	\renewcommand{\theequation}{S\arabic{equation}}
	\hypersetup{linkcolor=black,citecolor=blue,urlcolor=blue}  
	
}%

\begin{document}

\title{Observing Dynamical Currents in a Non-Hermitian Momentum Lattice}
\author{Rodrigo Rosa-Medina}
\thanks{These authors contributed equally to this work.}
\affiliation{Institute for Quantum Electronics, ETH Zürich, 8093 Zürich, Switzerland}
\author{Francesco Ferri}
\thanks{These authors contributed equally to this work.}
\affiliation{Institute for Quantum Electronics, ETH Zürich, 8093 Zürich, Switzerland}
\author{Fabian Finger}
\affiliation{Institute for Quantum Electronics, ETH Zürich, 8093 Zürich, Switzerland}
\author{Nishant Dogra}
\thanks{Present address: Cavendish Laboratory, University of Cambridge, J. J. Thomson Avenue, Cambridge CB3 0HE, United Kingdom.}
\affiliation{Institute for Quantum Electronics, ETH Zürich, 8093 Zürich, Switzerland}
\author{Katrin Kroeger}
\affiliation{Institute for Quantum Electronics, ETH Zürich, 8093 Zürich, Switzerland}
\author{Rui Lin}
\affiliation{Institute for Theoretical Physics, ETH Zürich, 8093 Zürich, Switzerland}
\author{R. Chitra}
\affiliation{Institute for Theoretical Physics, ETH Zürich, 8093 Zürich, Switzerland}
\author{Tobias Donner}
\email{donner@phys.ethz.ch}
\affiliation{Institute for Quantum Electronics, ETH Zürich, 8093 Zürich, Switzerland}
\author{Tilman Esslinger}
\affiliation{Institute for Quantum Electronics, ETH Zürich, 8093 Zürich, Switzerland}
\date{\today}

\begin{abstract}
	We report on the experimental realization and detection of dynamical currents in a spin-textured lattice in momentum space. Collective tunneling is implemented via cavity-assisted Raman scattering of photons by a spinor Bose-Einstein condensate into an optical cavity. The photon field inducing the tunneling processes is subject to cavity dissipation, resulting in effective directional dynamics in a non-Hermitian setting. We observe that the individual tunneling events are superradiant in nature and locally resolve them in the lattice by performing real-time, frequency-resolved measurements of the leaking cavity field. The results can be extended to a regime exhibiting a cascade of currents and simultaneous coherences between multiple lattice sites, where numerical simulations provide further understanding of the dynamics. Our observations showcase dynamical tunneling in momentum-space lattices and provide prospects to realize dynamical gauge fields in driven-dissipative settings.  
\end{abstract}
\maketitle

%\section{Introduction}

Experiments with quantum degenerate atomic gases have successfully realized a wide variety of many-body lattice models, facilitating the exploration of complex out-of-equilibrium phenomena in highly controlled settings ~\cite{Langen_2015,Gross_2017,PRXRoadmap_2021}. Engineering lattice bonds that dynamically depend on the local particle configuration is essential for simulating lattice gauge theories~\cite{Wiese_2013,Zohar_2015,Kasper_2020} and electron-phonon coupling~\cite{Bissbort_2013,Giustino_2017}. Specifically, systems exhibiting density-dependent tunneling hold the potential to realize correlated many-body phenomena, such as pair superfluidity~\cite{Eckholt_2009,Rapp_2012,Di_2014} and quantum scars~\cite{Zhao_2020,Hudomal_2020}. So far,  density-dependent tunneling in optical lattices has been implemented via periodic driving~\cite{Ma_2011,Meinert_2016,Xu_2018,Clark_2018,Gorg_2019,Schweizer_2019} or dipolar interactions~\cite{Baier_2016}, yet solely inferred from spectroscopic measurements or by comparison to theory. Here, we realize a complementary experimental scheme that allows us to engineer dynamical tunneling events in a momentum-space lattice and directly measure them in real-time. %They are mediated by an emerging photon field that self-consistently evolves with the atomic state.

\begin{figure}[thbp]
	\centering
	\includegraphics[width=0.93\columnwidth]{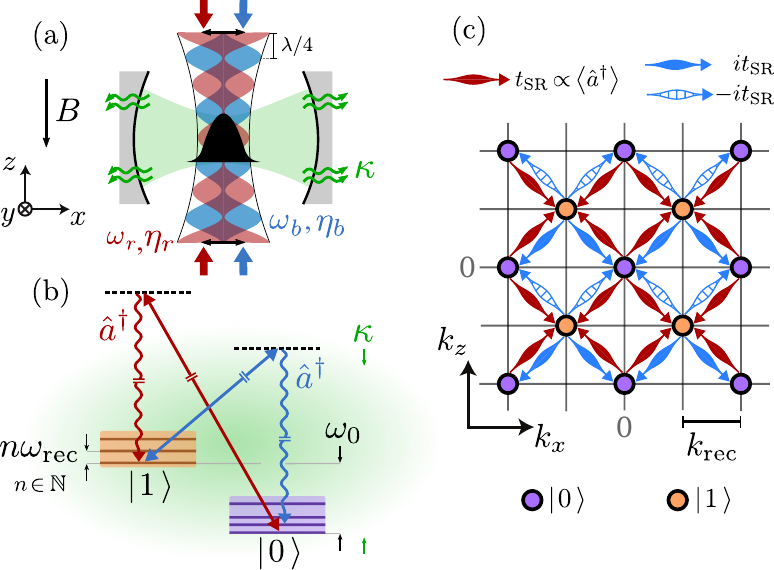}
	\caption{\textbf{Realization of dynamical currents in a momentum lattice}.  (a) Experimental setup. A BEC inside an optical cavity with decay rate $\kappa$ is illuminated by two $x$-polarized, retroreflected Raman drives (red and blue) with frequencies $\omega_{r,b}$ and coupling strengths $\eta_{r,b}$. Their standing-wave modulations are shifted by $\lambda/4$. (b) Coupling scheme in a rotating frame at frequency $\bar{\omega}$. Raman scattering, involving absorption from the drives (solid arrows) and net emission of photons in the cavity mode (wiggly arrows and creation operator $\hat{a}^\dagger$), induces population transfer between specific momentum states of two different spin manifolds $\ket{0}$ (purple) and $\ket{1}$ (orange lines). These states are offset by the two-photon detuning $\omega_0$ and by multiples of the recoil frequency $\omega_\text{rec}$, with $\omega_\text{rec}\leq \omega_0 \ll \kappa$. (c) Tunneling dynamics in a momentum grid. Raman scattering couples discrete momentum states of $\ket{0}$ (purple) and $\ket{1}$ (orange circles) differing by $\pm k_\text{rec}$ in $x$ and $z$-direction, giving rise to dynamical tunneling (red and blue arrows) with rate $t_\text{SR}\propto\expval{\hat{a}^\dagger}$.}
	\label{fig:Fig1}
\end{figure}

Our implementation employs a spinor Bose-Einstein condensate (BEC) coupled to the fundamental mode of a high-finesse optical cavity by two transverse laser beams, see Fig.~\ref{fig:Fig1}(a). Cavity-assisted Raman scattering transfers atoms between two spin levels ($\ket{0}$, $\ket{1}$), while imparting momentum to the BEC in multiples of the photon recoil [Fig.~\ref{fig:Fig1}(b,c)]. This engenders spin and particle dynamics in a two-dimensional momentum grid, which we interpret as photon-assisted tunneling events in a synthetic lattice ~\cite{Gadway_2015}. These events are mediated by an emergent cavity field, which self-consistently evolves with the atomic spin and density configuration. Hence, the tunneling rate dynamically depends on the buildup of coherences between neighboring sites, in contrast to experiments employing Bragg scattering from classical drives to control single-particle hopping rates in momentum lattices~\cite{Gadway_2016,An_2018,Yan_2020}. The underlying process is superradiant Raman scattering in an optical cavity~\cite{Kasevich_2011,Thompson_2012}, which is collectively enhanced by the number of participating emitters~\cite{Dicke_1954,Haroche_1982,Ketterle_2004,Kuga_2004,Piovella_2004,Yelin_2005}.  Since the resonator linewidth significantly exceeds site-to-site energy offsets,  the cavity mode can accept different spectral components and mediate tunneling in a large momentum grid. The inherent dissipation due to cavity losses stimulates the superradiant transfers and renders the dynamics non-Hermitian. Making use of the cavity leakage, we gain nondestructive, real-time access to the atomic currents, which is often challenging in analog quantum simulations~\cite{Zoller_2017,Zoller_2018,Geier_2021}. By performing frequency-resolved heterodyne measurements of the cavity field, we locally resolve the tunneling events in the momentum grid. A key feature of this implementation is that tunneling processes in opposite directions occur via different quantum paths and are independently controlled by the two drives. Our system constitutes a flexible platform to explore nonequilibrium lattice physics, thanks to the possibility to optically engineer dynamical currents and resolve them via the cavity field. 
%As opposed to collective Raman scattering in BECs from a single drive~\cite{}, our scheme relies on two classical laser fields and a cavity mode. 

In our experiments, we prepare a BEC formed by $N\approx10^5$ $^{87}$Rb atoms 
in the $m_F=-1$ magnetic sublevel of the $F=1$ hyperfine manifold. A magnetic field along the $z$-direction generates a Zeeman splitting of $\omega_z=2\pi\cdot 48~\text{MHz}$ between the sublevels $m_F=-1$ and $m_F=0$, which we label as $\ket{0}$ and $\ket{1}$, respectively. The atomic cloud is prepared inside an ultrahigh-finesse optical cavity with decay rate $\kappa=2\pi\cdot1.25~\text{MHz}$, and illuminated by two retro-reflected laser fields far red-detuned from the atomic resonance. Their wavelength $\lambda=784.7~\text{nm}$ is associated with a recoil frequency of $\omega_\text{rec}=2\pi\cdot3.73~\text{kHz}$. The frequencies of the drives $\omega_{r,b}$ lie on opposite sides of the dispersively shifted cavity resonance $\omega_c$, with $\omega_{b}-\omega_c\approx\omega_z$ and $\omega_{r}-\omega_c\approx-\omega_z$. As shown in Fig.~\ref{fig:Fig1}(a), their standing-wave modulations are shifted by $\lambda/4$ at the position of the atoms, such that the combined static lattice potential is fully erased for balanced laser powers~\cite{SI}.

This scheme realizes cavity-assisted Raman transitions between the spin states $\ket{0}$ and $\ket{1}$ with two-photon coupling rates $\eta_{r,b}$, cf. Fig.~\ref{fig:Fig1}(b).  We map the system to a  tight-binding model in a rotating frame defined by the intermediate frequency of the two drives $\bar{\omega}=(\omega_b+\omega_r)/2$~\cite{SI}, where the cavity detuning is defined as $\Delta_c=\bar{\omega}-\omega_c$. For balanced two-photon couplings $\eta=\eta_r=\eta_b$, the many-body Hamiltonian in momentum space $\hat{H}=\hat{H}_0 + \hat{H}_\mathrm{t_{SR}}$ contains a diagonal contribution
\begin{align}
&\hat{H}_0=-\hbar\Delta_c \hat{a}^\dagger \hat{a} \label{eq:HamiltonianKinetic}\\
&+\!\!\!\sum_{\substack{\{j,k\}\in\mathbb Z \\ \sigma\in\{0,1\}}} \!\!\!\hbar[\sigma\omega_0+\omega^\text{kin}_{(2j+\sigma,2k+\sigma)}]\hat{c}_{(2j+\sigma, 2k+\sigma)}^{\sigma\dagger} \hat{c}^\sigma_{(2j+\sigma, 2k+\sigma)},\nonumber
\end{align}

and a light-assisted tunneling term  
\begin{align}
	\hat{H}_\mathrm{t_{SR}} \!=\! -\frac{\hbar\eta}{\sqrt{8}}\hat{a}^\dagger  \!\!\!\!  \sum_{\substack{\{j,k\}\in\mathbb Z \\ s_{1,2}=\pm1}}& 
	 \Big[\hat{c}^{1\dagger}_{(2j+s_1,2k+s_2)}\hat{c}^{0}_{(2j,2k)} \label{eq:Hamiltonian}\\  -is_2& \hat{c}^{0\dagger}_{(2j,2k)}\hat{c}^{1}_{(2j+s_1,2k+s_2)}\Big] + \mathrm{H.C.}\Big. \nonumber 	
\end{align}

The bosonic operators $\hat{a}^\dagger$ and $\hat{c}^{\sigma\dagger}_{(l,m)}$ create photons in the fundamental mode of the cavity field and atoms in $\ket{\sigma}$ with $(l,m)$ units of recoil momentum $k_\text{rec}$ along $(x,z)$-directions. We indicate the corresponding atomic modes in the momentum grid as $\ket{l,m}_\sigma$, with $l,m$ being an even (odd) number for $\sigma=0~(1)$. The site-to-site energy offset results from a kinetic contribution $\omega^\text{kin}_{(l,m)}=(l^2+m^2)\omega_\mathrm{rec}$ and a global splitting between the spin manifolds $\omega_0=(\omega_b-\omega_r)/2-\omega_z$. This key feature allows us to resolve the emerging currents in the lattice by measuring the frequency of the corresponding cavity field.  The Hamiltonian in Eq.~\eqref{eq:Hamiltonian} describes photon-mediated tunneling between next neighbors in the momentum grid with self-consistent rates $t_\mathrm{SR}(t)=-\eta\expval{\hat{a}^\dagger(t)}/\sqrt{8}$.  The components of the atomic state tunneling in $\pm z$-direction acquire a phase of $\mp i$ when scattering from the  drive at $\omega_b$, as depicted in Fig.~\ref{fig:Fig1}(c). This is due to the relative spatial phase between the two standing-wave drives, which is also crucial for suppressing Bragg scattering within a spin sector along $z$-direction~\cite{Gadway_2015}, e.g., between $\ket{0,0}_0$ and $\ket{0,\pm2}_0$.%Within our tight-binding description, contact interactions give rise to occupation-dependent energy shifts of the lattice sites~\cite{SI}, which can lead to self-trapping in the initial momentum state for small tunneling rates~\cite{An_2018,An_2021}.

The system described by Eq.~\eqref{eq:Hamiltonian} is also a multilevel Tavis-Cummings model with collective coupling $\eta\sqrt{N/8}$. Since the experiment operates in an overdamped regime ($\kappa\gg\eta\sqrt{N/8}$), the system is strongly dissipative and decays through superradiant scattering when initialized in $\ket{0,0}_0$~\cite{SI,Fan_2020,Lin_2021}. In an illustrative picture, the evolution is primarily determined by Raman processes creating cavity photons ($\propto\hat{a}^\dagger$), as the opposite process of absorbing photons ($\propto\hat{a}$)  is hindered by cavity losses. As a consequence, the non-Hermitian dynamics in the momentum lattice are directional, with preferred tunneling directions illustrated by the arrows in Fig.~\ref{fig:Fig1}(c). The arising superradiant transfers are collectively enhanced, which results in tunneling rates evolving self-consistently with the coherences between the sites involved in each hopping process. This behavior is fundamentally different from the one observed in related experiments employing standing-wave Raman drives with equal spatial phase at the position of the BEC, where a low-momentum stable superradiant phase is created above a critical driving strength~\cite{Lev_2018,Ferri_2021,Mivehvar_2021}. The large cavity linewidth ($\kappa \gg \omega_0,\omega_\text{rec}$) facilitates tunneling in a large momentum grid, in contrast to potential implementations involving solely classical drives~\cite{Gadway_2015} or subrecoil cavities~\cite{Kessler_2014}, where multiple lasers would be required.

\begin{figure}[thbp]
	\centering
	\includegraphics[width=1\columnwidth]{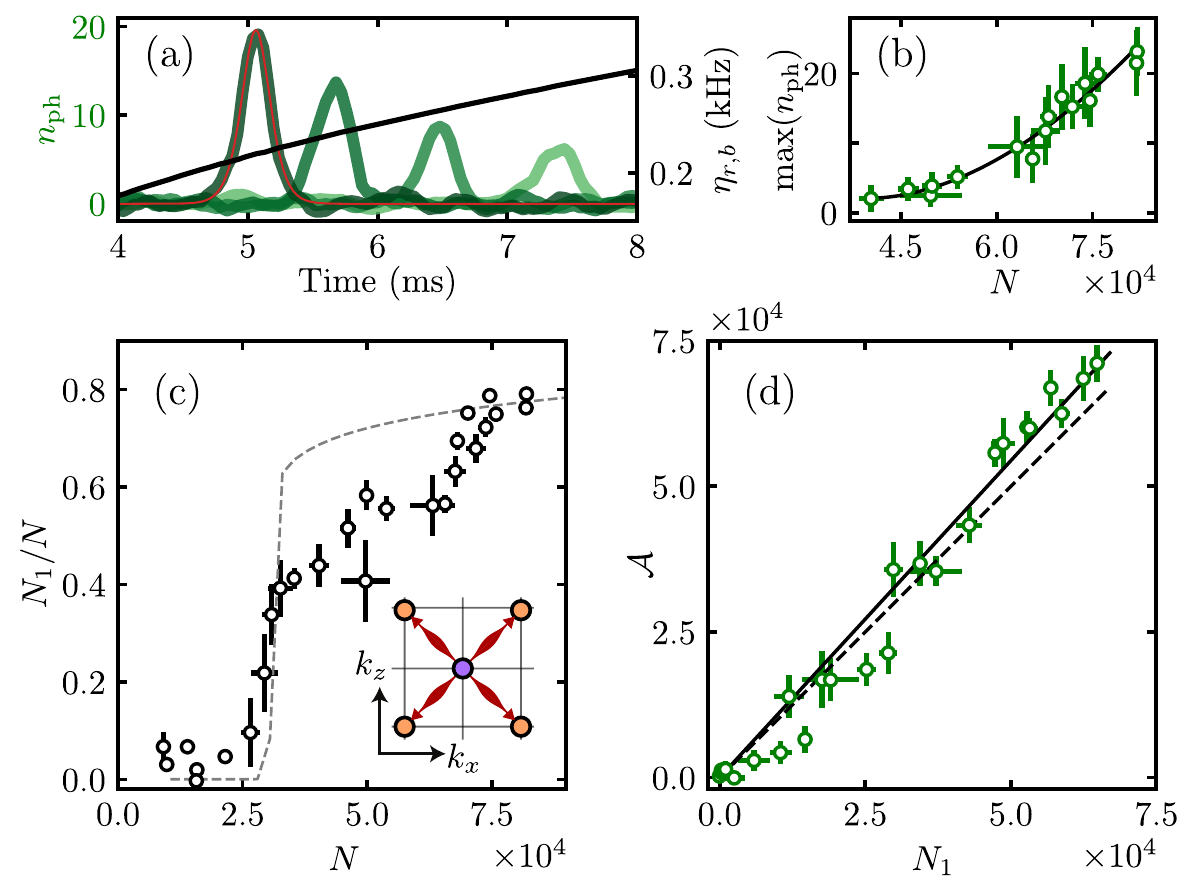}
	\caption{\textbf{Superradiant tunneling in the momentum lattice}. (a) Representative photon pulses for different atom numbers $N=(8.1,6.6,4.9,2.9)\cdot10^4$ (darker to lighter green curves), together with a typical fit (red)~\cite{SI} and coupling ramp $\eta$ (black line). (b) Pulse amplitude versus $N$, with a power-law fit (black line) yielding an exponent of 1.8(3). (c) Transfer efficiency  as a function of $N$. Dashed line: mean-field simulations~\cite{SI}. Inset: sketch of the observed $\ket{0,0}_0\rightarrow\ket{\pm1,\pm1}_1$ hopping. (d) One-to-one relation between the number of atoms $N_1$ in $\ket{\pm1,\pm1}_1$ and the photon pulse area $\mathcal{A}$, obtained by fitting the photon traces. A linear fit (solid) yields a slope of  $1.09(2)$, compatible to the expectation of 1 (dashed line) within the combined systematic uncertainty due to photon  (0.07) and atom number calibrations (0.04). Here, $\Delta_c= - 2\pi\cdot 1.4(1)~\text{MHz}$ and $\omega_0=2\pi\cdot 26(1)~\text{kHz}$. Throughout this work, the error bars represent the standard error of the mean.}
	\label{fig:Fig2}
\end{figure}

\begin{figure*} 
	\centering
	\includegraphics[width=1.9\columnwidth]{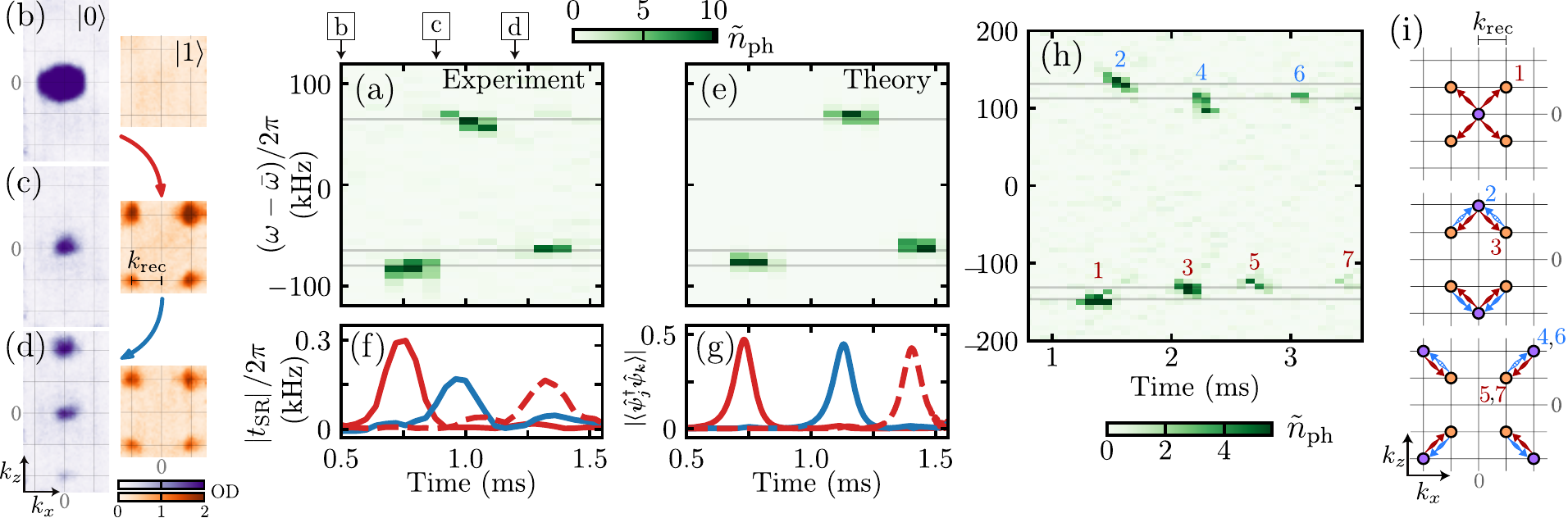}
	\caption{\textbf{Readout of lattice dynamics from the spectrum of the cavity field}. (a)~Representative photon number spectrogram $\tilde{n}(t,\omega)$ displaying time and frequency-resolved superradiant pulses. (b-d)~Spin-resolved time-of-flight (TOF) images of the momentum distribution at different stages of the evolution [see square labels in (a)]. (e) Spectrogram obtained from mean-field simulations. (f)~Experimental tunneling amplitudes $\abs{t_\text{SR}}$,  integrated in a $\delta\omega= 2\pi\cdot5~\text{kHz}$ window around the first (solid red), second (solid blue) and third photon peak (dashed red line) of the spectrogram in (a), respectively. (g)~Corresponding simulated two-mode coherences $\abs*{\ev*{\hat{\psi}^\dagger_1\hat{\psi}_0}}$, $\abs*{\ev*{\hat{\psi}^\dagger_3\hat{\psi}_1}}$ and $\abs*{\ev*{\hat{\psi}^\dagger_2\hat{\psi}_3}}$. (h) Spectrogram displaying multiple tunneling events, and (i) schematics of the inferred currents (1-7). The horizontal lines in the spectrograms indicate the pulse frequencies expected from Eq.~\eqref{eq:Frequency emission}. We prepare $N=1.06(2)\cdot10^5$ atoms in $\ket{0}$.  For (a-d), the couplings are increased to $\eta=2\pi\cdot0.62(2)~\text{kHz}$ within $t_\text{ramp}=1.5~\text{ms}$, while $\Delta_c= - 2\pi\cdot 0.7(2)~\text{MHz}$ and $\omega_0= 2\pi\cdot 72.5(5)~\text{kHz}$. For (h), $\eta=2\pi\cdot1.2(1)~\text{kHz}$, $t_\text{ramp}=5~\text{ms}$, $\Delta_c= - 2\pi\cdot 1.2(2)~\text{MHz}$ and $\omega_0= 2\pi\cdot 139(4)~\text{kHz}$.} %For the simulations, we set $\Gamma=2\pi\cdot 250~$Hz.
	\label{fig:Fig3}
\end{figure*} 

%(f) Simulated population dynamics of states $\ket{0,0}_0$ (solid purple), $\ket{\pm 1,\pm 1}_1$ (solid orange), $\ket{0,\mp 2}_0$ (dashed orange) and $\ket{\pm 1,\mp 1}_1$ (dashed purple curve).
In a first set of experiments, we prepare $N$ atoms in the central lattice site $\ket{0,0}_0$ and characterize the first tunneling event, which populates a symmetric superposition of nearest-neighboring sites $\ket{\pm1,\pm1}_1=1/2\sum_{l,m=\pm1}\ket{l,m}_1$. The strength of the drives is increased to $\eta=2\pi\cdot0.35(1)~\text{kHz}$  within 10~ms and population transfer is signaled by a single pulse of the cavity field, cf. Fig.~\ref{fig:Fig2}(a). We verify the superradiant nature of the scattered field by increasing the initial atom number $N$~\cite{Norcia_2016,Hemmerich_2019,Ferioli_2021}. As a result, the light pulse occurs at shorter times and its amplitude increases superlinearly,  cf. Fig.~\ref{fig:Fig2}(b)~\cite{Haroche_1982,Mandel_1995}. We infer peak tunneling strengths ranging between max$(\abs{t_\text{SR}}/2\pi)=0.2(1)$-$0.4(1)~\text{kHz}$ for different atom numbers. Tunneling occurs only above a finite atom number, which we attribute to the combined influence of residual spin dephasing and self-trapping due to finite contact interactions~\cite{An_2018,SI}. Accordingly, we reach transfer efficiencies of up to $0.8$ [Fig.~\ref{fig:Fig2}(c)], which are well captured by few-mode mean-field simulations~\cite{SI}. The conservation of total angular momentum in the light-matter system leads to a one-to-one correspondence between the number of transferred atoms $N_1$ and the photon pulse area $\mathcal{A}=2\kappa\int_0^\infty n_\mathrm{ph}dt$. We experimentally verify the relation $N_1=\mathcal{A}$~\cite{SI} in Fig.~\ref{fig:Fig2}(d). 
%The latter is extracted by fitting the data with hyperbolic secants, the functional form derived from a semi-classical model~\cite{SI}.

Next, we further leverage on the real-time readout of the cavity field and demonstrate how frequency-resolved measurements allow us to localize the currents in the momentum grid. Around its peak value, the frequency of the cavity field $\omega_{01}$ ($\omega_{10}$) associated with transitions from $\ket{0}$ to $\ket{1}$ ($\ket{1}$ to $\ket{0}$) follows from energy conservation
\begin{equation}
\omega_{\small\substack{01 \\ 10}}-\bar{\omega}=\mp\omega_0-[\omega^\text{kin}(l_f,m_f)-\omega^\text{kin}(l_i,m_i)],
\label{eq:Frequency emission}
\end{equation}
where the indices $l_i,m_i$ ($l_f,m_f$) label the initial (final) state of a given tunneling process~\cite{SI}. In the rotating frame  at $\bar{\omega}$, this corresponds to phase-modulated tunneling rates $t_\text{SR}\propto\sqrt{n_\text{ph}(t)}\exp(i\omega t)$ which remove ($\omega=\omega_{10}-\bar{\omega}$) or provide ($\omega=\omega_{01}-\bar{\omega}$) energy to reach different atomic configurations~\cite{Arimondo_2008}.

To assess this frequency dependence,  we expose the system to stronger drives which results in several superradiant tunneling events connecting multiple lattice sites. In Fig.~\ref{fig:Fig3}(a), we present a representative spectrogram of the cavity field displaying three superradiant pulses, which we attribute to specific tunneling events in the momentum lattice. These are $\ket{0,0}_0\rightarrow\ket{\pm1,\pm1}_1\rightarrow\ket{0,\mp2}_0\rightarrow\ket{\pm1,\mp1}_1$, with $\ket{0,\mp2}_0 =i/\sqrt{2}(\ket{0,-2}_0-\ket{0,2}_0$) and $\ket{\pm1,\mp1}_1=-i/2\sum_{l,m=\pm1}m\ket{\l,m}_1$, with the corresponding creation operators defined as $\hat{\psi}^\dagger_j$ ($j=\{0,1,3,2\}$)~\cite{SI}. The observed frequencies of emission agree with Eq.~\eqref{eq:Frequency emission} and with the results obtained from few-mode numerical simulations, see Fig.~\ref{fig:Fig3}(e).  We verify the involvement of the aforementioned states by performing spin-resolved measurements of the momentum distribution at different stages of the evolution [Fig.~\ref{fig:Fig3}(b-d)]. The observed population imbalance between states with $k_z>0$ and $k_z<0$ is attributed to spurious optical losses in the retroreflected path of the standing-wave drives ($-z$-direction). We further benchmark the dynamics with \textit{ab initio} Gross-Pitaevskii simulations (GPS) including the effects of the harmonic confinement and contact interactions~\cite{SI}. The presence of tunneling terms with opposite signs ($\pm it_\text{SR}$) can give rise to destructive path interference when hopping towards inner sites in $z$-direction. In particular, this effect is reflected in the suppressed hopping $\ket{\pm1,\pm1}_1\not\rightarrow\ket{0,0}_0$ [see Fig.~\ref{fig:Fig3}(d)]. The emerging cavity field and the overall tunneling strength depend, in principle, on the sum of two-site coherences in Eq.~\eqref{eq:Hamiltonian}~\cite{SI}. However, each tunneling event is associated with a well-defined spectral component of the cavity field fulfilling energy conservation. For sufficiently small tunneling rates ($\abs{t_\text{SR}}\ll \omega_\text{rec}$), this field solely induces coherences between the corresponding adjacent lattice sites and the system exhibits local dynamical tunneling. This is reflected by the simulations of the coherences [Fig.~\ref{fig:Fig3}(g)], which are compatible with the experimentally determined tunneling amplitudes $\abs{t_\text{SR}}$ associated with each of the superradiant pulses [see Fig.~\ref{fig:Fig3}(f)]. 

%$\ev*{\hat{\psi}^\dagger_j\hat{\psi}_k}$

%Starting from certain lattice configurations, the presence of tunneling rates with different signs ($\pm it_\text{SR}$) gives rise to destructive path interference, which prevents some lattice sites from being accessed. In particular, this effect is reflected in the suppressed population transfer from $\ket{\pm1,\pm1}_1$ to $\ket{0,0}_0$ [see Fig.~\ref{fig:Fig3}(d)].

%The overall evolution is compatible with the simulated population dynamics displayed in Fig.~\ref{fig:Fig3}(f), which exhibits increasingly less complete population transfers in the momentum lattice

The number of tunneling events can be extended by further increasing the coupling strength. As shown in Fig.~\ref{fig:Fig3}(h,i), we observe up to seven superradiant transfers involving outer lattice sites, such as $\ket{2,2}_0$, which we identify by reading out the frequency of the cavity field and employing Eq.~\eqref{eq:Frequency emission}. The tunneling events are not restricted to the shown processes as they arise from multiple competing quantum paths. A quantitative prediction in this regime goes beyond the scope of this work. We identify the following fundamental limitations to the number of tunneling events.  First, in the absence of confining lattice potentials, the momentum states move out of the grid nodes due to oscillatory motion in the trap~\cite{Gadway_2016,AnPhD_2020}. While for noninteracting systems this rate is solely determined by the trap frequencies ($\sim\!2\pi\cdot200~\text{Hz}$), our GPS indicate that repulsive contact interactions effectively increase the lifetime of the momentum lattice~\cite{SI}. Second, heating of the BEC from off-resonant spontaneous emission progressively melts the momentum lattice when approaching the recoil temperature. However, this effect is negligible within the duration of our experiments.
%$T_\mathrm{rec} \approx 180 ~\text{nK}$

%, resulting in a breakdown of the tight-binding description close to the recoil temperature

\begin{figure}[thbp]
	\centering
	\includegraphics[width=0.95\columnwidth]{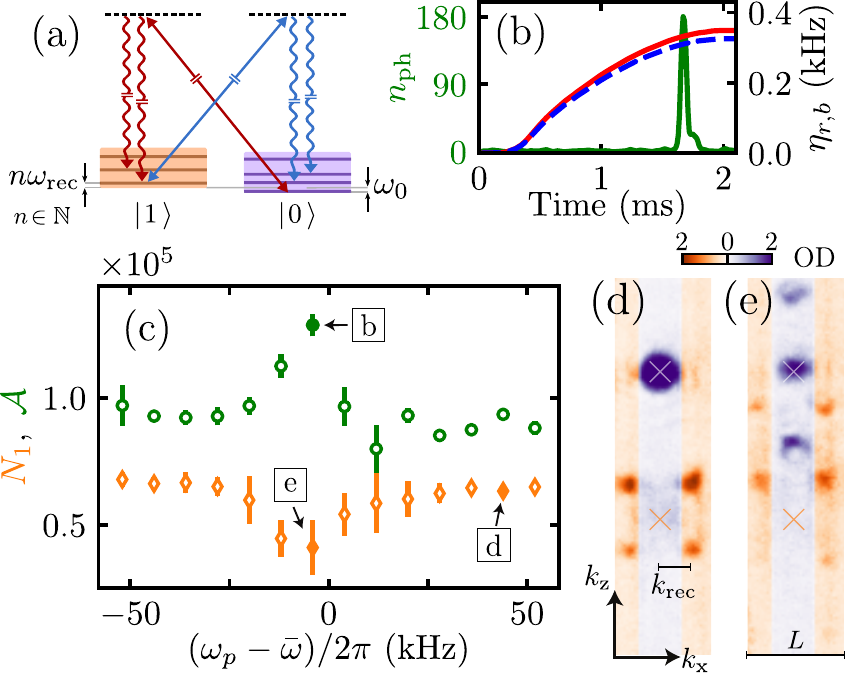}
	\caption{\textbf{Cascaded lattice dynamics}. (a) Coupling scheme. The global splitting $\omega_0$ is reduced below $\omega_\text{rec}$, shifting several lattice sites close to degeneracy in the rotating frame.  (b) Representative photon pulse (green). The couplings $\eta_{r,b}$ (red, blue curves) are increased with a small imbalance $\eta_r/(\eta_r+\eta_b)=0.034(3)$. (c) Photon excess measurement. Pulse area $\mathcal{A}$ (circles) and final number of atoms $N_1$ in $\ket{\pm 1, \pm 1}_1$ (diamonds) as functions of $\omega_{p}$. (d-e) Representative TOF images, with the white (orange) cross denoting the position of $\ket{0,0}_0$ ($\ket{0,0}_1$), which are separated by a Stern-Gerlach gradient along $z$. The purple (orange) color map indicates regions solely populated by atoms in $\ket{0}$ ($\ket{1}$). The small distance between the cavity mirrors $L\approx175~\mu\text{m}$ limits the both field of view along $x$-direction to $k_x \lesssim k_\text{rec}$. The square labels in (c) indicate the data points corresponding to panels (b), (d) and (e). For these measurements,  $\Delta_c= - 2\pi\cdot 3.4(2)~\text{MHz}$ and $N=9.1(1)\cdot10^4$.}
	\label{fig:Fig4}
\end{figure}

The observations discussed so far involve independent tunneling events occurring sequentially in time. Our scheme can be extended to generate cascaded dynamics, where the tunneling events between different sites stimulate each other. We reduce the offset between the two spin manifolds to values comparable to the recoil frequency, shifting multiple states in the momentum lattice close to degeneracy [Fig.~\ref{fig:Fig4}(a)]. In Fig.~\ref{fig:Fig4}(b), we observe a single strong emission in the cavity field that is accompanied by several tunneling events within the pulse duration. Different from the results in Fig.~\ref{fig:Fig2}(d),  we observe an  excess of detected photons in comparison to the population in $\ket{\pm1,\pm1}_1$, see Fig.~\ref{fig:Fig4}(c). This effect is amplified as the  emission frequency $\omega_{p}$ approaches the two-photon resonance ($\omega_{p} - \bar{\omega} \rightarrow 0$). Concurrently, we observe states with up to $10\hbar\omega_\mathrm{rec}$ kinetic energy in the time-of-flight images [see Fig.~\ref{fig:Fig4}(d,e)]. The GPS reproduce these results, helping to discern a cascade of hopping events towards outer lattice sites, in which the next tunneling starts before the previous one finishes~\cite{SI}. These findings indicate that, within the duration of the cavity pulse, finite coherences between multiple lattice sites are simultaneously established, in contrast to the subsequent tunneling events at larger $\omega_0$. 

%such as $\ev*{\hat{c}^{1\dagger}_{(1,1)}\hat{c}^{0}_{(0,0)}}$, $\ev*{\hat{c}^{0\dagger}_{(0,2)}\hat{c}^{1}_{(1,1)}}$ and $\ev*{\hat{c}^{1\dagger}_{(1,3)}\hat{c}^{0}_{(0,2)}}$,

We experimentally demonstrated a scheme  giving rise to self-consistent tunneling in a non-Hermitian momentum grid, engineered by superradiant Raman scattering of a spinor BEC coupled to an optical cavity. In particular, the tunneling rates evolve together with the coherences between the sites participating in each hopping event, which we locally resolve  via frequency-selective measurements of the leaking cavity field. As an extension, the combination of real-time probing and continuous feedback acting on the phase of the drives~\cite{Kroeger_2020} could facilitate the realization of nontrivial tunneling phases and pave the way to observe synthetic magnetic fields and topologically protected states in non-Hermitian settings~\cite{Gong_2018,Ozawa_2019,Yan_2020}. In particular, an extension to running-wave Raman drives can result in emergent spin-orbit coupling~\cite{Lev_2019,Kollath_2019,Farokh_2021}. Finally, exploring the interplay between cavity-assisted tunneling and Bose-Hubbard physics~\cite{Landig_2016} holds the potential to realize unconventional strongly-correlated phases and dynamics~\cite{Halati_2017,Chanda_2021,Colela_2021}. 
\newline

We are grateful to A. Frank for electronic support, to M. Landini for discussions and early contributions to the project,  and to  K. Viebahn, A.U.J. Lode and O. Zilberberg for fruitful discussions. We acknowledge computational time on the ETH Euler cluster. R.R-M., F. Ferri, F. Finger, T. D., and T. E. acknowledge funding from the Swiss National Science Foundation: Project No. 182650 and No. 175329 (NAQUAS QuantERA) and NCCR QSIT (Grant No. 51NF40-185902), from EU Horizon2020: ERC advanced grant TransQ (project Number 742579). R.L. and R.C. acknowledge funding from the ETH grants.
\newline

\let\oldaddcontentsline\addcontentsline% Store \addcontentsline
\renewcommand{\addcontentsline}[3]{}% Make \addcontentsline a no-op

%\color{black}
%\bibliography{References}
%\bibliographystyle{apsrev}

\let\addcontentsline\oldaddcontentsline% Restore \addcontentsline

\beginsupplement
\onecolumngrid
%\setstretch{1.25}
\newpage
\begin{center}
	\large
	\textbf{Supplemental Material}
\end{center}
\normalsize
\tableofcontents
\normalsize

\section{Experimental details}
In this section, we discuss the relevant experimental protocols and their calibration. In particular, we characterize the erased lattice configuration due to the two phase-shifted drives, and provide details on the heterodyne detection of the cavity field and the absorption imaging.   

\subsection{BEC preparation, B-field and transverse pumps characterization}\label{sec:BECprep}
We prepare a Bose-Einstein condensate (BEC) of $^{87}$Rb atoms in the $\ket{F=1,m_F=-1}$ magnetic sublevel of the $5^2$S$_{1/2}$ electronic ground state. The BEC is confined in the cavity mode by a crossed-beam optical dipole trap with frequencies  $[\omega_\text{hx},\omega_\text{hy}, \omega_\text{hz}]=2\pi\cdot[175(4),29(1),172(1)]\text{Hz}$. We apply a magnetic field along the z-direction $\bold{B}=B_z\bold{e}_z$ ($B_z <0$) and characterize the Zeeman splitting $\omega_z$ between the $m_F=-1$ and $m_F=0$ sublevels using cavity-assisted Raman transtions~\cite{Ferri_2021}.

We derive the two transverse pump (TP) drives from the same laser and adjust their frequencies ($\omega_{r,b}$) via independent double-pass acousto optical modulators. A small fraction of each TP is split before recombining both beams, in order to regulate the intensity in each path. The lattice depth associated to each drive is calibrated by means of Kapitza-Dirac diffraction~\cite{Gadway_2009}. For all measurements discussed in this work, we increase the pump powers via s-shaped ramps of the form $V_{r,b}(t)=\tilde{V}_{r,b}\left[3(t/t_r)^2-2(t/t_r)^3\right]$, with  $t_r$ and $\tilde{V}_{r,b}$ being the ramp duration and the final power of each drive, respectively. The frequencies of the two drives differ by $\omega_b-\omega_r\approx 2\omega_z=~2\pi\cdot96~\text{MHz}$; their wavelength is $\lambda=784.7~\text{nm}$ which is associated to a recoil frequency of $\omega_\text{rec} = 2\pi \cdot 3.73 ~\text{kHz} $ for $^{87}$Rb atoms. Since the drives are Gaussian beams which are red detuned with respect to the effective atomic resonance, they increase the harmonic confinement in the xy-plane. For representative experimental lattice depths of $\sim 15~\hbar \omega_\text{rec}$ per drive, we measure the trap frequencies $[\omega_\text{hx},\omega_\text{hy}]=2\pi\cdot[218(4),165(2)]~\text{Hz}.$ More details on the experimental setup can be found in Ref.~\cite{Ferri_2021}.

\subsection{Erased lattice configuration}\label{Sec:ErasedLattice}
\begin{figure}[thbp]
	\centering
	\includegraphics[width=0.36\columnwidth]{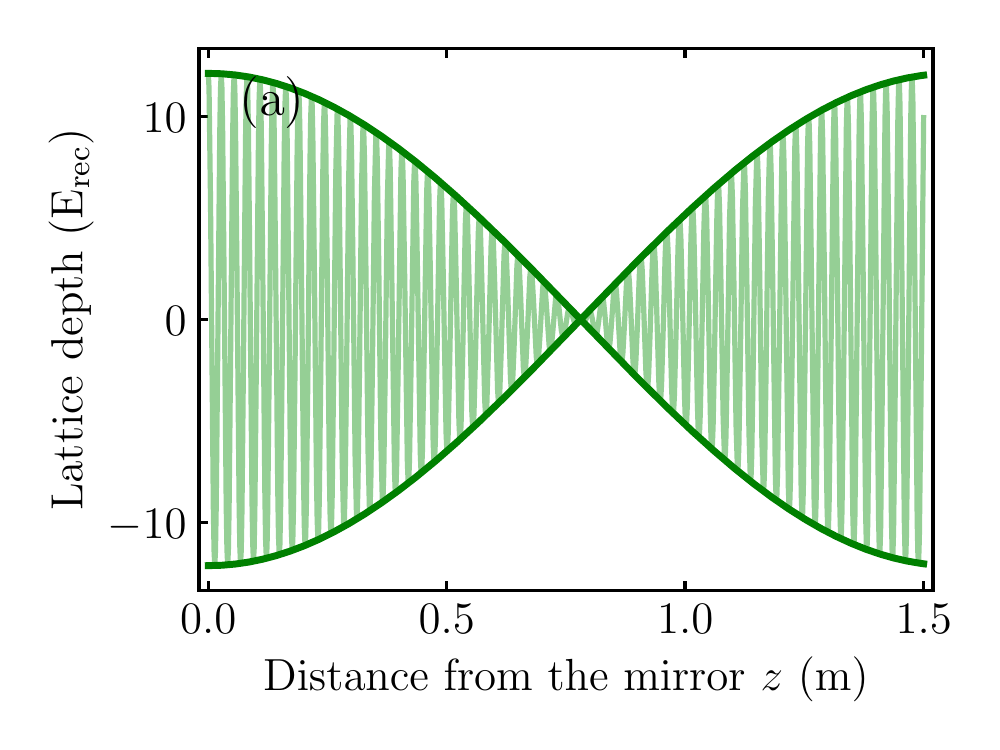}
	\includegraphics[width=0.36\columnwidth]{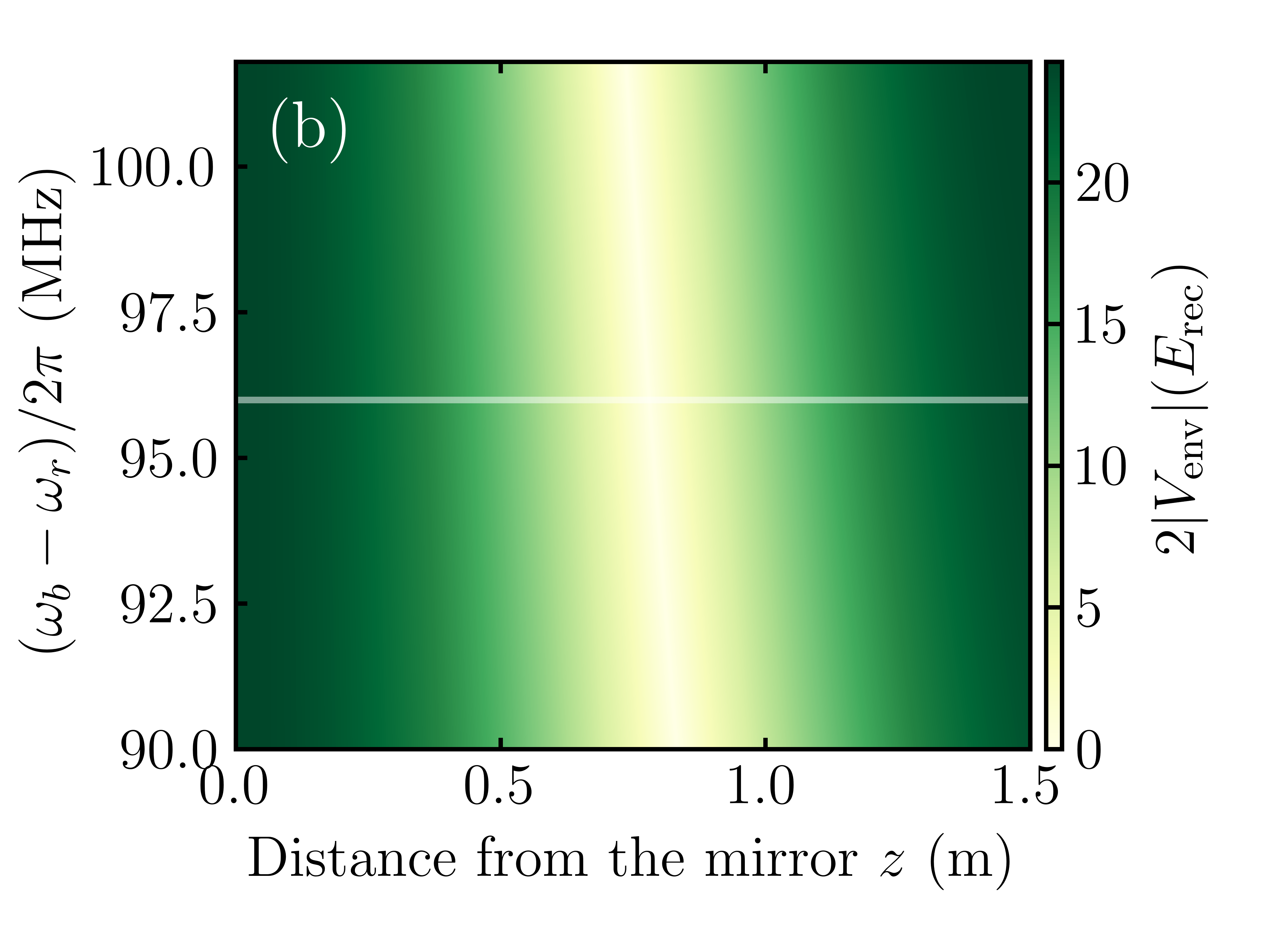}
	\caption{Principle and parameter dependence of the erased lattice configuration. (a) Combined lattice generated by both transverse pumps (light green) and the lattice envelope (dark green) at a frequency difference between the drives of $\omega_b-\omega_r=2\pi\cdot\SI{96}{\mega\hertz}$ and  $V=12~E_\text{rec}$. The physically irrelevant offset $V$ in Eq.~\eqref{eq:beat_note} is removed and the combined lattice is undersampled for better visibility. For a distance of $z\approx\SI{0.78}{\meter}$ the envelope vanishes and the combined lattice is constant over the extent of the atomic cloud. (b) Lattice envelope as a function of mirror distance and frequency difference (shifted by an offset of $V$). The gray line marks the experimentally relevant cut shown in (a).} 
	\label{figSI:ErasedLatticeTheory}
\end{figure}

We adjust the distance between the retro-reflecting mirror and the atomic cloud, such that the standing-wave modulations of the two laser drives are opposite. In this configuration, the corresponding lattice potentials are erased if the power of the two drives are balanced. To obtain the optimal distance, we consider the two $x$-polarized transverse drives as classical standing waves with field amplitudes $E_{r,b}$, frequencies $\omega_{r,b}$ and wavevectors $k_{r,b}=\omega_{r,b}/c$ propagating in $z$-direction.
The spatial phase reference for both fields is given by the position of the retro-reflecting mirror.
The negative part of the combined electric field $\bold{E^{(-)}}$ is given by
\begin{equation}
	\bold{E^{(-)}}=\frac{E_{r}}{2}\cos(k_rz)\bold{e}_xe^{-i\omega_rt} + \frac{E_{b}}{2}\cos(k_bz)\bold{e}_x e^{-i\omega_bt}.
\end{equation}
%with unit vector $\bold{e}_x$.
As derived below in Eq.~\eqref{eqSI:H1_scalar}, this electric field results in an optical lattice potential with a maximal amplitude modulated in space by the beat-note between the drives
\begin{equation}
	\begin{aligned}
		V_\text{tot}(z) &= \frac{\alpha_s}{4}\qty[E_r^2\cos[2](\frac{\omega_r}{c}z) + E_b^2\cos[2](\frac{\omega_b}{c}z)] = \underbrace{V\cos(\frac{\omega_b-\omega_r}{c}z)}_{V_\text{env}}\cos(\frac{\omega_b+\omega_r}{c}z) + V,
		\label{eq:beat_note}
	\end{aligned}
\end{equation}
where $\alpha_s$ is the scalar polarizability at the frequency of the driving lasers~\cite{Le_2013,Landini_2018} and  $V=(\alpha_s/4)E^2$ is the maximal lattice depth per drive in a balanced configuration, i.e., $E_r^2=E_b^2=E^2$. The expression in Eq.~\eqref{eq:beat_note} comprises a rapidly varying $\lambda/2$-periodic lattice potential, with $\lambda=2\pi c/(\omega_b+\omega_r)$, and a slowly changing envelope $V_\text{env}$. In Fig.~\ref{figSI:ErasedLatticeTheory}(a), we plot the resulting potential as a function of the distance $z$ between the retro-reflecting mirror and the atomic cloud for the experimentally relevant frequency difference between the drives $\omega_b-\omega_r=2\pi\cdot\SI{96}{\mega\hertz}$.

For an optimal distance of $z_\text{opt}=\SI{0.78}{\meter}$, the envelope  vanishes and the lattice potential is suppressed.  Since variations of the envelope are negligible within the extent of the atomic cloud ($\sim \SI{10}{\micro\meter}$), the potential can be considered to be  constant within this scale. The resulting erased lattice motivates the definition of spatial mode profiles for both pumps at the position of the atoms $f_r(\bold{x})=\cos(kz)$ and $f_b(\bold{x})=\sin(kz)$, with $k=(k_r+k_b)/2$, such that $f_r(\bold{x})^2 + f_b(\bold{x})^2=1 $. For completeness, we show the full parameter dependence of the lattice envelope on the frequency difference between the drives and the distance of the atoms from the mirror in Fig.~\ref{figSI:ErasedLatticeTheory}(b).

\begin{figure}[thbp]
	\centering
	\includegraphics[width=0.36\columnwidth]{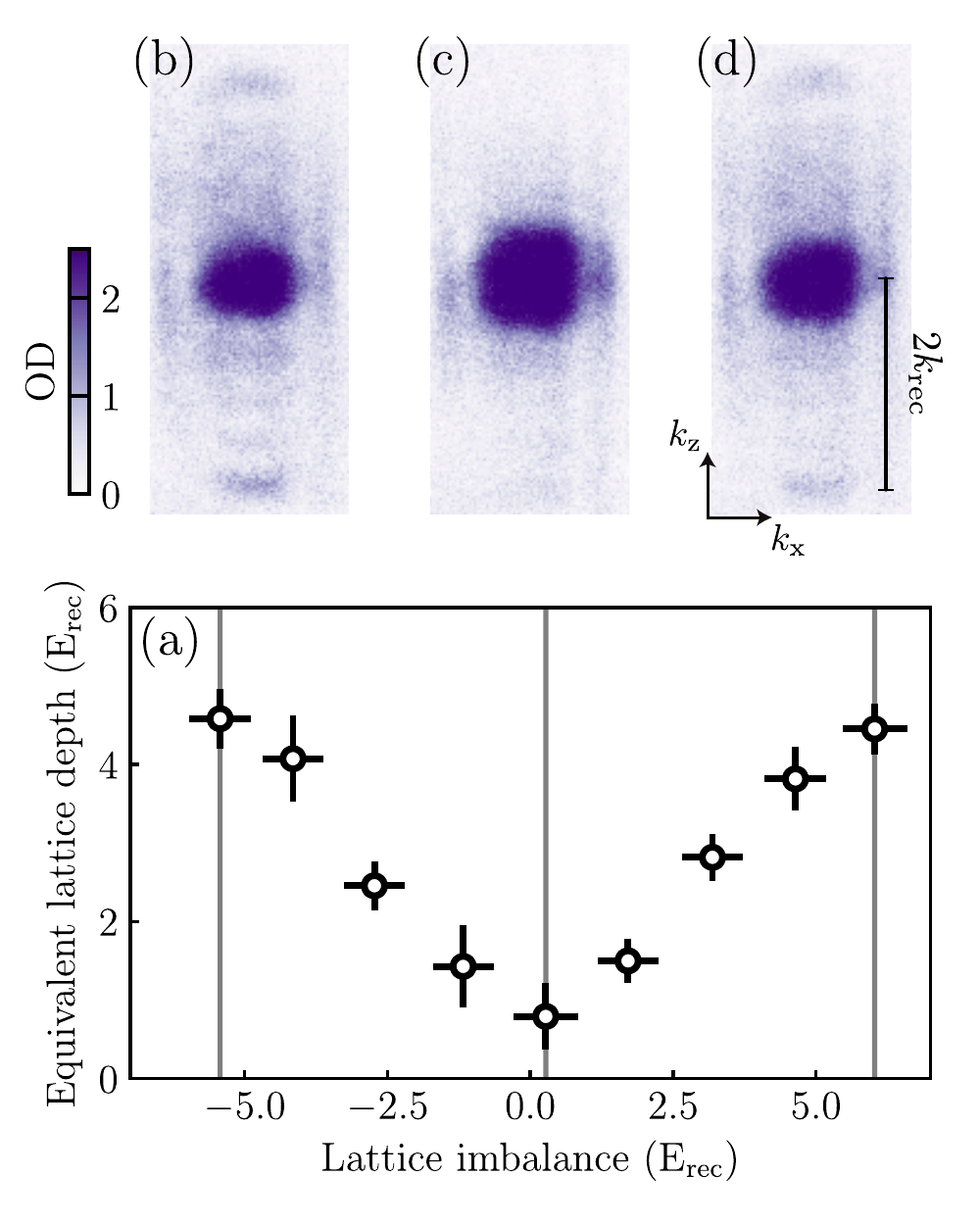}
	\caption{(a) Equivalent lattice depth in the erased lattice configuration as a function of pump imbalance. The vertical error bars include the standard error of the mean and a systematic uncertainty of the absorption imaging, while the horizontal ones represent the estimated accuracy for setting the lattice imbalance ($0.5~E_\text{rec}$). (b-d) Representative time-of-flight images of the atomic cloud showing population in the momentum states $k_z=0,~-2k_\text{rec}$ and $+2k_\text{rec}$ of a BEC in $m_F=-1$.  The corresponding imbalances are marked with gray lines in (a).} 
	\label{figSI:ErasedLatticeExperiment}
\end{figure}

To experimentally assess the quality of the erased lattice configuration at the optimal distance of $z_\text{opt}\approx 0.78~\text{m}$, we ramp up both drives to a total lattice of  $24.2(1)~$E$_\text{rec}$ within $\SI{20}{\milli\second}$ and vary the imbalance between the drives. We measure the population in the momentum states $k_z=\pm2k_\text{rec}$ after a sudden switch-off of all confining potentials, with $k_\text{rec}\defeq k$ being the recoil momentum.
By performing an analogous protocol with a single drive, we can convert the measured populations into equivalent lattice depths.
The equivalent lattice depth for different nominal imbalances is shown in Fig.~\ref{figSI:ErasedLatticeExperiment}. For optimally balanced drives with a small nominal imbalance of $0.5(4)~$E$_\text{rec}$, we measure a small residual lattice depth of $0.8(4)~$E$_\text{rec}$ which indicates a suppression of the lattice modulation by a factor $>30$. The nominal and measured lattice imbalances deviate more at larger values, which we attribute to non-perfect atom counting in increasingly heated clouds.

%We attribute the small residual lattice to deviations from the optimal distance in the experimental setup, as well as to systematic uncertainties in the evaluation of the absorption images (see section \hyperref[sec:Absorption_imaging]{\textit{Absorption imaging}}).
%	The small deviation from a one-to-one correspondence between the equivalent lattice depth and the lattice imbalance can be attributed to finite power fluctuations during each measurement.

\FloatBarrier
\subsection{Heterodyne detection and photon field spectrograms}\label{sec:Heterodyne}
The leaking cavity field is monitored by two polarization-selective heterodyne setups, which measure the $y$- and $z$-polarization modes, respectively. While the former is employed to probe the cavity resonance after each experimental run, the latter is used to measure the cavity-mediated Raman transfers discussed in this work. Its high bandwidth ($250~\text{MS/s}$) allows for an all-digital demodulation of the beat-note between the cavity field and an optical local oscillator. We estimate a systematic uncertainty in the photon number calibration on the order $\Delta n_\text{ph,s}/n_\text{ph}=0.068$, which arises from the relative fluctuations in the heterodyne detector ($3.8 \% $) and the nominal accuracy of the power sensor used for the calibration ($5.0 \% $). 

The intra-cavity field $\alpha(t)=X(t)-iY(t)$ is obtained from the real $X(t)$ and imaginary  quadrature $Y(t)$ after digital demodulation. We compute spectrograms of the cavity field using fast Fourier-transforms $\text{FFT}[\alpha](f)=dt/\sqrt{\tilde{N}}\sum_i \alpha^*(t_i) e^{-i2\pi f t_i}$ \cite{Dogra_2019}, where $t_i$ is the time of the $i^\text{th}$ step and $\tilde{N}$ is the total number of steps in an integration window. The traces are divided in time intervals of $T=\SI{150}{\micro\second}$ with $\SI{50}{\percent}$ overlap to subsequent intervals. Finally, the photon number spectrograms are calculated as $\tilde{n}_\text{ph}(f) =|\text{FFT}[\alpha](f)|^2/T$.

%power spectral density PSD($f$)=$|\text{FFT}(\alpha)|^2(f)$ 

\subsection{Absorption imaging}\label{sec:Absorption_imaging}

We measure the momentum space distribution of the atoms by imaging the cloud after $8~\text{ms}$ of free time-of-flight expansion (TOF). In order to spatially resolve atoms populating different magnetic sublevels in $F=1$, we apply a magnetic field gradient along $z$-direction during TOF (Stern-Gerlach separation). We perform high intensity absorption imaging~\cite{Reinaudi_2007} and estimate a systematic uncertainty in the atom number calibration on the order $\Delta N/N\approx~\!0.043$. This arises from uncertainties in the magnification of the imaging setup ($1.0\%$) and in the determination of the effective saturation intensity ($4.1\%$).

\section{Derivation of the Hamiltonian}

In this section, we derive both the single particle and many-body Hamiltonian. The latter constitutes a tight-binding description of the system in momentum space. We discuss the role of contact interactions within this framework.

\label{sec:Hamiltonian}
\subsection{Single particle Hamiltonian}
\label{sec:SP_Hamiltonian}

The Hamiltonian of a single atom coupled to the cavity mode reads
\begin{equation}
	\hat{H}'_\text{1}=\hat{H}'_\text{at} + \hat{H}'_\text{cav} + \hat{H}'_\text{int},
	\label{eq:H_SP}
\end{equation}
with the bare cavity Hamiltonian described by

\begin{equation}
	\hat{H}'_\text{cav} = \hbar\omega_c \hat{a}^\dagger \hat{a},
\end{equation}
where the operator $\hat{a}^\dagger$ creates photons in the TEM$_{00}$ $z$-polarized  cavity mode with resonance frequency $\omega_c$. Since the experiment operates in the dispersive regime \cite{Goldman_2014},  we adiabatically eliminate the excited electronic states of the atoms. The atoms are initialized in $\ket{F=1, m_F=-1}$ and coupled to $\ket{F=1, m_F=0}$ by near-resonant cavity-assisted Raman transitions. We neglect transitions to the $F=2$ manifold as they are detuned by the hyperfine splitting $\omega_\mathrm{HF}=2\pi\cdot6.834~$GHz. Thus, we can write the atomic Hamiltonian in terms of the spin operator $\bold{\hat{F}}=(\hat{F}_x,\hat{F}_y,\hat{F}_z)^T$ in $F=1$
\begin{equation}
	\hat{H}'_\text{at} = \frac{\hat{p}^2}{2M} + \hbar{\omega_z^{(1)}}\hat{F}_z + \hbar\omega_z^{(2)}\hat{F}_z^2.
	\label{eqSI:H_at}
\end{equation}
The energy difference between the different Zeeman sublevels is determined by the first- and second-order Zeeman shifts $\hbar\omega_z^{(1)}<0$ and $\hbar\omega_z^{(2)}>0$~\cite{Stamper-Kurn_2013}. Moreover, we neglect the effect of external confining potentials at this point.

The atom-light interactions in the dispersive regime are given by  
\begin{equation}
	\hat{H}'_\text{int} = \alpha_s\bold{\hat{E}^{(+)}}\cdot\bold{\hat{E}^{(-)}} - i\frac{\alpha_v}{2F}\left[\bold{\hat{E}^{(+)}}\times\bold{\hat{E}}^{(-)}\right]\cdot\bold{\hat{F}},
	\label{eqSI:H_int}
\end{equation}
where $\bold{\hat{E}^{(\pm)}}$ are the positive and negative components of the electric field at the position of the atoms and $\alpha_s$ ($\alpha_v$) is the scalar (vectorial) polarizability of the atoms at the frequency of the driving lasers \cite{Goldman_2014,Le_2013,Landini_2018,Ferri_2021}. We consider classical fields for the standing-wave transverse pumps and a quantized field for the cavity mode. The negative part of the total electric field $\bold{\hat{E}^{(-)}}$ is given by
\begin{equation}
	\bold{\hat{E}^{(-)}}=\frac{E_{r}}{2}f_r(\bold{x})\bold{e}_xe^{-i\omega_rt} + \frac{E_{b}}{2}f_b(\bold{x})\bold{e}_x e^{-i\omega_bt} + E_{0}g(\bold{x})\hat{a}\bold{e}_z,%e^{-i\bar{\omega}t},
	\label{eqSI:total_electric_field}
\end{equation}
with unit vectors $\bold{e}_j$ ($j\in\{x,y,z\}$) and spatial mode profiles $f_r(\bold{x})$, $f_b(\bold{x})$, $g(\bold{x})$. The two laser drives with electric field amplitudes $E_r$ and $E_b$ give rise to standing wave modulations which are phase shifted by $\lambda/4$ at the position of the atoms, as discussed in the subsection~\hyperref[Sec:ErasedLattice]{\textit{Erased lattice configuration}}. Given their small frequency difference $\omega_b-\omega_r=2\pi\cdot96~$MHz, we consider the same wavevector $k=\bar{\omega}/c$ for the two drives, with $\bar{\omega}=(\omega_b + \omega_r)/2$.  By neglecting the transverse Gaussian modulation of the optical beams, we assume that the mode profile of the drive at frequency $\omega_r$ [$\omega_b$] is given by $f_r(\bold{x})=\cos(kz)$ [$f_b(\bold{x})=\sin(kz)$], while the cavity mode profile is $g(\bold{x})=\cos(kx)$. The cavity field amplitude per photon $E_{0}=403~$V/m is determined by the resonance frequency and the volume of the mode. 

We introduce the unitary transformation $\hat{U}=\exp\qty(\frac{i}{\hbar}\hat{H}_\text{rot} t)$ with $\hat{H}_\text{rot}=\hbar\bar{\omega}\hat{a}^\dagger \hat{a} - \hbar\omega_z'\hat{F}_z$ and  $\omega_z'=(\omega_b-\omega_r)/2$. By employing the rotating wave approximation, we obtain a time independent single particle Hamiltonian 
\begin{equation}
	\hat{H}_\text{1} = \hat{H}_\text{at} + \hat{H}_\text{cav} + \hat{H}_\text{s} + \hat{H}_\text{v},
	\label{eqSI:H1}
\end{equation}
where
\begin{equation}
	\hat{H}_\text{at} = \frac{\hat{p}^2}{2M} + \hbar\delta_z\hat{F}_z + \hbar\omega_z^{(2)}\hat{F}_z^2,
	\label{eqSI:H1_at}
\end{equation}
\begin{equation}
	\hat{H}_\text{cav} =-\hbar\Delta_c \hat{a}^\dagger \hat{a},
\end{equation}
with cavity detuning $\Delta_c=\bar{\omega}-\omega_c$ and effective linear Zeeman shift $\delta_z=\omega_z^{(1)}+\omega_z'$. The light-matter interactions have a scalar (vectorial) contribution $\hat{H}_\text{s}$ ($\hat{H}_\text{v}$) given by
\begin{equation}
	\hat{H}_\text{s}=\frac{\alpha_s}{4}\left[E_b^2f_b(\bold{x})^2+E_r^2f_r(\bold{x})^2\right] + \alpha_s E_0^2\hat{a}^\dagger\hat{a}g(\bold{x})^2,
	\label{eqSI:H1_scalar}
\end{equation}

\begin{align}
	\hat{H}_\text{v} &= -\frac{\alpha_v E_r E_0}{8}g(\bold{x})f_r(\bold{x})\left[\hat{a}\hat{F}_- +\hat{a}^\dagger\hat{F}_+ \right]+\frac{\alpha_v E_b E_0}{8} g(\bold{x})f_b(\bold{x})\left[\hat{a}\hat{F}_+ + \hat{a}^\dagger\hat{F}_-\right] \nonumber \\ 
	&= -\hbar\eta_r\cos(kx)\cos(kz)\left[\hat{a}\hat{F}_- +\hat{a}^\dagger\hat{F}_+ \right]+\hbar\eta_b\cos(kx)\sin(kz)\left[\hat{a}\hat{F}_+ +\hat{a}^\dagger\hat{F}_-\right].
	\label{eqSI:H1_vectorial}
\end{align}
In the limit of balanced drives ($E_r^2=E_b^2$), their combined static lattice is erased at the position of the atoms since $f_r(\bold{x})^2+f_b(\bold{x})^2=1$. In Eq.~\eqref{eqSI:H1_vectorial}, we introduce the Raman couplings  $\eta_{r,b}$ arising from the photons scattered from the drive at frequency $\omega_{r,b}$  into the cavity. We relate these couplings to the experimentally measured lattice depth $V_{r,b}=-\alpha_sE^2_{r,b}/4$ of the drives at frequency $\omega_{r,b}$, and find
\begin{equation}
	\eta_{r,b}=\frac{\alpha_v}{8\hbar}E_0E_{r,b}=\frac{\alpha_v}{4\sgn{\alpha_s}\cdot\alpha_s}\sqrt{-\frac{U_0 V_{r,b}}{\hbar}}.
	\label{eqSI:mapping_coupling}
\end{equation}
The maximal dispersive shift per cavity photon is  $U_0=\alpha_sE_0^2/\hbar=- 2\pi\cdot 56.3~\text{Hz}$. For the typical photon numbers $n_\text{ph}=\expval{\hat{a}^\dagger\hat{a}}	\lesssim20$ observed in this work, we estimate a small residual intracavity lattice depth $V_\text{c}=\hbar U_0n_\text{ph}\lesssim 0.3~\hbar\omega_\text{rec}$ which has negligible influence on the dynamics in the momentum lattice. 
%\todo{Fabian $\rightarrow$ please double check that everything is consistent with your notes}

%\hat{H}_\text{v} = \frac{\alpha_v}{8}E_0\left[\left(E_{b} + E_{r}\right)\left(\hat{a}+\hat{a}^\dagger\right)\hat{F}_x + \left(E_{b} - E_{r}\right)i\left(\hat{a}-\hat{a}^\dagger\right)\hat{F}_y\right]f(\hat{\bold{x}})g(\hat{\bold{x}}),

\subsection{Many-body Hamiltonian in momentum space}
We derive a tight-binding description of the system in momentum space. We expand the spinor BEC in a discrete set of two-dimensional plane waves  
\begin{align}
	\psi_{(l,m)}^\sigma (\hat{\bold{x}})= \frac{k}{2\pi}e^{ik(lx+mz)} \otimes \ket{\sigma=m_F+1},
	\label{eqSI:PlaneWaves}
\end{align} 
with the average wavevector of the two drives $k$ determining the recoil momentum of the atoms $k_\text{rec}$ ($k=k_\text{rec}$). The indices $l,m\in \mathbb Z$ and $\sigma\in\{0,1\}$ label a discrete set of plane waves and the spin state associated to the Zeeman sublevel $m_F\in\{-1,0\}$ of $F=1$, respectively. The states are normalized within a unit cell $ (x,z) \in [-\pi/k,\pi/k)\otimes [-\pi/k,\pi/k) \eqdef R$ in real space. As in the main text, we refer to these single particle states as $\ket{l,m}_\sigma$. We neglect cavity-assisted Raman transitions to $m_F=+1$, since they are detuned by $\Delta_{+1}\approx 2\omega_z^{(2)}=2\pi \cdot 0.7~\text{MHz}$ at the magnetic field we operate. The plane waves in Eq.~\eqref{eqSI:PlaneWaves} are eigenstates of the kinetic energy operator with $(\hat{p}^2/2M)~\psi_{(l,m)}^\sigma = (l^2+m^2)\hbar\omega_\text{rec}~\psi_{(l,m)}^\sigma$, where $\omega_\text{rec}=2\pi \cdot 3.73~\text{kHz}$ is the recoil frequency.

In order to obtain a many-body description of the light-matter system, we exploit the fact that the atoms are initialized in a BEC in $m_F=-1$, i.e., in $\psi_{(0,0)}^{0}$, and that each cavity-assisted two-photon scattering event simultaneously changes the spin ($\sigma$) and motional states of the atoms ($l,m$) by $\pm 1$, cf. Hamiltonian in Eq.~\eqref{eqSI:H1_vectorial}. Resorting to second quantization, we expand the atomic field operator as 
\begin{align}
	\hat{\Psi}(\hat{\bold{x}})= \sum_{\{j,k\}\in \mathbb Z} \psi_{(2j,2k)}^0(\hat{\bold{x}})\hat{c}^0_{(2j,2k)} + \psi_{(2j+1,2k+1)}^1(\hat{\bold{x}})\hat{c}^1_{(2j+1,2k+1)} ,
	\label{eqSI:FieldOperator}
\end{align} 
with the bosonic operators $\hat{c}^\sigma_{(2j+\sigma,2k+\sigma)}$ annihilating particles in the mode $\psi_{(2j+\sigma,2k+\sigma)}^\sigma$. We consider the limit of balanced Raman couplings, $\eta\defeq\eta_r=\eta_b$, and derive the following many-body Hamiltonian 
\begin{align}
	\hat{H}=\hat{H}_0 + \hat{H}_\mathrm{int}.
	\label{eqSI:H_MB}
\end{align}
The diagonal term is given by
\begin{align}
	\hat{H}_0 &=\hat{H}_\text{cav} +\int_{R}\hat{\Psi}^\dagger(\hat{\bold{x}})\left[\hat{H}_\text{s} + \hat{H}_\text{at}\right] \hat{\Psi}(\hat{\bold{x}}) d\bold{x} \nonumber \\
	&=-\hbar\tilde{\Delta}_c \hat{a}^\dagger\hat{a} + \sum_{\substack{\{j,k\}\in\mathbb Z \\ \sigma\in\{0,1\}}} \hbar[\sigma\omega_0+\omega^\text{kin}_{(2j+\sigma,2k+\sigma)}]\hat{c}_{(2j+\sigma, 2k+\sigma)}^{\sigma\dagger} \hat{c}^\sigma_{(2j+\sigma, 2k+\sigma)} , 
\end{align}
where we introduce the dispersively shifted cavity detuning $\tilde{\Delta}_c=\Delta_c-NU_0/2$, where $U_0$ is the average dispersive shift per atom. Since the cavity detuning $\Delta_c$ is large enough, we neglect the dependence of the dispersive shift on the momentum distribution. In the main text, we redefine the detuning as $\tilde{\Delta}_c\rightarrow\Delta_c$ for better clarity. The energy offset between the different atomic modes arises from a kinetic contribution $\omega^\text{kin}_{(2j+\sigma,2k+\sigma)}=[(2j+\sigma)^2+(2k+\sigma)^2]\omega_\mathrm{rec}$ and a global splitting between the two spin manifolds $\omega_0=\omega_z'+\omega_z^{(1)}-\omega_z^{(2)}$, which is on the order of $\sim 2\pi \cdot 100~\text{kHz}$. Moreover, the light-matter interactions

\begin{align}
	\hat{H}_\mathrm{t_{SR}} =& \int_{R}\hat{\Psi}^\dagger(\hat{\bold{x}})\hat{H}_\text{v}  \hat{\Psi}(\hat{\bold{x}}) d\bold{x} \nonumber \\ 
	=&-\frac{\hbar\eta}{\sqrt{8}} \sum_{\substack{\{j,k\}\in\mathbb Z \\ s_{1,2}=\pm1}} 
	\Big\{\hat{a}^\dagger \Big[\hat{c}^{1\dagger}_{(2j+s_1,2k+s_2)}\hat{c}^{0}_{(2j,2k)} -i s_2 \hat{c}^{0\dagger}_{(2j,2k)}\hat{c}^{1}_{(2j+s_1,2k+s_2)}\Big] + \mathrm{h.c.}\Big\} \nonumber \\
	\label{eq:Hamiltonian}
\end{align}
give rise to Raman-assisted tunnelings between neighboring states in a spin-textured two-dimensional momentum grid. Hence, even (odd) sites in the momentum lattice are exclusively populated by atoms in the spin state $\ket{0}$ ($\ket{1}$). Each tunneling process changes the spin state of the atoms and is associated with the creation ($\propto \hat{a}^\dagger $) or annihilation of cavity photons ($\propto \hat{a} $). This motivates the introduction of a time-dependent self-consistent tunneling amplitude $t_\text{SR}(t)=-\eta/\sqrt{8}\expval{\hat{a}^\dagger(t)}$. %In Eq.~\eqref{eq:Hamiltonian}, we apply a global phase transformation $(\hat{a},\hat{c}^0_{(2j,2k)})\rightarrow(\hat{a},\hat{c}^0_{(2j,2k)})\exp(-i\pi/4)$ in order to obtain a Hamiltonian with real coefficients.

%\tocheck{Fabian, let's check this derivation and how to map it to the Hamiltonian in the main text (i.e. get rid of the i's!)}
\subsection{Role of contact interactions}
Here, we discuss the role of contact interactions within the momentum grid picture. We assume that all atoms in a given momentum state occupy the same spatial mode and resort to the mode expansion of Eq.~\eqref{eqSI:FieldOperator}. We obtain a momentum space representation of the Hamiltonian describing contact interactions
\begin{align}
	\hat{H}_u & = g\int_{R}\hat{\Psi}^\dagger(\hat{\bold{x}})\hat{\Psi}^\dagger(\hat{\bold{x}}) \hat{\Psi}(\hat{\bold{x}})\hat{\Psi}(\hat{\bold{x}}) d\bold{x} = u \sum_{\bold{j_1},\bold{j_2},\bold{j_3},\bold{j_4}} \hat{c}^\dagger_\bold{j_1}\hat{c}_\bold{j_2}^\dagger\hat{c}_\bold{j_3}\hat{c}_\bold{j_4},
\end{align}  
where we introduce the short-hand notation $\hat{c}_\bold{j}=\hat{c}^\sigma_{(2j+\sigma,2k+\sigma)}$, with $\{j,k\} \in \mathbb{Z}$ and $\sigma\in\{0,1\}$ for the operators in the momentum grid. The strength of the contact interactions $g=4\pi\hbar^2a_s/m$ depends on the s-wave scattering length $a_s$, while $u=g\rho/N$ also depends on the average atomic density $\rho$. Closely following the approach of previous works on momentum-space lattices \cite{An_2018,Chen_2021,An_2021}, we neglect four-wave mixing processes \cite{Deng_1999} and retain only mode-conserving contributions of the form $\hat{c}_\bold{j} ^\dagger\hat{c}_\bold{k}^\dagger\hat{c}_\bold{j}\hat{c}_\bold{k}$, $\hat{c}^\dagger_\bold{j}\hat{c}_\bold{k}^\dagger\hat{c}_\bold{k}\hat{c}_\bold{j}$ and $\hat{c}^\dagger_\bold{j}\hat{c}_\bold{j}^\dagger\hat{c}_\bold{j}\hat{c}_\bold{j}$. By employing the standard bosonic commutation relations, we can obtain a simplified Hamiltonian

% (i.e. $[\hat{c}^\dagger_\bold{j},\hat{c}_\bold{k}]=\delta_{\bold{j},\bold{k}}$)

\begin{align}
	\hat{H}_u &\approx  u \sum_\bold{j}  \Big[\hat{n}_\bold{j}(\hat{n}_\bold{j}-1)/2 + \sum_{\bold{k}\neq\bold{j}} \hat{n}_\bold{j}\hat{n}_\bold{k}\Big] = u\hat{N}(\hat{N}-1/2) -u/2\sum_{\bold{j}}\hat{n}_\bold{j}^2,
	\label{eqSI:HamiltonianU}
\end{align}  
where we introduce the density operator $\hat{n}_\bold{j}=\hat{c}^\dagger_\bold{j}\hat{c}_\bold{j}$ and enforce particle number conservation $\hat{N}=\sum_\bold{j}\hat{n}_\bold{j}$. For repulsive contact interactions $u>0$, as it is the case for the $F=1$ manifold of $^{87}$Rb,  the Hamiltonian of Eq.~\eqref{eqSI:HamiltonianU} yields effective on-site attractive interactions in the momentum grid. This term can induce dephasing of the population dynamics~\cite{Chen_2021} or give rise to self-trapping in the initial state of the momentum lattice~\cite{An_2018,An_2021}, if it becomes dominant over the effective tunneling strength $J$ of the system ($uN > 4J)$. In our experiment, we estimate the contact interactions to be on the order of $uN/h=g\rho/h\approx \SI{0.8}{\kilo\hertz}$, by assuming an average number density of $\rho = \SI{2.1e20}{\per\meter\tothe{3}}$ and a scattering length of $a_s\approx 100a_0$ for $^{87}$Rb atoms~\cite{Stamper-Kurn_2013}, with $a_0$ being the Bohr radius. As the self-consistent tunneling rates in Eq.~\eqref{eqSI:H_int} can reach larger values, i.e., $\hbar\cdot\text{max}(\abs{t_\text{SR}})> uN/4  $, we expect that the BEC does not remain self-trapped in the initial momentum state, which is consistent with the experimental observations.  
%(\hbar\sqrt{2}\eta/4) \cdot\text{max}(\sqrt{n_\text{ph}})
%as present in $^{}87W$Rb,   
\section{Non-Hermitian Dynamics in the momentum lattice}
This section is dedicated to the theoretical description of the non-Hermitian dynamics in the momentum lattice. First, we present a two-mode analytical model which captures the main features of superradiant Raman scattering.  Moreover, we derive few-mode equations of motion which are employed in the simulations in the main text.   
\subsection{Time and frequency characteristics of superradiant pulses}
We gain insights into the mechanism and characteristics of the cavity-assisted superradiant transfers by studying a simplified scheme involving only two atomic modes and a single Raman drive. They are coupled by a Raman process driven by a single classical field with frequency $\omega_p$ and a quantized cavity mode with resonance frequency $\omega_c$. This scheme provides a good description for the dynamics of our system in the regime where the superradiant transfers driven by each of the two pump lasers at $\omega_{r,b}$ are well separated in time. In the next section, we will describe the numerical solutions for a more elaborated model including both drives and several atomic modes. 

The two atomic modes $\ket{\uparrow}$ (initial) and $\ket{\downarrow}$ (final state) are separated by an energy offset $\omega_A$ that includes both Zeeman and kinetic contributions. The bosonic operator $\hat{c}_{\uparrow}$ ($\hat{c}_{\downarrow}$) describes the annihilation of a particle in mode $\ket{\uparrow}$ ($\ket{\downarrow}$). Moreover, we enforce particle number conservation, i.e., $\langle\hat{c}_{\uparrow}^\dagger\hat{c}_{\uparrow}\rangle+\langle\hat{c}_{\downarrow}^\dagger\hat{c}_{\downarrow}\rangle=N_\uparrow+N_\downarrow=N$. In presence of a single-frequency drive, it is convenient to describe the dynamics in the rotating frame defined by the auxiliary Hamiltonian $\hat{H}_\text{rot}=\hbar\tilde{\omega}\hat{a}^\dagger \hat{a} - \hbar\omega_A\hat{F}_z$, with
\begin{equation}
	\tilde{\omega}=\omega_p+\omega_A
	\label{eqSI:omega_tilde}
\end{equation}	 
being the frequency of the photon field which ensures energy conservation in a Raman transfer from $\ket{\uparrow}$ to $\ket{\downarrow}$. The effective Hamiltonian of the closed system is then derived in complete analogy to the results presented in the section~\hyperref[sec:Hamiltonian]{\textit{Derivation of the Hamiltonian}}. It reads
\begin{equation}
	\hat{H}=-\hbar\tilde{\Delta}_c\hat{a}^\dagger\hat{a} + \hbar\eta_p\left(\hat{a}^\dagger \hat{J}_{-} + \hat{a}\hat{J}_{+}\right),
	\label{eq:single_pump_Hamiltonian}
\end{equation}
where we introduced a pseudo-spin $N/2$ operator $\hat{\bold{J}}$, with $\hat{J}_z=(\hat{c}_{\uparrow}^\dagger\hat{c}_{\uparrow}-\hat{c}_{\downarrow}^\dagger\hat{c}_{\downarrow})/2$, $\hat{J}_+=\hat{c}_\uparrow^\dagger\hat{c}_\downarrow$ and $\hat{J}_-=\hat{c}_\downarrow^\dagger\hat{c}_\uparrow$. The detuning between the expected frequency of emission $\tilde{\omega}$ and the dispersively-shifted cavity resonance is $\tilde{\Delta}_c$, and the effective coupling $\eta_p$ can be found in the same way as Eq.~\eqref{eqSI:mapping_coupling}. The Hamiltonian in Eq.~\eqref{eq:single_pump_Hamiltonian} is an effective Tavis-Cummings model with degenerate atomic levels. 

We study the mean-field dynamics of the system by deriving semiclassical equations of motion for the cavity field and the collective spin, including the cavity decay $\kappa$. Since cavity dissipation dominates over the coherent coupling ($\kappa\gg \sqrt{N}\eta_p$), we study the effective dynamics of the collective spin $\bold{j}=[j_x,j_y,j_z]^T=\langle\hat{\bold{J}}\rangle$ after adiabatic elimination of the cavity field. We first consider the simplest case of zero cavity detuning $\tilde{\Delta}_c=0$. The system is prepared in mode $\ket{\uparrow}$ at $t=0$. This is an unstable steady state of the system: even in presence of infinitesimally small fluctuations, i.e., $\bold{j}(t=0)=N/2[\epsilon\cos\theta,\epsilon\sin\theta,\sqrt{1-\epsilon^2}]^T$, with $\epsilon\ll 1$ and $\theta\in[0,2\pi)$, the spin spontaneously evolves towards mode $\ket{\downarrow}$. The dynamics follows the equation of motion 
\begin{equation}
	\frac{d}{dt}j_z=-\frac{2\eta_p^2}{\kappa}\left(\frac{N^2}{4}-j_z^2\right),
	\label{eqSI:SR_decay_EOM}
\end{equation}
where we made use of the total spin conservation $\vert\bold{j}(t)\vert=N/2$. This equation can be solved analytically, and gives the following solution:
\begin{align}
	\begin{split}
		j_z(t)=-\frac{N}{2}\tanh\left(\frac{t-t_0}{\tau}\right),\quad\quad\quad\quad
		j_\perp(t)=\frac{N}{2}\sech\left(\frac{t-t_0}{\tau}\right),
	\end{split}
	\label{eqSI:SR_decay_spin}
\end{align}
where $j_\perp$ is the transverse spin projection $j_\perp=\sqrt{j_x^2+j_y^2}$. From the associated amplitude of the cavity field  $\alpha=\langle\hat{a}\rangle$, we derive the average photon number $n_\text{ph}=\vert\alpha\vert^2$ as 
\begin{equation}
	n_\text{ph}(t)=\left(\frac{\eta_p}{\kappa} j_\perp\right)^2=\frac{N^2\eta_p^2}{4\kappa^2}\sech^2\left(\frac{t-t_0}{\tau}\right).
	\label{eqSI:SR_decay_photons}
\end{equation} 
Eqs.~\eqref{eqSI:SR_decay_spin} and \eqref{eqSI:SR_decay_photons} describe a superradiant decay where the initial spin $\ket{\uparrow}$ fully inverts to $\ket{\downarrow}$ in the presence of a collectively-enhanced pulse of the cavity field. The duration of the superradiant transfer is determined by $\tau=\kappa/(N\eta_p^2)$, while the time $t_0$ at which the pulse reaches its maximum depends on the initial fluctuations $t_0=-\tau/2\log(\epsilon^2/4)$. Moreover, the super-linear scaling of the maximal photon number with the atom number $\text{max}(n_\text{ph})=\eta_p^2N^2/4\kappa^2$ is a hallmark of a collectively-enhanced superradiant decay~\cite{Haroche_1982,Mandel_1995}. The total number of photons scattered into the cavity is independent of the timescales of the transfer, and is determined by the number of atoms participating in the dynamics
\begin{equation}
	2\kappa\int_0^\infty n_\text{ph}(t)dt=N.
\end{equation}
This one-to-one correspondence is a direct consequence of total angular momentum conservation in the system: while a complete transfer from  $\ket{\uparrow}$ to $\ket{\downarrow}$ changes the atomic angular momentum by $\pm \hbar N$, this can be compensated by absorbing $N$ $\sigma_\pm$-polarized pump photons and re-emitting them into the $\pi$-polarized cavity field. The scattered cavity field is stationary in the rotating frame, with a frequency of $\omega_{\uparrow,\downarrow}=\tilde{\omega}$ in the lab frame. The phase of the  cavity field is imprinted by the initial phase of the fluctuations. This phase also determines the azimuthal angle $\phi=\arctan(j_y/j_x)$ at which the spin transfer occurs, which stays constant during the dynamics.

We now turn to the more general case of finite cavity detuning $\tilde{\Delta}_c\neq0$. By following an analogous procedure, we find that the spin dynamics can be decomposed in two parts. First, a spin transfer from $\ket{\uparrow}$ to $\ket{\downarrow}$ occurs, as described in Eqs.~\eqref{eqSI:SR_decay_spin} and ~\eqref{eqSI:SR_decay_photons}. The duration $\tau$ of the corresponding superradiant pulse is now given by $\tau=(\tilde{\Delta}_c^2+\kappa^2)/(N\eta_p^2\kappa)$. Concurrently, the collective spin precesses about $j_z$ at a variable rate $\omega_\text{rot}(t)=\dot{\phi}(t)\propto j_z(t)$. This behavior is signaled by a varying frequency of the cavity field $\omega_{\uparrow,\downarrow}(t)=\tilde{\omega}+\omega_\text{rot}(t)$ in the lab frame. Around the maximum of the field amplitude, i.e., for $j_\perp=N/2$ and $j_z=0$, the frequency is $\omega_{\uparrow,\downarrow}(t=t_0)=\tilde{\omega}$ and coincides with the result in the resonant case ($\tilde{\Delta}_c=0$). By combining this finding with the definition of $\tilde{\omega}$ in Eq.~\eqref{eqSI:omega_tilde}, we obtain
\begin{equation}
	\omega_{\uparrow,\downarrow}(t=t_0)=\omega_p+\omega_A.
	\label{eqSI:energy_conservation_pulse}
\end{equation} 
Thus, near the maximum of the superradiant pulse, the frequency of the scattered field is the one expected from the conservation of energy in a two-photon process connecting the initial and final bare atomic states.

The result in Eq.~\eqref{eqSI:energy_conservation_pulse} allows us to exploit the readout of the frequency of the cavity field to assess the energy difference $\omega_A$ between the initial and final states of the superradiant transfer, as stated in Eq.~(4) in the main text. Specifically, we apply this to a transfer in the momentum space lattice driven by the pump at frequency $\omega_r$, i.e., from $\ket{\uparrow}=\ket{l_i,m_i}_0$ to $\ket{\downarrow}=\ket{l_f,m_f}_1$. We substitute $\omega_p\rightarrow\omega_r$ and $\omega_{\uparrow,\downarrow}\rightarrow\omega_{01}$, set the energy splitting between the two states to $\omega_A=\omega_z+\omega^\text{kin}_{(l_i, m_i)}-\omega^\text{kin}_{(j_f, k_f)}$, and obtain
\begin{align}
	\begin{split}
		\omega_{01}&=\omega_r+\omega_z+\omega^\text{kin}_{(l_i, m_i)}-\omega^\text{kin}_{(l_f, m_f)}\\
		&=\bar{\omega}-\omega_0-[\omega^\text{kin}_{(l_f, m_f)}-\omega^\text{kin}_{(l_i, m_i)}],
	\end{split}
	\label{eqSI:omega01}
\end{align}
which is equivalent to the expression of Eq.~(4) in the main text. The derivation of the frequency for the opposite process $\omega_{10}$ is analogous.

As a final note, we consider in more detail the effect of the dispersive shift. When the atomic mode $\ket{\uparrow}$ ($\ket{\downarrow}$) is populated, the cavity resonance gets dispersively shifted by a frequency $N_{\uparrow} U_{\uparrow}$ ($N_{\downarrow} U_{\downarrow}$), where $U_{\uparrow,\downarrow}$ and $N_{\uparrow,\downarrow}$ are the dispersive shift per particle and atom number in the corresponding mode, respectively. We label the average and differential dispersive shift as $\bar{U}=(U_\uparrow+U_\downarrow)/2$ and $\delta U=(U_\uparrow-U_\downarrow)/2$, respectively. With these definitions, the effective cavity detuning reads $\tilde{\Delta}_c=\Delta_c-N\bar{U}-\delta U j_z$, with $\Delta_c=\omega_p-\omega_c$. The differential dispersive shift $\delta U$ is responsible for a dynamical $j_z$-dependent shift of the cavity resonance, which modifies the frequency evolution of the cavity field during the superradiant transfer. However, close to the pulse maximum ($j_z=0$), the frequency of the field is only slightly affected by this shift. More specifically, the deviation from $\tilde{\omega}$ is $\omega_\text{rot}(t=t_0)\approx\delta U N \eta_p^2/[2(\tilde{\Delta}_c^2+\kappa^2)]$. Since  for our typical parameters this shift amounts to less than $2\pi\cdot 1~$Hz and is below our frequency resolution, we neglect the effect of the differential dispersive shift in the frequency analysis of the superradiant pulses obtained in the experiment.

\subsection{Few-mode expansion and mean-field equations of motion}\label{sec:FewModeExpansion}
%\todo{The dynamics of the open system is EOM in Quantum Lagevin approach, spin-dephasing, and MATLAB ode45 solver...}

In this section, we extend the notion of the superradiant decay to describe the non-Hermitian dynamics  in the momentum grid in the presence of cavity dissipation.	The open system dynamics is determined by the master equation

%and allows to continuously monitor its evolution.

\begin{align}
	\frac{d\hat{\rho}}{dt}&= -\frac{i}{\hbar}[\hat{H},\hat{\rho}] + \mathcal{L}[\hat{a}],
	%+\Gamma_\phi [2 \hat{J}_{z,eff} \hat{\rho} \hat{J}_{z,eff} - \{\hat{J}_{z,eff} \hat{J}_{z,eff}, \hat{\rho}\}]\, 
	\label{eqSI:ME}	
\end{align}
with the Lindblad operator $\mathcal{L}[\hat{a}] = \kappa[2\hat{a} \hat{\rho} \hat{a}^\dagger - \{ \hat{a}^\dagger \hat{a} , \hat{\rho}\}]$ capturing the evolution arsing from photon loss at a rate $\kappa$. As discussed in previous sections, we expect the superradiant dynamics to be primarily determined by the interplay of cavity leakage and Hamiltonian terms creating cavity photons. In order to efficiently simulate the dynamics and capture the first few superradiant processes, we identify four low-energy atomic modes

%Since the experiment operates in the bad cavity limit ($\kappa > \omega_0, \eta\sqrt{N} $) and the interacting Hamiltonian in Eq.~\eqref{eq:Hamiltonian} cannot be casted into a Dicke-type Hamiltonian with a finite number of atomic states, photons scattered into the cavity are very likely to dissipate before they can get reabsorbed by the atoms \tocheck{How can we make this formulation more concise?}. Hence, the open system dynamics is primarily determined by the interplay between cavity leakage ($\propto \kappa$) and Hamiltonian terms scattering photons into the cavity ($\propto H_\text{int}(\hat{a}^\dagger)$, see Eq.~\eqref{eqSI:HamiltonianINT})
%		where we refrain from applying the phase transformation used for Eq.~\eqref{eq:Hamiltonian}.

\begin{align}
	\psiM &= \hat{c}^0_{(0,0)}, \nonumber \\  
	\psiD &= (\hat{c}^1_{(1,1)}+\hat{c}^1_{(1,-1)}+\hat{c}^1_{(-1,1)}+\hat{c}^1_{(-1,-1)})/2 , \nonumber \\
	\psiB &= i(\hat{c}^1_{(1,-1)}+\hat{c}^1_{(-1,-1)}-\hat{c}^1_{(1,1)}-\hat{c}^1_{(-1,1)})/2, \nonumber \\
	\psiF &= i(\hat{c}^0_{(0,-2)} - \hat{c}^0_{(0,2)})/\sqrt{2},
	\label{eqSI:FewModeExpansion}	
\end{align}
which are coupled by terms creating cavity photons in Eq.~\eqref{eq:Hamiltonian}, if the system is initialized in the mode associated to $\psiM$. The single particle wave functions of these orthonormal modes are $\psiMNH \propto 1$, $\psiDNH \propto \cos(kx)\cos(kz)$, $\psiBNH \propto \cos(kx)\sin(kz)$ and $\psiFNH \propto \sin(2kz)$. Within this expansion, the Hamiltonian in Eq.~\eqref{eqSI:H_MB} can be simplified to

\begin{align}
	\hat{H}= -\hbar \tilde{\Delta}_c \hat{a}^\dagger\hat{a} + \hbar (\omega_0+2\omega_\text{rec})(\psiD^\dagger\psiD+\psiB^\dagger\psiB)+4\hbar\omega_\text{rec}\psiF^\dagger\psiF - \frac{\hbar\eta}{\sqrt{2}} \hat{a}^\dagger\left[\psiD^\dagger\psiM - \frac{1}{\sqrt{2}}\psiF^\dagger\psiD  + \frac{1}{\sqrt{2}}\psiB^\dagger\psiF -\psiM^\dagger\psiB   \right] + ~\text{h.c.}~.
	\label{eqSI:H_FewMode}
\end{align}
We derive Langevin equations of motion (EOM) for the expectation values of the cavity field $\alpha=\expval{\hat{a}}/\sqrt{N}$, atomic populations $\rho_{jj}=\expval{\hat{\psi}^\dagger_j\hat{\psi}_j}/N$ and atomic coherences $\rho_{jk}=\expval{\hat{\psi}^\dagger_j\hat{\psi}_k}/N$, with $\{j,k\}\in\{0,1,2,3\}$. We obtain a set of eleven complex coupled EOM 
\begin{align}
	\frac{d}{dt}\alpha &= -(\kappa-i\tilde{\Delta}_c)\alpha+ i\sqrt{N}\eta\left(\frac{1}{\sqrt{2}}\rhoMD^*-\frac{1}{\sqrt{2}}\rhoMB-\frac{1}{2}\rhoDF^*+\frac{1}{2}\rhoBF\right), \nonumber \\
	\frac{d}{dt}\rhoMM &=i\sqrt{\frac{N}{2}}\eta \left(\alpha\rhoMD-\alpha^*\rhoMD^*+\alpha\rhoMB^*-\alpha^*\rhoMB \right),  \nonumber \\
	\frac{d}{dt}\rhoDD &=i\sqrt{N}\eta\left(-\frac{1}{\sqrt{2}}\alpha\rhoMD+\frac{1}{\sqrt{2}}\alpha^*\rhoMD^*-\frac{1}{2}\alpha\rhoDF+\frac{1}{2}\alpha^*\rhoDF^* \right),  \nonumber \\
	\frac{d}{dt}\rhoBB &=i\sqrt{N}\eta \left(-\frac{1}{\sqrt{2}}\alpha\rhoMB^*+\frac{1}{\sqrt{2}}\alpha^*\rhoMB-\frac{1}{2}\alpha\rhoBF^*+\frac{1}{2}\alpha^*\rhoBF \right),  \nonumber \\
	\frac{d}{dt}\rhoFF &=i\frac{\sqrt{N}}{2}\eta \left(\alpha\rhoDF-\alpha^*\rhoDF^*+\alpha\rhoBF^*-\alpha^*\rhoBF \right), \nonumber \\
	\frac{d}{dt}\rhoMD &=-[\Gamma+i(\omega_0+2\omega_\text{rec})]\rhoMD +i\sqrt{N}\eta\left[\frac{1}{\sqrt{2}}\alpha^*\left(\rhoMM-\rhoDD\right)+\frac{1}{\sqrt{2}}\alpha\rhoDB^*-\frac{1}{2}\alpha\rhoMF\right],  \\
	\frac{d}{dt}\rhoMB &=-[\Gamma+i(\omega_0+2\omega_\text{rec})]\rhoMB +i\sqrt{N}\eta\left[-\frac{1}{\sqrt{2}}\alpha\left(\rhoMM-\rhoBB\right)-\frac{1}{\sqrt{2}}\alpha^*\rhoDB+\frac{1}{2}\alpha^*\rhoMF\right],  \nonumber \\
	\frac{d}{dt}\rhoMF &=-(\Gamma+i4\omega_\text{rec})\rhoMF +i\sqrt{N}\eta\left(-\frac{1}{2}\alpha^*\rhoMD+\frac{1}{2}\alpha\rhoMB -\frac{1}{\sqrt{2}}\alpha^*\rhoDF + \frac{1}{\sqrt{2}}\alpha\rhoBF\right), \nonumber \\
	\frac{d}{dt}\rhoDB &=i \sqrt{N}\eta\left(-\frac{1}{\sqrt{2}}\alpha\rhoMB - \frac{1}{\sqrt{2}}\alpha\rhoMD^* + \frac{1}{2}\alpha^*\rhoDF + \frac{1}{2}\alpha^*\rhoBF^*  \right), \nonumber \\
	\frac{d}{dt}\rhoDF &= -[\Gamma+i(2\omega_\text{rec}-\omega_0)]\rhoDF + i\sqrt{N}\eta\left[\frac{1}{2}\alpha^*\left(\rhoFF-\rhoDD\right)-\frac{1}{\sqrt{2}}\alpha^*\rhoMF+\frac{1}{2}\alpha\rhoDB\right], \nonumber \\
	\frac{d}{dt}\rhoBF &= -[\Gamma+i(2\omega_\text{rec}-\omega_0)]\rhoBF +i\sqrt{N}\eta\left[-\frac{1}{2}\alpha\left(\rhoFF-\rhoBB\right)+\frac{1}{\sqrt{2}}\alpha^*\rhoMF-\frac{1}{2}\alpha^*\rhoDB^*\right], \nonumber  
	\label{eqSI:EOMs}
\end{align}
where we employ the mean-field decoupling $\expval{\hat{a}\hat{\psi}^\dagger_j\hat{\psi}_k}\approx N^{3/2}\alpha\rho_{jk}$ and set  $\rho_{jk}^* = \rho_{kj}$. Moreover, we include a phenomenological spin dephasing rate, which we estimate to be on the order of $\Gamma/2\pi \sim 200~\text{Hz}$ in our experiment. We attribute it to the combined effect of atomic collisions and magnetic field fluctuations~\cite{Ferri_2021}, which effectively damp atomic coherences between states in different spin manifolds at rate $\Gamma$.

%\footnote{The spin dephasing is compatible with a Lindblad  term of the form $\mathcal{L}[\hat{J}_\text{z,eff}]=\Gamma\left(2\hat{J}_\text{z,eff}\hat{\rho}\hat{J}_\text{z,eff} + \{\hat{J}_\text{z,eff}\hat{J}_\text{z,eff},\hat{\rho}\}\right)$, with  \tocheck{$\hat{J}_\text{z,eff}=1/2 \left(\psiD^\dagger\psiD+ \psiB^\dagger\psiB - \psiM^\dagger\psiM- \psiF^\dagger\psiF\right)$}

%The many-body Hamiltonian introduced in Eq.~\eqref{eqSI:HamiltonianINT} cannot be cast into a Dicke type Hamiltonian, as it involves a infinite number states that are coupled via the creation or anhilation of cavity photons \tocheck{make formulation more concise, relate to previous subsection ?}

\subsection{Numerical simulations}
In order to model the dynamics of the system, we numerically evaluate the mean-field EOM derived in the previous section. We employ the MATLAB solver `ode45' which is based on a Runge-Kutta (4,5) method \cite{Shampine_1997}. We employ adaptive time steps and constrain the relative error tolerance in each step to $10^{-8}$. In order to seed the mean-field dynamics, we sample small fluctuations on top of the expectation value of the cavity field. This sampling ensures an initial cavity field at $t=0$ compatible with a coherent vacuum state~\cite{Ferri_2021}. For the simulations presented in Fig.~2 and 3 in the main text, we initialize the atoms in a zero-momentum BEC in $m_F=-1$ ($\rhoMM=1$) and increase the couplings via s-shaped ramps $\eta(t)$ compatible to the experimental ones. Moreover, the simulated photon number spectrograms $\tilde{n}_\text{ph}(t,\omega)$ are constructed using the same method as the experimental ones (see section~\hyperref [sec:Heterodyne]{\textit{Heterodyne detection and photon field spectrograms}}). For the simulations presented in Fig.~2 and 3 in the main text, we choose spin dephasing rates of $\Gamma=2\pi\cdot150~\text{Hz}$ and $2\pi\cdot250~\text{Hz}$, respectively. 

\section{Gross-Pitaevskii simulations}

In this section, we present ab intio Gross-Pitaevskii equation simulations to benchmark the dynamics in the momentum lattice: After deriving the equations of motion, we present simulations for the parameters of Fig.~3 and 4 in the main text. Moreover, we provide evidence for the simultaneous build-up of coherences in the cascaded tunneling regime, and elucidate the lifetime of the momentum lattice due to oscillatory dynamics in the harmonic trap.

\subsection{Equations of motion}
We complement our experimental results and few-mode simulations with simulations of the Gross-Pitaevskii equations (GPS) using the Multiconfigurational Time-Dependent Hartree Method for Indistinguishable Particles~\cite{axel16,alon08,fasshauer16,lin20,axel20}, which is implemented in the MCTDH-X software~\cite{ultracold}. We solve the time evolution of the two-component mean-field Hamiltonian 
\begin{eqnarray}
	\hat{\mathcal{H}} = N\int \Phi^\dagger \hat{H}^{(1)} \Phi dxdz + \frac{g_0}{2}N(N-1)\int |\phi_0+\phi_1|^4 dxdz
\end{eqnarray}
with $\Phi = (\phi_0,\phi_1)^\mathrm{T}$, where $\phi_0(x,z)$ and $\phi_1(x,z)$ are the mean-field wave functions of the two spin levels $\ket{0}$ and $\ket {1}$ with normalization $\int \Phi^\dagger \Phi dxdz = 1$. The second term describes contact interactions between the atoms, where we assume identical inter- and intra-spin coupling constants $g_0$. This is a good approximation for $^{87}$Rb atoms in the $F=1$ manifold~\cite{Stamper-Kurn_2013}. Moreover, the first term integrates over the single-particle Hamiltonian, which is given by
\begin{eqnarray}\label{eqSI:MCTDHX_hamil}
	\hat{H}^{(1)} &=& \left[\frac{\hat{p}^2}{2M} + \frac{M}{2}(\omega_{\mathrm{hx}}^2x^2+\omega_{\mathrm{hz}}^2z^2)\right] \mathbbm{1} - \omega_0\sigma_z \nonumber \\
	&& + \eta(\alpha+\alpha^*) \cos(k_\mathrm{rec}x)\cos(k_\mathrm{rec}z) \sigma_y + i\eta (\alpha-\alpha^*) \cos(k_\mathrm{rec}x)\sin(k_\mathrm{rec}z) \sigma_x,
\end{eqnarray}
where  $\sigma_j$ refer to the Pauli matrices, with $j\in\{x,y,z\}$.
This Hamiltonian contains the same contributions as the one presented in Eq.~\eqref{eqSI:H1}, but further considers the harmonic confinement and omits the constant term arising from scalar light-matter interactions [cf. Eq.~\eqref{eqSI:H1_scalar}]. More specifically, the first line includes the kinetic term, the harmonic trap with typical experimental trapping frequencies $[\omega_\text{hx},\omega_\text{hz}]=2\pi\cdot[218,172]~\text{Hz}$,
and the total splitting between the two levels ($\propto\omega_0$), cf. Eq.~\eqref{eqSI:H1_at}. Moreover, the second line describes the cavity-assisted Raman transitions between the two spin levels, and coincides with the Hamiltonian in Eq.~\eqref{eqSI:H1_vectorial} in the limit of balanced drives $\eta=\eta_r=\eta_b$. The contact interaction strength $Ng_0=1210\hbar^2/m$ is chosen such that the initial Thomas-Fermi radii coincide with the experimental values $[r_{\mathrm{TF},x},r_{\mathrm{TF},z}]=[4.3,5.5]\,\mu\text{m}$. The cavity field is treated as a coherent light field and represented by a complex number $\alpha$, whose evolution follows
\begin{subequations}
	\begin{eqnarray}
		\partial_t \alpha &=& [i\tilde{\Delta}_c-\kappa]\alpha - i\eta N \theta, \label{eq:Sim_eom_alpha}\\
		\theta &=& \int \Phi^\dagger \left[\cos(k_\mathrm{rec}x)\cos(k_\mathrm{rec}z) \sigma_y + i\cos(k_\mathrm{rec}x)\sin(k_\mathrm{rec}z) \sigma_x \right] \Phi dxdz.
	\end{eqnarray}
\end{subequations}
\subsection{Numerical solution of the dynamics}
Using MCTDH-X, we employ a variational method and evolve the wave function $\Phi(x,z;t)$ to numerically solve the dynamics. The system is prepared in a slightly perturbed BEC state in a harmonic trap. This perturbation represents the noise in the system, and is empirically set such that the first superradiant pulse occurs at a time comparable to the one observed in the experiment. We then activate the cavity field and evolve the system under two different sets of cavity parameters, each corresponding to the observations reported in Fig.~3 and Fig.~4 in the main text, respectively. 

\begin{figure}[thbp]
	\centering
	\includegraphics[width=0.82\columnwidth]{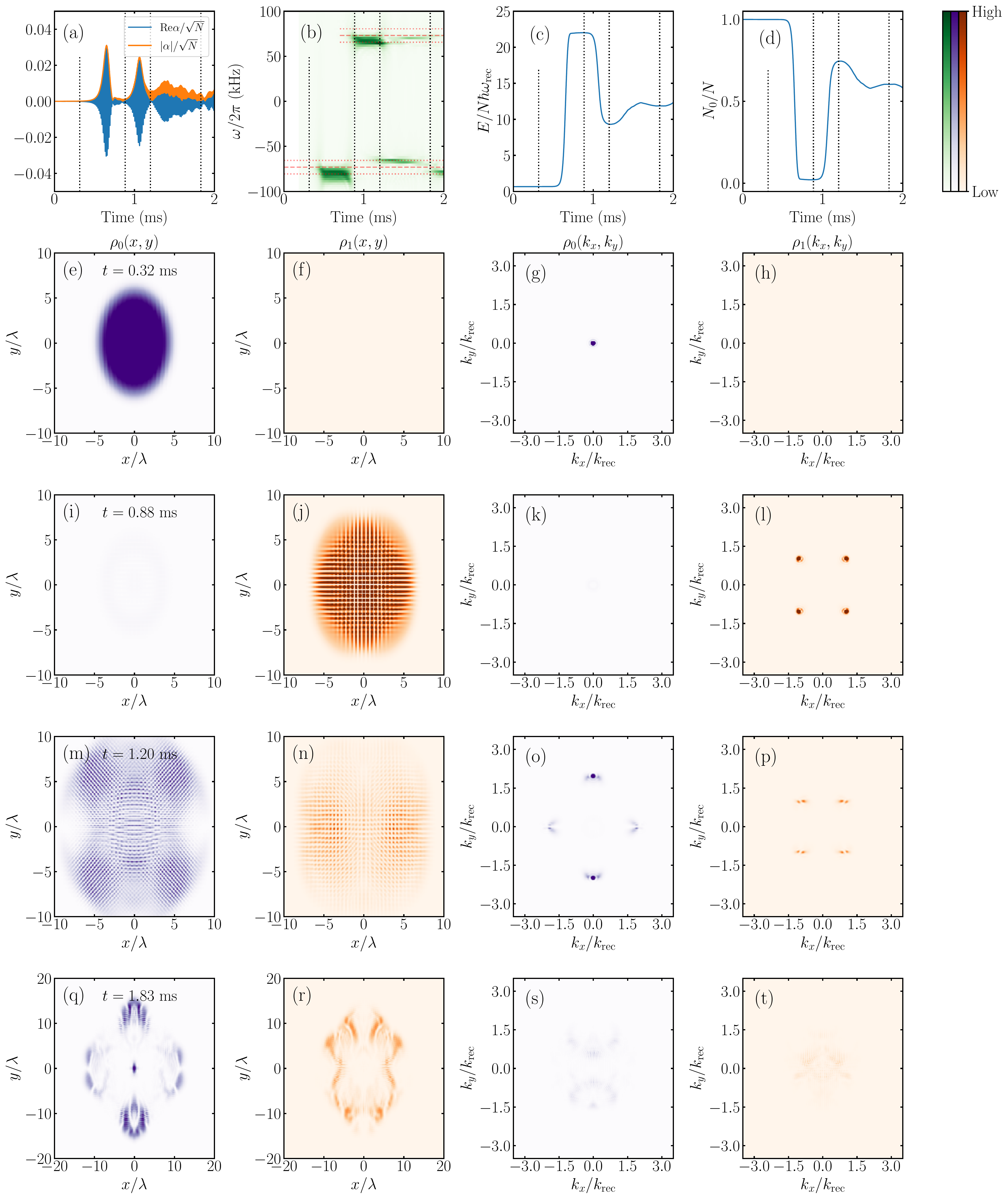}
	\caption{Simulations of Gross-Pitaevskii equations reproducing results from Fig.~3 in the main text. (a-d) Time evolution of the real part [(a), blue], magnitude [(a), orange] and spectrogram (b) of the cavity field, (c) system energy, and (d) occupation of the $\ket{0}$ mode. In panel (b), the thick dashed lines indicate $\omega=\pm\omega_0$, whereas the thin dotted lines indicate $\omega=\pm\omega_0\pm2\omega_\mathrm{rec}$. (e-t) The real space and momentum space density distributions are shown at four representative time points $t=\{0.32,~0.88, ~1.20,~1.83\}~\text{ms}$ for atoms in the spin states $\ket{0}$ (purple) and $\ket{1}$ (orange colormaps). These four time points are indicated as vertical dashed points in panels (a-d). }
	\label{figSI:Sim_Fig3}
\end{figure}

The first simulation  is performed with $\tilde{\Delta}_c=-2\pi\cdot0.7~\text{MHz}$ and $\omega_0=2\pi\cdot72.5~\text{kHz}$, while the coupling is increased to $\eta_\mathrm{max}=2\pi\cdot0.62~\text{kHz}$ within $t_r=1.5~\text{ms}$ using an s-shaped ramp as in the experiment. The simulation results are presented in Fig.~\ref{figSI:Sim_Fig3}. The behavior of the cavity field [Fig.~\ref{figSI:Sim_Fig3}(a)] and its spectrogram [Fig.~\ref{figSI:Sim_Fig3}(b)] reproduce the experimental results presented in Fig.~3 in the main text. Three strong photon pulses are observed, whose frequencies are determined by the atomic splitting and the recoil frequency according to Eq.~\eqref{eqSI:omega01}. Accompanying each photon pulse, the energy of the atomic state $E=\expval{\hat{\mathcal{H}}}$ [Fig.~\ref{figSI:Sim_Fig3}(c)] and the atomic occupation of the $\ket{0}$ spin manifold  [Fig.~\ref{figSI:Sim_Fig3}(d)] $N_0 = N\int|\phi_{0,1}(x,z)|^2dxdz$ change drastically. To better understand the atomic dynamics induced by the emerging cavity field, we look at four representative spin and density distributions taken between the photon bursts. The snap-shots of the real-space $\rho_{0,1}(x,z)=|\phi_{0,1}(x,z)|^2$ and momentum-space distributions $\rho_{0,1}(k_x,k_z)=|\phi_{0,1}(k_x,k_z)|^2$ at different points in time are shown in Fig.~\ref{figSI:Sim_Fig3}(e-t), where $\phi_0(k_x,k_z)$ and  $\phi_1(k_x,k_z)$ are the Fourier transforms of $\phi_0(x,z)$ and $\phi_1(x,z)$, respectively.
The atomic transfers in the momentum-space lattice are clearly visible, and the momentum space densities Fig.~\ref{figSI:Sim_Fig3}(g,h,k,l,o,p) qualitatively reproduce the experimental results as shown in Figs.~3(b-d) in the main text. Through the first and second bursts, the majority of the atoms undergo the transfer $|0,0\rangle_0\to|\pm1,\pm1\rangle_1\to|0,\pm2\rangle_0$. This dynamics is also reflected in real space by the formation of the corresponding density waves. At long times $t>1.83~\text{ms}$ [Fig.~\ref{figSI:Sim_Fig3}(q-t)], the momentum distribution becomes washed out and starts to deviate from a tight-binding description. This is due to the combined effect of the harmonic trap and contact interactions, which induces complex dynamics in momentum space. In the next subsection, we characterize this effect in detail. 

\begin{figure}[thbp]
	\centering
	\includegraphics[width=0.82\columnwidth]{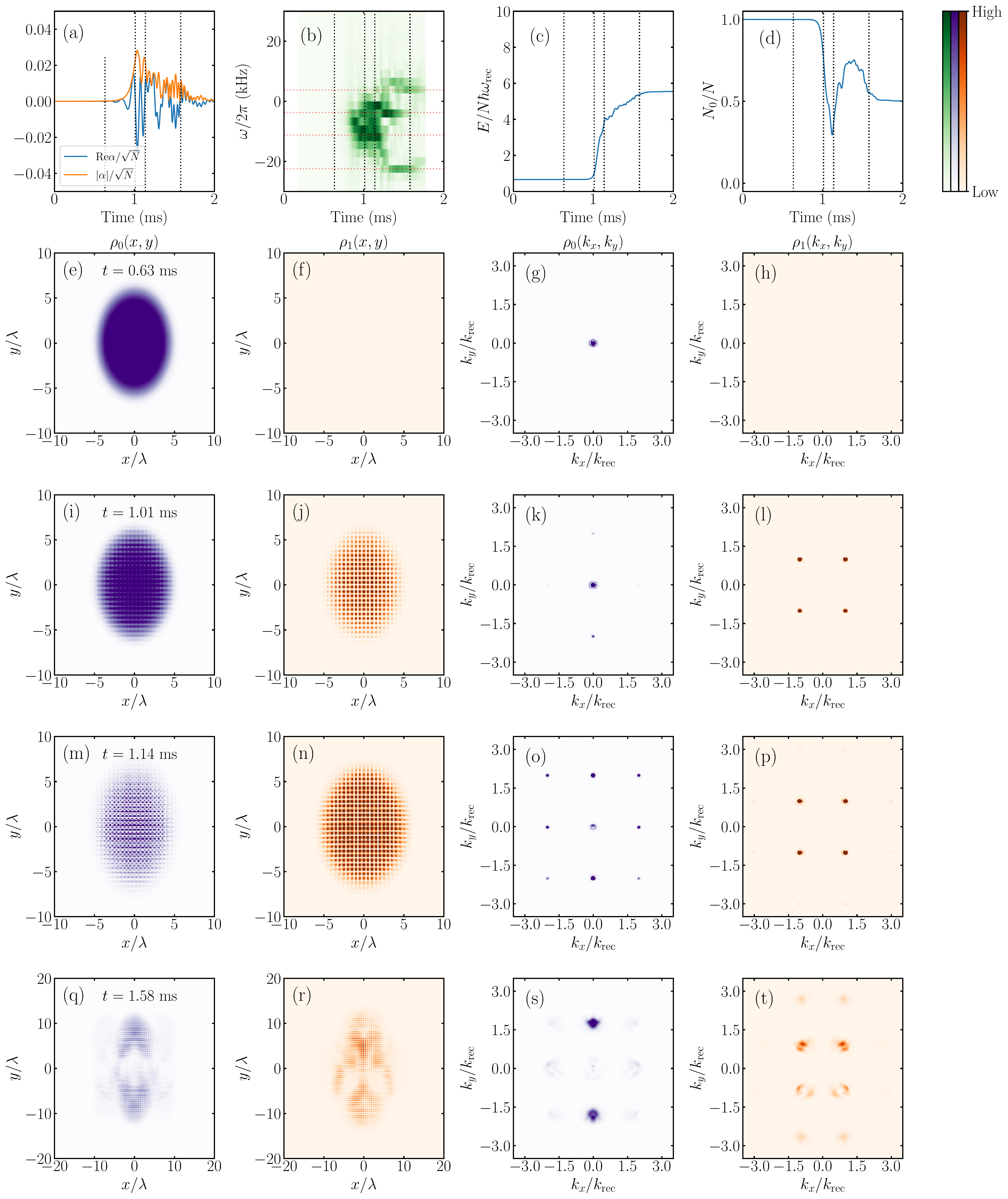}
	\caption{Simulations of Gross-Pitaevskii equations reproducing results from Fig.~4 in the main text. (a-d) The time evolution of the real part [(a), blue], magnitude [(a), orange] and spectrogram (b) of the cavity field, and (d) occupation of the $\phi_0$ mode. In panel (b), the thick dashed lines indicate $\omega=\omega_\text{rec}$, $-\omega_\text{rec}$, $-3\omega_\text{rec}$ and $-6\omega_\text{rec}$, respectively. (e-t) The real space and momentum space density distributions at four representative time points $t=\{0.63,~1.01, ~1.14,~1.58\}~\text{ms}$ for atoms in the spin states $\ket{0}$ (purple) and $\ket{1}$ (orange colormaps). These four time points are indicated as vertical dashed points in panels (a-d). } 
	\label{figSI:Sim_Fig4}
\end{figure}

The second simulation  is performed with $\tilde{\Delta}_c=-2\pi\cdot1.26~\text{MHz}$ and $\omega_0=2\pi\cdot3.7~\text{kHz}\approx\omega_\mathrm{rec}$, while the coupling is increased up to $\eta_\mathrm{max}=2\pi\cdot0.50~\text{kHz}$ within $t_r=2~\text{ms}$ using the experimental protocol. The simulations are shown in Fig.~\ref{figSI:Sim_Fig4} and reproduce the experimental results presented in Fig.~4 in the main text. Compared to the first simulation, here we observe a strong single photon burst lasting for a relatively long time [Fig.~\ref{figSI:Sim_Fig4}(a)]. During this pulse, the spectrogram [Fig.~\ref{figSI:Sim_Fig4}(b)], system energy [Fig.~\ref{figSI:Sim_Fig4}(c)], and atomic occupation of the $\phi_0$ level [Fig.~\ref{figSI:Sim_Fig4}(c)] all show complex, rapidly changing behaviors. To better understand the system dynamics, we again choose four representative time points and show the corresponding density distributions in Fig.~\ref{figSI:Sim_Fig4}(e-t). From the momentum space densities in Fig.~\ref{figSI:Sim_Fig4}(g,h,k,l,o,p), we notice a ``cascaded atomic transfer'' taking place only between nearest neighboring sites in the momentum lattice. Remarkably, the  next tunneling event starts before the previous one is fully completed. This is also evidenced in the population dynamics of $\ket{0}$ [Fig.~\ref{figSI:Sim_Fig4}(c)], which keeps decreasing as the mode $\ket{0,\pm2}_0$ gets populated. As the time elapses, the atoms gradually occupy modes with larger momentum within the same photon burst, giving rise to qualitatively different dynamics from the one presented in Fig.~\ref{figSI:Sim_Fig3}. We can thus understand this single pulse as a conjunction of several bursts, where  due to the small energy difference between the two spin manifolds ($\omega_0\approx\omega_\text{rec}$), the succeeding pulse  is stimulated by the preceding one and starts before the latter finishes. As a result,  we show that multiple sites in the momentum lattice can be occupied simultaneously, see Fig.~\ref{figSI:Sim_Fig4}(o,p). These features indicate the presence of finite coherences between multiple pairs of adjacent sites during the extension of the photon pulse. Moreover, in Fig.~\ref{figSI:Sim_Fig4}(o,p), we show that before oscillatory dynamics in the harmonic trap completely blurs the momentum lattice, the highest order momentum states occupied are $\ket{1,3}_1$, $\ket{1,-3}_1$, $\ket{-1,3}_1$ and $\ket{-1,-3}_1$. This is compatible to the experimental observations discussed in Fig.~\ref{figSI:PEandMomPeaks}.

\subsection{Cascaded tunneling regime: Relation between site coherences and the cavity field}
As discussed in the main text, the emerging cavity field and the associated tunneling events dynamically depend on the build-up of coherences between adjacent sites in the momentum grid. In particular, our experiment operates in an overdamped regime ($\kappa\gg \omega_0,\omega_\text{rec}$), where the cavity field follows the atomic configuration adiabatically  [cf. Eq.~\eqref{eq:Sim_eom_alpha}], resulting in
%This can be elucidated by comparing the dynamics of the different two-site coherences and the cavity field
\begin{eqnarray}
	\partial_t \alpha = 0 \implies \alpha = \frac{\eta N}{\tilde{\Delta}_c+i\kappa}\theta.
\end{eqnarray}
In the discrete representation of the momentum lattice [cf. Eq.~\eqref{eq:Hamiltonian}], the order parameter $\theta$ can be evaluated as 
\begin{subequations}
	\begin{eqnarray}
		\theta &=& \sum_{\substack{j,k\in\mathbb Z \\ s_{1,2}=\pm1}} \theta_{j,k,s_1,s_2}\\
		N\theta_{j,k,s_1,s_2} &=& -\frac{1}{\sqrt{8}} 
		\Big[\left\langle\hat{c}^{1\dagger}_{(2j+s_1,2k+s_2)}\hat{c}^{0}_{(2j,2k)}\right\rangle -i s_2 \left\langle\hat{c}^{0\dagger}_{(2j,2k)}\hat{c}^{1}_{(2j+s_1,2k+s_2)}\right\rangle\Big],
	\end{eqnarray}
\end{subequations}
which is a sum of local two-site coherences $\left\langle \hat{c}^{1\dagger}_{(2j+s_1,2k+s_2)}\hat{c}^{0}_{(2j,2k)}\right\rangle$. Since the GPS work in the continuum, we approximate them by integrals over the corresponding Brillouin zones:
\begin{eqnarray}
	\left\langle \hat{c}^{1\dagger}_{(2j+s_1,2k+s_2)}\hat{c}^{0}_{(2j,2k)}\right\rangle \approx \int_{-k_\mathrm{rec}}^{k_\mathrm{rec}} dk_x dk_z \phi_1^*(k_x-(2j+s_1)k_\mathrm{rec},k_z-(2k+s_2)k_\mathrm{rec}) \phi_0(k_x-2jk_\mathrm{rec},k_z-2kk_\mathrm{rec}).\nonumber \\
\end{eqnarray}

\begin{figure}[thbp]
	\centering
	\includegraphics[width=0.6\columnwidth]{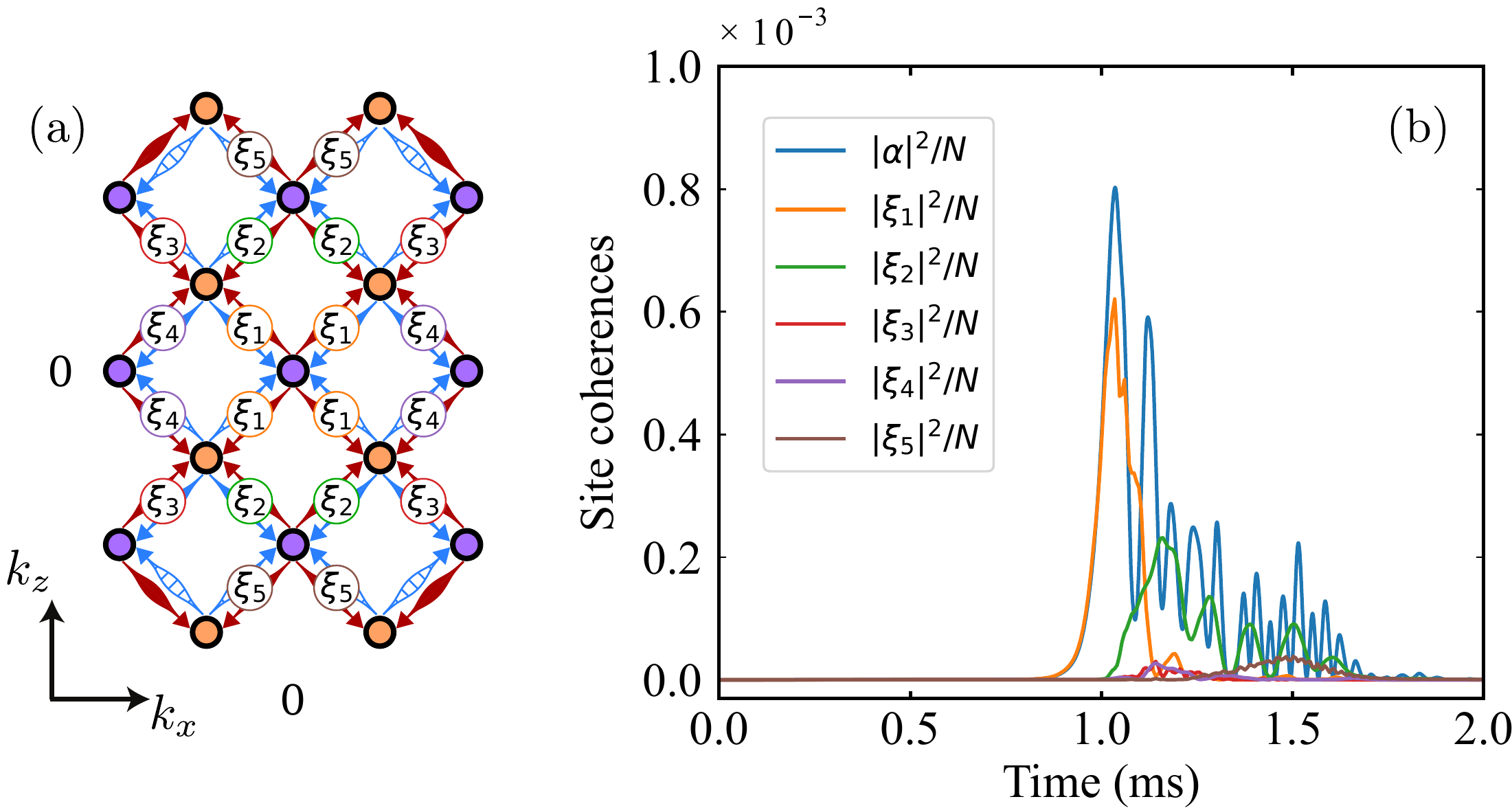}
	\caption{Cascaded currents in the momentum lattice: (a) Schematic representation of the two-site coherences $\xi_j$   associated with the different tunneling events in the lattice, and (b) GPS comparing the evolution of the cavity field $|\alpha|^2$ and $|\xi_j|^2$ for the simulations shown in Fig.~\ref{figSI:Sim_Fig4}. We note that the cavity field is related to the coherences as $\alpha = \sum_i \xi_i$, such that $|\alpha|^2\neq \sum_i|\xi_i|^2$ due to interference terms.}
	\label{figSI:Sim_hopping}
\end{figure}

Moreover, we scale the coherences by a factor of $\eta N/(\tilde{\Delta}_c+i\kappa)$ and group them according to the associated tunneling events in the momentum lattice 
\begin{equation}
	\begin{split}
		\xi_1 &= \frac{\eta N}{\tilde{\Delta}_c+i\kappa}\left(\theta_{0,0,+1,+1}+\theta_{0,0,+1,-1}+\theta_{0,0,-1,+1}+\theta_{0,0,-1,-1}\right), \\
		\xi_2 &= \frac{\eta N}{\tilde{\Delta}_c+i\kappa}\left(\theta_{0,2,+1,-1}+\theta_{0,2,-1,-1}+\theta_{0,-2,+1,+1}+\theta_{0,-2,-1,+1}\right), \\
		\xi_3 &= \frac{\eta N}{\tilde{\Delta}_c+i\kappa}\left(\theta_{2,2,-1,-1}+\theta_{2,-2,-1,+1}+\theta_{-2,2,+1,-1}+\theta_{-2,-2,+1,+1}\right), \\
		\xi_4 &= \frac{\eta N}{\tilde{\Delta}_c+i\kappa}\left(\theta_{2,0,-1,+1}+\theta_{2,0,-1,-1}+\theta_{-2,0,+1,+1}+\theta_{-2,0,+1,-1}\right), \\
		\xi_5 &= \frac{\eta N}{\tilde{\Delta}_c+i\kappa}\left(\theta_{0,2,+1,+1}+\theta_{0,2,-1,+1}+\theta_{0,-2,+1,-1}+\theta_{0,-2,-1,-1}\right),
	\end{split}
\end{equation}
as shown in the schematics in Fig.~\ref{figSI:Sim_hopping}(a).

In Fig.~\ref{figSI:Sim_hopping}(b), we present the dynamics of $\xi_j$ ($j \in\{1,$...$, 5\}$)  to elucidate the build-up of simultaneous two-site coherences in the parameter regime of Fig.~4 in the main text. Indeed, we observe a cascaded build-up of coherences between adjacent lattice sites accompanied by a strong photon pulse, indicating that the next tunneling event starts before the previous one is fully finished.

% with main contributions from $\xi_1$, $\xi_2$, and $\xi_5$ in different time period. These results again confirm the dynamical hopping process in the momentum lattice, and confirm that the photon bursts are collective effects of the local correlations. 

%		and show them in Fig.~\ref{figSI:Sim_hopping} as functions of time for the two simulations described above. For the first simulation [Fig.~\ref{figSI:Sim_hopping}(a)], we observe that indeed the first photon burst is mainly contributed by $\xi_1$, while the second and third photon bursts are mainly contributed by $\xi_2$. For the second simulation [Fig.~\ref{figSI:Sim_hopping}(b)],

%and show them in Fig.~\ref{figSI:Sim_hopping} as functions of time for the two simulations described above. For the first simulation [Fig.~\ref{figSI:Sim_hopping}(a)], we observe that indeed the first photon burst is mainly contributed by $\xi_1$, while the second and third photon bursts are mainly contributed by $\xi_2$. For the second simulation [Fig.~\ref{figSI:Sim_hopping}(b)], we observe that the correlations are indeed activated in sequence, with main contributions from $\xi_1$, $\xi_2$, and $\xi_5$ in different time period. These results again confirm the dynamical hopping process in the momentum lattice, and confirm that the photon bursts are collective effects of the local correlations. 

\subsection{Dynamics due to harmonic confinement and contact interactions}
\begin{figure}[thbp]
	\centering
	\includegraphics[width=0.36\columnwidth]{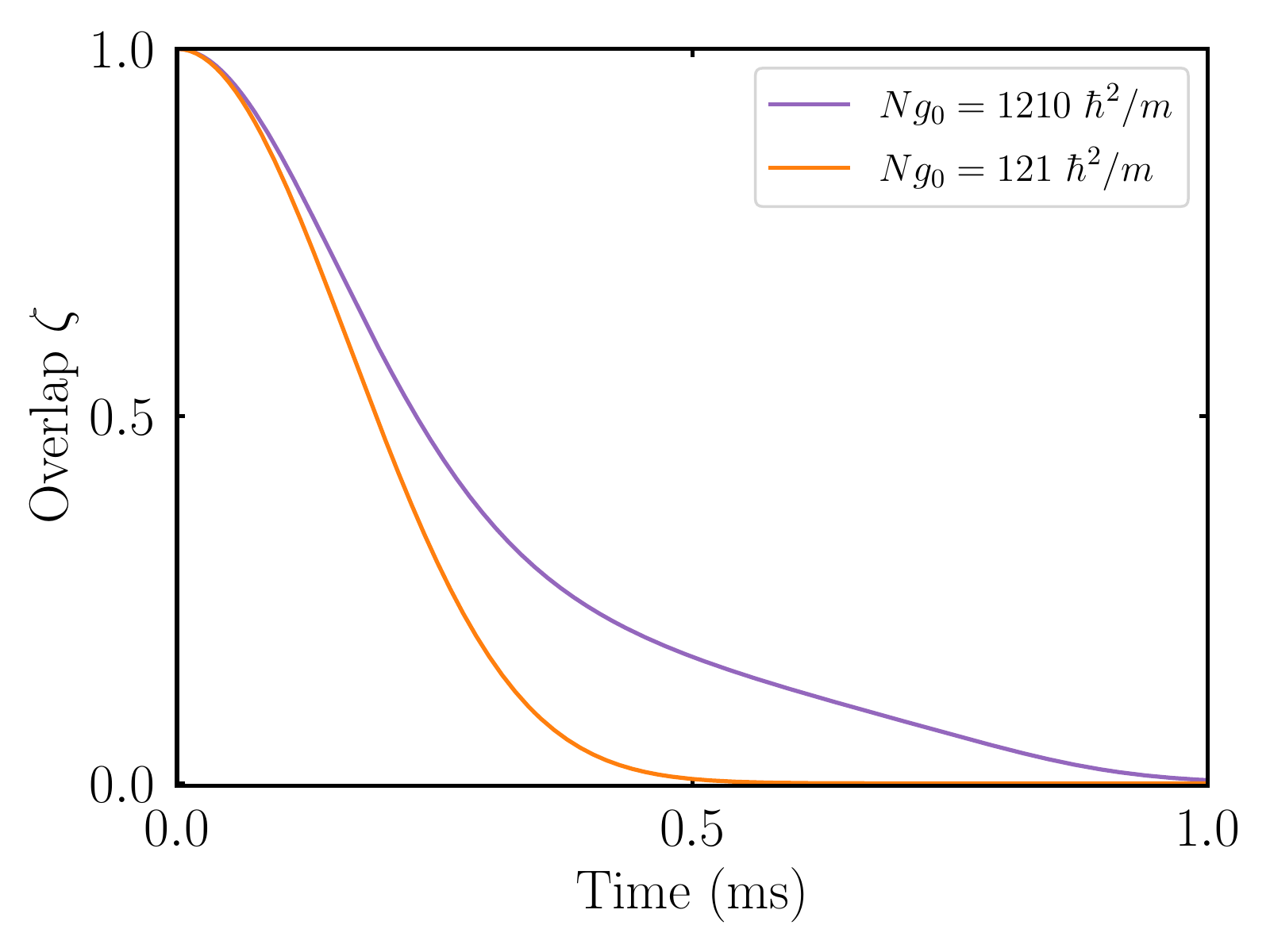}
	\caption{The overlap $\zeta$ [cf. Eq.~\eqref{eqSI:overlap}] as a function of time for the evolution of the state Eq.~\eqref{eqSI:overlap_initial} in a harmonic trap with strong (purple) and weak contact interactions (orange curve).} 
	\label{figSI:Sim_overlap1}
\end{figure}

\begin{figure}[thbp]
	\centering
	\includegraphics[width=0.8\columnwidth]{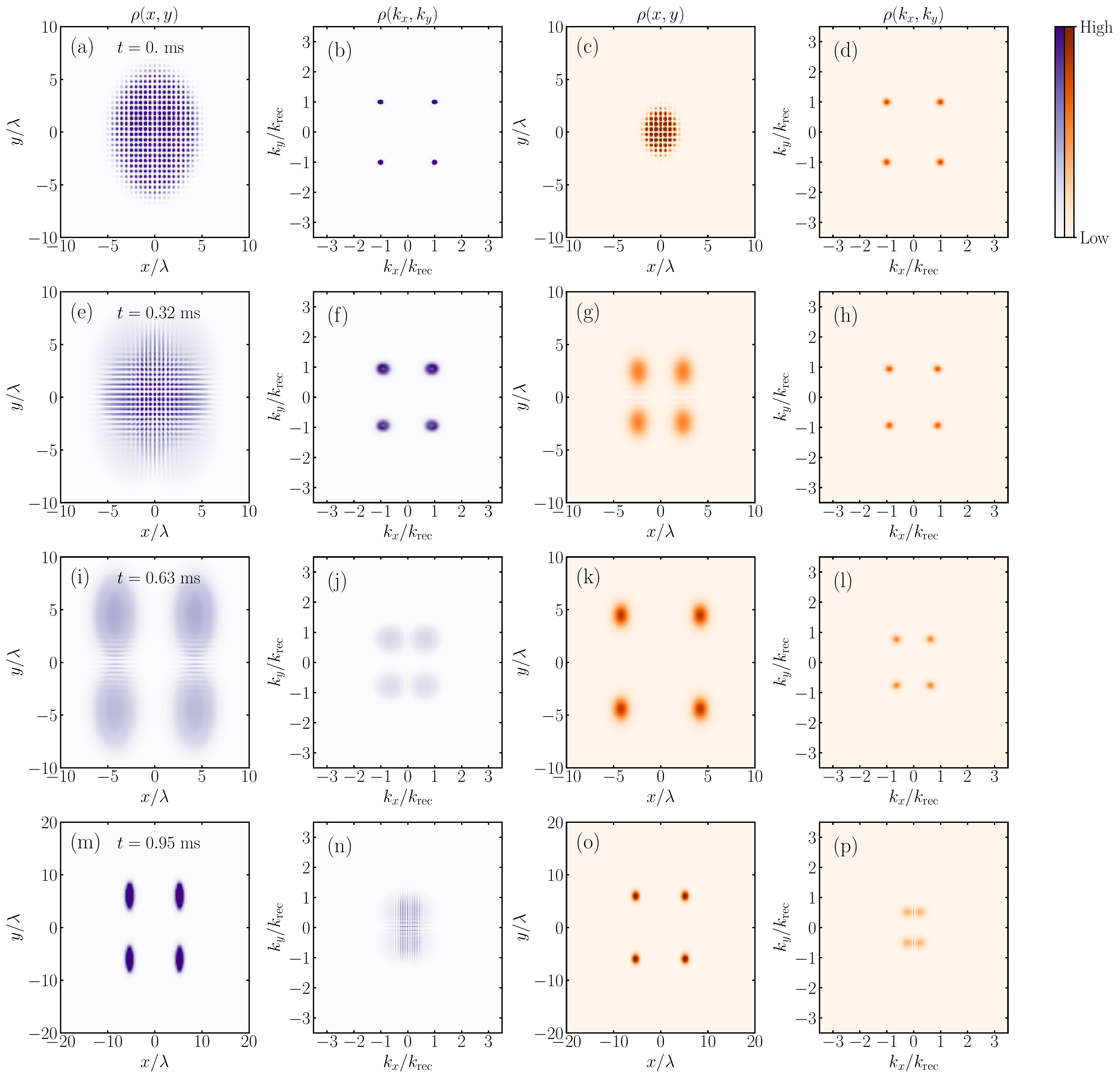}
	\caption{Representative snapshots of the evolution of the real and momentum space distributions for a 2D harmonically confined atomic cloud initialized in $\phi(x,z)$ with trap frequencies  $[\omega_\text{hx},\omega_\text{hz}]=2\pi\cdot[218,172]~\text{Hz}$  [cf.  Eq.~\eqref{eqSI:overlap_initial}]. The purple (orange) colormap corresponds to the experimentally relevant (smaller) contact interactions, which are on the order of  $Ng_0 = 1210~\hbar^2/m$ ($Ng_0 = 121~\hbar^2/m$).} 
	\label{figSI:Sim_overlap2}
\end{figure}

Harmonically confined Bose-Einstein condensates exhibit oscillatory motion when prepared away from their equilibrium configuration~\cite{Ketterle_1996,Dalfovo_1998}, for example through excited breathing modes~\cite{Chevy_2002}.  As the states in the momentum lattice generally differ from the equilibrium Thomas-Fermi distribution, we expect them to oscillate in real space in the trap. This moves the momentum components out of the grid nodes, progressively rendering the tight-binding picture invalid. In particular, we observed in the simulations in Fig.~\ref{figSI:Sim_Fig3}, that the lifetime of the momentum lattice ($\sim1~\text{ms}$)  is roughly on the same order as the inverse trap frequency. Since this time scale is on the same order of magnitude as the dynamics in the momentum lattice, it constitutes one of the main limitations of our scheme. Nevertheless, we observe in our simulations that contact interactions can increase this lifetime. To quantify this process, we perform Gross-Pitaevskii simulations for a spinless system, where we prepare an initial wave function
\begin{eqnarray}
	\phi(x,z) = \psi(x,z) \cos(k_\mathrm{rec}x)\cos(k_\mathrm{rec}z).
	\label{eqSI:overlap_initial}
\end{eqnarray}
The envelop function $\psi$  describes a Thomas-Fermi profile or a Gaussian profile, depending on whether contact interactions are considered or not. This state resembles the atomic state after the first tunneling event in the momentum lattice, i.e., $\ket{\pm1,\pm1}_1$ [cf. Figs.~\ref{figSI:Sim_Fig3}(j,l)]. 	We then propagate the state freely in the harmonic trap while enforcing the cavity field to be zero. The same simulation is performed for both the experimentally relevant contact interaction strength $Ng_0 = 1210~\hbar^2/m$ and for a smaller value $Ng_0 = 121~\hbar^2/m$. During the simulation, we measure the overlap between the instantaneous wave function and the initial one
\begin{eqnarray}
	\zeta = \left\vert\int dk_xdk_z \phi^*(k_x,k_z;t=0)\phi(k_x,k_z;t) \right\vert = \left\vert \int dxdz \phi^*(x,z;t=0)\phi(x,z;t) \right\vert,
	\label{eqSI:overlap}
\end{eqnarray}
and show it in Fig.~\ref{figSI:Sim_overlap1}. For the experimentally relevant interaction strength, the lifetime is roughly 1~ms, which is approximately twice longer than for a system with weak contact interactions. The real and momentum space densities of the two simulations are also shown in Fig.~\ref{figSI:Sim_overlap2}. We observe that strong contact interactions effectively diffuse the lattice peaks in momentum space, which slows down the evolution of the atomic distribution away from the lattice sites in momentum space.	We note that this lifetime is also consistent with the simulation results in Figs.~\ref{figSI:Sim_Fig3} and \ref{figSI:Sim_Fig4}, where oscillatory motion in the trap washes out the momentum lattice roughly $1~\text{ms}$ after the first photon pulse.

As a conclusion, these simulations of the Gross-Pitaevskii equations capture the dynamics observed in the experiment and validate the tight-binding description of the momentum lattice at sufficiently small times. In addition, they allow us to estimate the lifetime due to the dynamics in the harmonic trap in the presence of contact interactions. 

%\section{Further experimental characterization of spin-currents}

\newpage

\section{Complementary experimental results}   
In this section, we present additional experimental observations. They complement the superradiant tunneling events and the cascaded dynamics discussed in Fig.~2 and Fig.~4 in the main text, respectively.

\subsection{Delay of superradiant tunneling}

\begin{figure}[thbp]
	\centering
	\includegraphics[width=0.72\columnwidth]{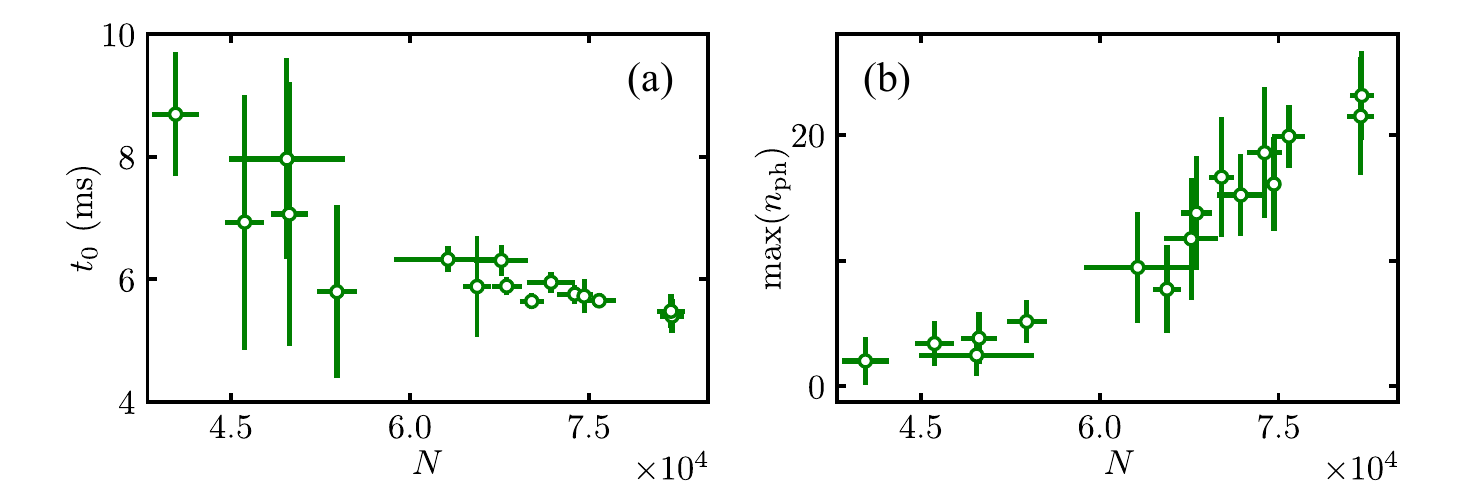}
	\caption{Superradiant tunneling in the momentum lattice. (a)  Average pulse delay $t_0$ versus atom number $N$. (b) Corresponding photon peak amplitude max($n_\text{ph}$), which is also shown in Fig.~2(b) in the main text.  The quantities are fitted using Eq.~\eqref{eqSI:SR_decay_photons} and the error bars represent the standard error of the mean. Here, $\tilde{\Delta}_c= - 2\pi\cdot 1.4(1)~\text{MHz}$ and $\omega_0=2\pi\cdot 26(1)~\text{kHz}$.}
	\label{fig:FigRR1}
\end{figure}

Here, we elaborate on the time delay of the photon pulses discussed in Fig.~2 in the main text. From our model in Eq.~\eqref{eqSI:SR_decay_photons}, we expect the delay $t_0$ to monotonically decrease with increasing atom number ($\propto 1/N$ ), if the dynamics is collectively enhanced. In Fig.~\ref{fig:FigRR1}(a), we display the average delay time $t_0$ and observe an overall monotonically decreasing trend with $N$. However, large fluctuations are clearly visible, especially at small atom numbers. From our model, we expect indeed a monotonically decreasing $\propto 1/N$ dependence with atom number, i.e., $t_0=-(\kappa/2N\eta_p^2)\ln(\epsilon^2/4)$. However, the exact value of the delay is directly affected by fluctuations ($\epsilon$) of the occupation of the initial spin state and cavity field, such as vacuum and classical fluctuations on top of the initially empty cavity mode. We attribute the large spread of $t_0$ at small atom numbers to this effect. In  contrast, the maximal pulse amplitude is not appreciably affected by these fluctuations  [Fig.~\ref{fig:FigRR1}(b) and Fig.~2 (b) in the main text], as it is expected to be independent of $\epsilon$, i.e, $max(n_\text{ph})=N^2\eta_p^2/4\kappa^2$. Hence, the latter appears to be the more robust observable, which allows to reliably fit the power-law scaling and experimentally diagnose superradiance in our experiment. 

\subsection{Spectrogram in cascaded dynamics regime}           

\begin{figure}[thbp]
	\centering
	\includegraphics[width=0.36\columnwidth]{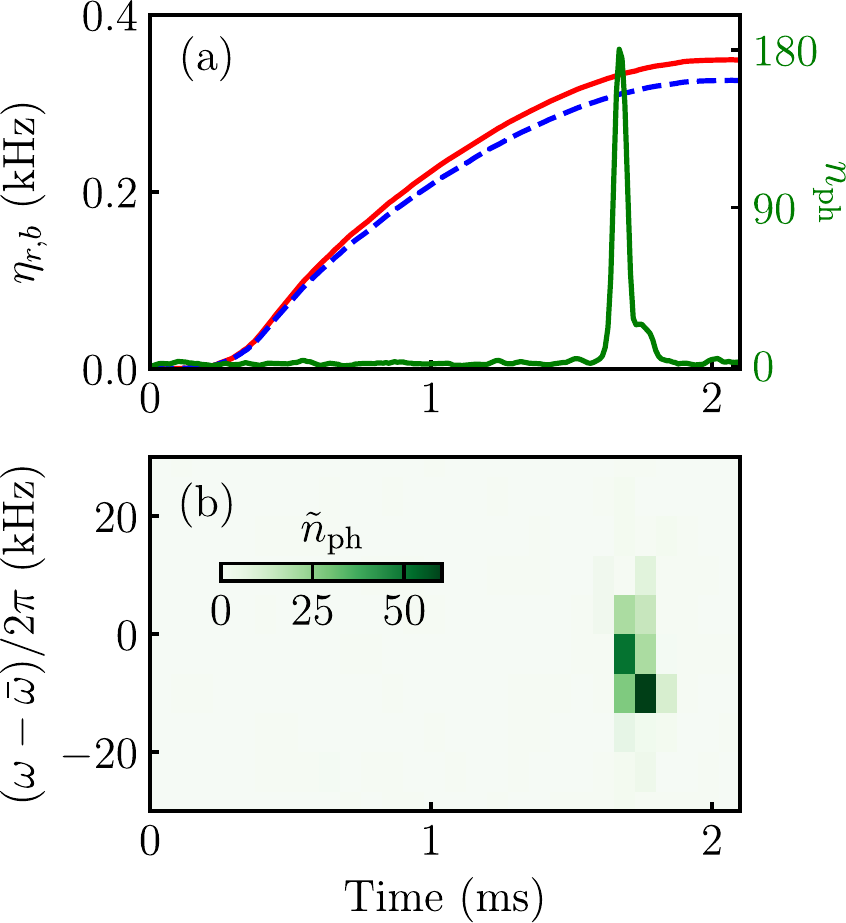}
	\caption{Cavity field spectrogram in the regime exhibiting  cascaded dynamics in the momentum lattice. (a) Representative realization, showing the coupling ramps and a single strong photon pulse and (b) corresponding photon number spectrogram. For these measurements, $N=9.1(1)\cdot10^4$ and $\tilde{\Delta}_c=-2\pi\cdot3.4(2)~\text{MHz}$, while the peak frequency of the emitted photon field is $\omega_p-\bar{\omega}=-2\pi\cdot10~\text{kHz}$.} 
	\label{figSI:PEandMomPeaks}
\end{figure}

In Fig.~\ref{figSI:PEandMomPeaks}(b), we present a typical photon number spectrogram showcasing a single photon pulse in the parameter regime of the cascaded dynamics discussed in Fig.~4 in the main text  [Fig.~\ref{figSI:PEandMomPeaks}(a)]. The spectrogram extends over $\sim7~\omega_\text{rec}$ in the frequency domain,  which is significantly larger than the resolution of the FFT ($\sim\omega_\text{rec}$) and suggests that multiple Raman transfers  between different states of the momentum grid can occur within the duration of a single photon pulse. This is confirmed by the absorption images showing the occupation of the momentum lattice in Fig.~4(e) in the main text.

\FloatBarrier
\newpage

%and (c) time-of-flight image, showing  the momentum space distribution of the spin states $\ket{0}$ (purple) and $\ket{1}$ (orange colormap). The red and black crosses denote the position of the modes $\ket{0,0}_0$ and $\ket{0,0}_1$ in the absorption image, respectively. The shift between these two positions is due to the Stern-Gerlach magnetic field gradient along $z$,  which is used to resolve the different $m_F$ components of the gas during TOF.

\bibliographystyle{apsrev}

\begin{thebibliography}{58}%
	\makeatletter
	\providecommand \@ifxundefined [1]{%
		\@ifx{#1\undefined}
	}%
	\providecommand \@ifnum [1]{%
		\ifnum #1\expandafter \@firstoftwo
		\else \expandafter \@secondoftwo
		\fi
	}%
	\providecommand \@ifx [1]{%
		\ifx #1\expandafter \@firstoftwo
		\else \expandafter \@secondoftwo
		\fi
	}%
	\providecommand \natexlab [1]{#1}%
	\providecommand \enquote  [1]{``#1''}%
	\providecommand \bibnamefont  [1]{#1}%
	\providecommand \bibfnamefont [1]{#1}%
	\providecommand \citenamefont [1]{#1}%
	\providecommand \href@noop [0]{\@secondoftwo}%
	\providecommand \href [0]{\begingroup \@sanitize@url \@href}%
	\providecommand \@href[1]{\@@startlink{#1}\@@href}%
	\providecommand \@@href[1]{\endgroup#1\@@endlink}%
	\providecommand \@sanitize@url [0]{\catcode `\\12\catcode `\$12\catcode
		`\&12\catcode `\#12\catcode `\^12\catcode `\_12\catcode `\%12\relax}%
	\providecommand \@@startlink[1]{}%
	\providecommand \@@endlink[0]{}%
	\providecommand \url  [0]{\begingroup\@sanitize@url \@url }%
	\providecommand \@url [1]{\endgroup\@href {#1}{\urlprefix }}%
	\providecommand \urlprefix  [0]{URL }%
	\providecommand \Eprint [0]{\href }%
	\providecommand \doibase [0]{https://doi.org/}%
	\providecommand \selectlanguage [0]{\@gobble}%
	\providecommand \bibinfo  [0]{\@secondoftwo}%
	\providecommand \bibfield  [0]{\@secondoftwo}%
	\providecommand \translation [1]{[#1]}%
	\providecommand \BibitemOpen [0]{}%
	\providecommand \bibitemStop [0]{}%
	\providecommand \bibitemNoStop [0]{.\EOS\space}%
	\providecommand \EOS [0]{\spacefactor3000\relax}%
	\providecommand \BibitemShut  [1]{\csname bibitem#1\endcsname}%
	\let\auto@bib@innerbib\@empty
	%</preamble>
	\bibitem [{\citenamefont {Langen}\ \emph {et~al.}(2015)\citenamefont {Langen},
		\citenamefont {Geiger},\ and\ \citenamefont {Schmiedmayer}}]{Langen_2015}%
	\BibitemOpen
	\bibfield  {author} {\bibinfo {author} {\bibfnamefont {T.}~\bibnamefont
			{Langen}}, \bibinfo {author} {\bibfnamefont {R.}~\bibnamefont {Geiger}},\
		and\ \bibinfo {author} {\bibfnamefont {J.}~\bibnamefont {Schmiedmayer}},\
	}\bibfield  {title} {\bibinfo {title} {{Ultracold Atoms Out of
				Equilibrium}},\ }\href
	{https://doi.org/10.1146/annurev-conmatphys-031214-014548} {\bibfield
		{journal} {\bibinfo  {journal} {Annual Review of Condensed Matter Physics}\
		}\textbf {\bibinfo {volume} {6}},\ \bibinfo {pages} {201} (\bibinfo {year}
		{2015})}\BibitemShut {NoStop}%
	\bibitem [{\citenamefont {Gross}\ and\ \citenamefont
		{Bloch}(2017)}]{Gross_2017}%
	\BibitemOpen
	\bibfield  {author} {\bibinfo {author} {\bibfnamefont {C.}~\bibnamefont
			{Gross}}\ and\ \bibinfo {author} {\bibfnamefont {I.}~\bibnamefont {Bloch}},\
	}\bibfield  {title} {\bibinfo {title} {{Quantum simulations with ultracold
				atoms in optical lattices}},\ }\href
	{https://doi.org/10.1126/science.aal3837} {\bibfield  {journal} {\bibinfo
			{journal} {Science}\ }\textbf {\bibinfo {volume} {357}},\ \bibinfo {pages}
		{995} (\bibinfo {year} {2017})}\BibitemShut {NoStop}%
	\bibitem [{\citenamefont {Altman}\ \emph {et~al.}(2021)\citenamefont {Altman},
		\citenamefont {Brown}, \citenamefont {Carleo}, \citenamefont {Carr},
		\citenamefont {Demler}, \citenamefont {Chin}, \citenamefont {DeMarco},
		\citenamefont {Economou}, \citenamefont {Eriksson}, \citenamefont {Fu},
		\citenamefont {Greiner}, \citenamefont {Hazzard}, \citenamefont {Hulet},
		\citenamefont {Koll\'ar}, \citenamefont {Lev}, \citenamefont {Lukin},
		\citenamefont {Ma}, \citenamefont {Mi}, \citenamefont {Misra}, \citenamefont
		{Monroe}, \citenamefont {Murch}, \citenamefont {Nazario}, \citenamefont {Ni},
		\citenamefont {Potter}, \citenamefont {Roushan}, \citenamefont {Saffman},
		\citenamefont {Schleier-Smith}, \citenamefont {Siddiqi}, \citenamefont
		{Simmonds}, \citenamefont {Singh}, \citenamefont {Spielman}, \citenamefont
		{Temme}, \citenamefont {Weiss}, \citenamefont {Vu\ifmmode \check{c}\else
			\v{c}\fi{}kovi\ifmmode~\acute{c}\else \'{c}\fi{}}, \citenamefont
		{Vuleti\ifmmode~\acute{c}\else \'{c}\fi{}}, \citenamefont {Ye},\ and\
		\citenamefont {Zwierlein}}]{PRXRoadmap_2021}%
	\BibitemOpen
	\bibfield  {author} {\bibinfo {author} {\bibfnamefont {E.}~\bibnamefont
			{Altman}}, \bibinfo {author} {\bibfnamefont {K.~R.}\ \bibnamefont {Brown}},
		\bibinfo {author} {\bibfnamefont {G.}~\bibnamefont {Carleo}}, \bibinfo
		{author} {\bibfnamefont {L.~D.}\ \bibnamefont {Carr}}, \bibinfo {author}
		{\bibfnamefont {E.}~\bibnamefont {Demler}}, \bibinfo {author} {\bibfnamefont
			{C.}~\bibnamefont {Chin}}, \bibinfo {author} {\bibfnamefont {B.}~\bibnamefont
			{DeMarco}}, \bibinfo {author} {\bibfnamefont {S.~E.}\ \bibnamefont
			{Economou}}, \bibinfo {author} {\bibfnamefont {M.~A.}\ \bibnamefont
			{Eriksson}}, \bibinfo {author} {\bibfnamefont {K.-M.~C.}\ \bibnamefont {Fu}},
		\bibinfo {author} {\bibfnamefont {M.}~\bibnamefont {Greiner}}, \bibinfo
		{author} {\bibfnamefont {K.~R.}\ \bibnamefont {Hazzard}}, \bibinfo {author}
		{\bibfnamefont {R.~G.}\ \bibnamefont {Hulet}}, \bibinfo {author}
		{\bibfnamefont {A.~J.}\ \bibnamefont {Koll\'ar}}, \bibinfo {author}
		{\bibfnamefont {B.~L.}\ \bibnamefont {Lev}}, \bibinfo {author} {\bibfnamefont
			{M.~D.}\ \bibnamefont {Lukin}}, \bibinfo {author} {\bibfnamefont
			{R.}~\bibnamefont {Ma}}, \bibinfo {author} {\bibfnamefont {X.}~\bibnamefont
			{Mi}}, \bibinfo {author} {\bibfnamefont {S.}~\bibnamefont {Misra}}, \bibinfo
		{author} {\bibfnamefont {C.}~\bibnamefont {Monroe}}, \bibinfo {author}
		{\bibfnamefont {K.}~\bibnamefont {Murch}}, \bibinfo {author} {\bibfnamefont
			{Z.}~\bibnamefont {Nazario}}, \bibinfo {author} {\bibfnamefont {K.-K.}\
			\bibnamefont {Ni}}, \bibinfo {author} {\bibfnamefont {A.~C.}\ \bibnamefont
			{Potter}}, \bibinfo {author} {\bibfnamefont {P.}~\bibnamefont {Roushan}},
		\bibinfo {author} {\bibfnamefont {M.}~\bibnamefont {Saffman}}, \bibinfo
		{author} {\bibfnamefont {M.}~\bibnamefont {Schleier-Smith}}, \bibinfo
		{author} {\bibfnamefont {I.}~\bibnamefont {Siddiqi}}, \bibinfo {author}
		{\bibfnamefont {R.}~\bibnamefont {Simmonds}}, \bibinfo {author}
		{\bibfnamefont {M.}~\bibnamefont {Singh}}, \bibinfo {author} {\bibfnamefont
			{I.}~\bibnamefont {Spielman}}, \bibinfo {author} {\bibfnamefont
			{K.}~\bibnamefont {Temme}}, \bibinfo {author} {\bibfnamefont {D.~S.}\
			\bibnamefont {Weiss}}, \bibinfo {author} {\bibfnamefont {J.}~\bibnamefont
			{Vu\ifmmode \check{c}\else \v{c}\fi{}kovi\ifmmode~\acute{c}\else
				\'{c}\fi{}}}, \bibinfo {author} {\bibfnamefont {V.}~\bibnamefont
			{Vuleti\ifmmode~\acute{c}\else \'{c}\fi{}}}, \bibinfo {author} {\bibfnamefont
			{J.}~\bibnamefont {Ye}},\ and\ \bibinfo {author} {\bibfnamefont
			{M.}~\bibnamefont {Zwierlein}},\ }\bibfield  {title} {\bibinfo {title}
		{{Quantum Simulators: Architectures and Opportunities}},\ }\href
	{https://doi.org/10.1103/PRXQuantum.2.017003} {\bibfield  {journal} {\bibinfo
			{journal} {PRX Quantum}\ }\textbf {\bibinfo {volume} {2}},\ \bibinfo {pages}
		{017003} (\bibinfo {year} {2021})}\BibitemShut {NoStop}%
	\bibitem [{\citenamefont {Wiese}(2013)}]{Wiese_2013}%
	\BibitemOpen
	\bibfield  {author} {\bibinfo {author} {\bibfnamefont {U.-J.}\ \bibnamefont
			{Wiese}},\ }\bibfield  {title} {\bibinfo {title} {{Ultracold quantum gases
				and lattice systems: quantum simulation of lattice gauge theories}},\ }\href
	{https://doi.org/https://doi.org/10.1002/andp.201300104} {\bibfield
		{journal} {\bibinfo  {journal} {Annalen der Physik}\ }\textbf {\bibinfo
			{volume} {525}},\ \bibinfo {pages} {777} (\bibinfo {year}
		{2013})}\BibitemShut {NoStop}%
	\bibitem [{\citenamefont {Zohar}\ \emph {et~al.}(2015)\citenamefont {Zohar},
		\citenamefont {Cirac},\ and\ \citenamefont {Reznik}}]{Zohar_2015}%
	\BibitemOpen
	\bibfield  {author} {\bibinfo {author} {\bibfnamefont {E.}~\bibnamefont
			{Zohar}}, \bibinfo {author} {\bibfnamefont {J.~I.}\ \bibnamefont {Cirac}},\
		and\ \bibinfo {author} {\bibfnamefont {B.}~\bibnamefont {Reznik}},\
	}\bibfield  {title} {\bibinfo {title} {{Quantum simulations of lattice gauge
				theories using ultracold atoms in optical lattices}},\ }\href
	{https://doi.org/10.1088/0034-4885/79/1/014401} {\bibfield  {journal}
		{\bibinfo  {journal} {Reports on Progress in Physics}\ }\textbf {\bibinfo
			{volume} {79}},\ \bibinfo {pages} {014401} (\bibinfo {year}
		{2015})}\BibitemShut {NoStop}%
	\bibitem [{\citenamefont {Kasper}\ \emph {et~al.}(2020)\citenamefont {Kasper},
		\citenamefont {Juzeli{\={u}}nas}, \citenamefont {Lewenstein}, \citenamefont
		{Jendrzejewski},\ and\ \citenamefont {Zohar}}]{Kasper_2020}%
	\BibitemOpen
	\bibfield  {author} {\bibinfo {author} {\bibfnamefont {V.}~\bibnamefont
			{Kasper}}, \bibinfo {author} {\bibfnamefont {G.}~\bibnamefont
			{Juzeli{\={u}}nas}}, \bibinfo {author} {\bibfnamefont {M.}~\bibnamefont
			{Lewenstein}}, \bibinfo {author} {\bibfnamefont {F.}~\bibnamefont
			{Jendrzejewski}},\ and\ \bibinfo {author} {\bibfnamefont {E.}~\bibnamefont
			{Zohar}},\ }\bibfield  {title} {\bibinfo {title} {{From the
				Jaynes{\textendash}Cummings model to non-abelian gauge theories: a guided
				tour for the quantum engineer}},\ }\href
	{https://doi.org/10.1088/1367-2630/abb961} {\bibfield  {journal} {\bibinfo
			{journal} {New Journal of Physics}\ }\textbf {\bibinfo {volume} {22}},\
		\bibinfo {pages} {103027} (\bibinfo {year} {2020})}\BibitemShut {NoStop}%
	\bibitem [{\citenamefont {Bissbort}\ \emph {et~al.}(2013)\citenamefont
		{Bissbort}, \citenamefont {Cocks}, \citenamefont {Negretti}, \citenamefont
		{Idziaszek}, \citenamefont {Calarco}, \citenamefont {Schmidt-Kaler},
		\citenamefont {Hofstetter},\ and\ \citenamefont {Gerritsma}}]{Bissbort_2013}%
	\BibitemOpen
	\bibfield  {author} {\bibinfo {author} {\bibfnamefont {U.}~\bibnamefont
			{Bissbort}}, \bibinfo {author} {\bibfnamefont {D.}~\bibnamefont {Cocks}},
		\bibinfo {author} {\bibfnamefont {A.}~\bibnamefont {Negretti}}, \bibinfo
		{author} {\bibfnamefont {Z.}~\bibnamefont {Idziaszek}}, \bibinfo {author}
		{\bibfnamefont {T.}~\bibnamefont {Calarco}}, \bibinfo {author} {\bibfnamefont
			{F.}~\bibnamefont {Schmidt-Kaler}}, \bibinfo {author} {\bibfnamefont
			{W.}~\bibnamefont {Hofstetter}},\ and\ \bibinfo {author} {\bibfnamefont
			{R.}~\bibnamefont {Gerritsma}},\ }\bibfield  {title} {\bibinfo {title}
		{{Emulating Solid-State Physics with a Hybrid System of Ultracold Ions and
				Atoms}},\ }\href {https://doi.org/10.1103/PhysRevLett.111.080501} {\bibfield
		{journal} {\bibinfo  {journal} {Phys. Rev. Lett.}\ }\textbf {\bibinfo
			{volume} {111}},\ \bibinfo {pages} {080501} (\bibinfo {year}
		{2013})}\BibitemShut {NoStop}%
	\bibitem [{\citenamefont {Giustino}(2017)}]{Giustino_2017}%
	\BibitemOpen
	\bibfield  {author} {\bibinfo {author} {\bibfnamefont {F.}~\bibnamefont
			{Giustino}},\ }\bibfield  {title} {\bibinfo {title} {{Electron-phonon
				interactions from first principles}},\ }\href
	{https://doi.org/10.1103/RevModPhys.89.015003} {\bibfield  {journal}
		{\bibinfo  {journal} {Rev. Mod. Phys.}\ }\textbf {\bibinfo {volume} {89}},\
		\bibinfo {pages} {015003} (\bibinfo {year} {2017})}\BibitemShut {NoStop}%
	\bibitem [{\citenamefont {Eckholt}\ and\ \citenamefont
		{Garc{\'{\i}}a-Ripoll}(2009)}]{Eckholt_2009}%
	\BibitemOpen
	\bibfield  {author} {\bibinfo {author} {\bibfnamefont {M.}~\bibnamefont
			{Eckholt}}\ and\ \bibinfo {author} {\bibfnamefont {J.~J.}\ \bibnamefont
			{Garc{\'{\i}}a-Ripoll}},\ }\bibfield  {title} {\bibinfo {title} {{Correlated
				hopping of bosonic atoms induced by optical lattices}},\ }\href
	{https://doi.org/10.1088/1367-2630/11/9/093028} {\bibfield  {journal}
		{\bibinfo  {journal} {New Journal of Physics}\ }\textbf {\bibinfo {volume}
			{11}},\ \bibinfo {pages} {093028} (\bibinfo {year} {2009})}\BibitemShut
	{NoStop}%
	\bibitem [{\citenamefont {Rapp}\ \emph {et~al.}(2012)\citenamefont {Rapp},
		\citenamefont {Deng},\ and\ \citenamefont {Santos}}]{Rapp_2012}%
	\BibitemOpen
	\bibfield  {author} {\bibinfo {author} {\bibfnamefont {A.}~\bibnamefont
			{Rapp}}, \bibinfo {author} {\bibfnamefont {X.}~\bibnamefont {Deng}},\ and\
		\bibinfo {author} {\bibfnamefont {L.}~\bibnamefont {Santos}},\ }\bibfield
	{title} {\bibinfo {title} {{Ultracold Lattice Gases with Periodically
				Modulated Interactions}},\ }\href
	{https://doi.org/10.1103/PhysRevLett.109.203005} {\bibfield  {journal}
		{\bibinfo  {journal} {Phys. Rev. Lett.}\ }\textbf {\bibinfo {volume} {109}},\
		\bibinfo {pages} {203005} (\bibinfo {year} {2012})}\BibitemShut {NoStop}%
	\bibitem [{\citenamefont {Liberto}\ \emph {et~al.}(2014)\citenamefont
		{Liberto}, \citenamefont {Creffield}, \citenamefont {Japaridze},\ and\
		\citenamefont {Smith}}]{Di_2014}%
	\BibitemOpen
	\bibfield  {author} {\bibinfo {author} {\bibfnamefont {M.~D.}\ \bibnamefont
			{Liberto}}, \bibinfo {author} {\bibfnamefont {C.~E.}\ \bibnamefont
			{Creffield}}, \bibinfo {author} {\bibfnamefont {G.~I.}\ \bibnamefont
			{Japaridze}},\ and\ \bibinfo {author} {\bibfnamefont {C.~M.}\ \bibnamefont
			{Smith}},\ }\bibfield  {title} {\bibinfo {title} {{Quantum simulation of
				correlated-hopping models with fermions in optical lattices}},\ }\href
	{https://doi.org/10.1103/PhysRevA.89.013624} {\bibfield  {journal} {\bibinfo
			{journal} {Phys. Rev. A}\ }\textbf {\bibinfo {volume} {89}},\ \bibinfo
		{pages} {013624} (\bibinfo {year} {2014})}\BibitemShut {NoStop}%
	\bibitem [{\citenamefont {Zhao}\ \emph {et~al.}(2020)\citenamefont {Zhao},
		\citenamefont {Vovrosh}, \citenamefont {Mintert},\ and\ \citenamefont
		{Knolle}}]{Zhao_2020}%
	\BibitemOpen
	\bibfield  {author} {\bibinfo {author} {\bibfnamefont {H.}~\bibnamefont
			{Zhao}}, \bibinfo {author} {\bibfnamefont {J.}~\bibnamefont {Vovrosh}},
		\bibinfo {author} {\bibfnamefont {F.}~\bibnamefont {Mintert}},\ and\ \bibinfo
		{author} {\bibfnamefont {J.}~\bibnamefont {Knolle}},\ }\bibfield  {title}
	{\bibinfo {title} {{Quantum Many-Body Scars in Optical Lattices}},\ }\href
	{https://doi.org/10.1103/PhysRevLett.124.160604} {\bibfield  {journal}
		{\bibinfo  {journal} {Phys. Rev. Lett.}\ }\textbf {\bibinfo {volume} {124}},\
		\bibinfo {pages} {160604} (\bibinfo {year} {2020})}\BibitemShut {NoStop}%
	\bibitem [{\citenamefont {Hudomal}\ \emph {et~al.}(2020)\citenamefont
		{Hudomal}, \citenamefont {Vasi{\'{c}}}, \citenamefont {Regnault},\ and\
		\citenamefont {Papi{\'{c}}}}]{Hudomal_2020}%
	\BibitemOpen
	\bibfield  {author} {\bibinfo {author} {\bibfnamefont {A.}~\bibnamefont
			{Hudomal}}, \bibinfo {author} {\bibfnamefont {I.}~\bibnamefont
			{Vasi{\'{c}}}}, \bibinfo {author} {\bibfnamefont {N.}~\bibnamefont
			{Regnault}},\ and\ \bibinfo {author} {\bibfnamefont {Z.}~\bibnamefont
			{Papi{\'{c}}}},\ }\bibfield  {title} {\bibinfo {title} {{Quantum scars of
				bosons with correlated hopping}},\ }\href
	{https://doi.org/10.1038/s42005-020-0364-9} {\bibfield  {journal} {\bibinfo
			{journal} {Communications Physics}\ }\textbf {\bibinfo {volume} {3}},\
		\bibinfo {pages} {99} (\bibinfo {year} {2020})}\BibitemShut {NoStop}%
	\bibitem [{\citenamefont {Ma}\ \emph {et~al.}(2011)\citenamefont {Ma},
		\citenamefont {Tai}, \citenamefont {Preiss}, \citenamefont {Bakr},
		\citenamefont {Simon},\ and\ \citenamefont {Greiner}}]{Ma_2011}%
	\BibitemOpen
	\bibfield  {author} {\bibinfo {author} {\bibfnamefont {R.}~\bibnamefont
			{Ma}}, \bibinfo {author} {\bibfnamefont {M.~E.}\ \bibnamefont {Tai}},
		\bibinfo {author} {\bibfnamefont {P.~M.}\ \bibnamefont {Preiss}}, \bibinfo
		{author} {\bibfnamefont {W.~S.}\ \bibnamefont {Bakr}}, \bibinfo {author}
		{\bibfnamefont {J.}~\bibnamefont {Simon}},\ and\ \bibinfo {author}
		{\bibfnamefont {M.}~\bibnamefont {Greiner}},\ }\bibfield  {title} {\bibinfo
		{title} {{Photon-Assisted Tunneling in a Biased Strongly Correlated Bose
				Gas}},\ }\href {https://doi.org/10.1103/PhysRevLett.107.095301} {\bibfield
		{journal} {\bibinfo  {journal} {Phys. Rev. Lett.}\ }\textbf {\bibinfo
			{volume} {107}},\ \bibinfo {pages} {095301} (\bibinfo {year}
		{2011})}\BibitemShut {NoStop}%
	\bibitem [{\citenamefont {Meinert}\ \emph {et~al.}(2016)\citenamefont
		{Meinert}, \citenamefont {Mark}, \citenamefont {Lauber}, \citenamefont
		{Daley},\ and\ \citenamefont {N\"agerl}}]{Meinert_2016}%
	\BibitemOpen
	\bibfield  {author} {\bibinfo {author} {\bibfnamefont {F.}~\bibnamefont
			{Meinert}}, \bibinfo {author} {\bibfnamefont {M.~J.}\ \bibnamefont {Mark}},
		\bibinfo {author} {\bibfnamefont {K.}~\bibnamefont {Lauber}}, \bibinfo
		{author} {\bibfnamefont {A.~J.}\ \bibnamefont {Daley}},\ and\ \bibinfo
		{author} {\bibfnamefont {H.-C.}\ \bibnamefont {N\"agerl}},\ }\bibfield
	{title} {\bibinfo {title} {{Floquet Engineering of Correlated Tunneling in
				the Bose-Hubbard Model with Ultracold Atoms}},\ }\href
	{https://doi.org/10.1103/PhysRevLett.116.205301} {\bibfield  {journal}
		{\bibinfo  {journal} {Phys. Rev. Lett.}\ }\textbf {\bibinfo {volume} {116}},\
		\bibinfo {pages} {205301} (\bibinfo {year} {2016})}\BibitemShut {NoStop}%
	\bibitem [{\citenamefont {Xu}\ \emph {et~al.}(2018)\citenamefont {Xu},
		\citenamefont {Morong}, \citenamefont {Hui}, \citenamefont {Scarola},\ and\
		\citenamefont {DeMarco}}]{Xu_2018}%
	\BibitemOpen
	\bibfield  {author} {\bibinfo {author} {\bibfnamefont {W.}~\bibnamefont
			{Xu}}, \bibinfo {author} {\bibfnamefont {W.}~\bibnamefont {Morong}}, \bibinfo
		{author} {\bibfnamefont {H.-Y.}\ \bibnamefont {Hui}}, \bibinfo {author}
		{\bibfnamefont {V.~W.}\ \bibnamefont {Scarola}},\ and\ \bibinfo {author}
		{\bibfnamefont {B.}~\bibnamefont {DeMarco}},\ }\bibfield  {title} {\bibinfo
		{title} {{Correlated spin-flip tunneling in a {F}ermi lattice gas}},\ }\href
	{https://doi.org/10.1103/PhysRevA.98.023623} {\bibfield  {journal} {\bibinfo
			{journal} {Phys. Rev. A}\ }\textbf {\bibinfo {volume} {98}},\ \bibinfo
		{pages} {023623} (\bibinfo {year} {2018})}\BibitemShut {NoStop}%
	\bibitem [{\citenamefont {Clark}\ \emph {et~al.}(2018)\citenamefont {Clark},
		\citenamefont {Anderson}, \citenamefont {Feng}, \citenamefont {Gaj},
		\citenamefont {Levin},\ and\ \citenamefont {Chin}}]{Clark_2018}%
	\BibitemOpen
	\bibfield  {author} {\bibinfo {author} {\bibfnamefont {L.~W.}\ \bibnamefont
			{Clark}}, \bibinfo {author} {\bibfnamefont {B.~M.}\ \bibnamefont {Anderson}},
		\bibinfo {author} {\bibfnamefont {L.}~\bibnamefont {Feng}}, \bibinfo {author}
		{\bibfnamefont {A.}~\bibnamefont {Gaj}}, \bibinfo {author} {\bibfnamefont
			{K.}~\bibnamefont {Levin}},\ and\ \bibinfo {author} {\bibfnamefont
			{C.}~\bibnamefont {Chin}},\ }\bibfield  {title} {\bibinfo {title}
		{Observation of density-dependent gauge fields in a {Bose-Einstein}
			condensate based on micromotion control in a shaken two-dimensional
			lattice},\ }\href {https://doi.org/10.1103/PhysRevLett.121.030402} {\bibfield
		{journal} {\bibinfo  {journal} {Phys. Rev. Lett.}\ }\textbf {\bibinfo
			{volume} {121}},\ \bibinfo {pages} {030402} (\bibinfo {year}
		{2018})}\BibitemShut {NoStop}%
	\bibitem [{\citenamefont {G{\"o}rg}\ \emph {et~al.}(2019)\citenamefont
		{G{\"o}rg}, \citenamefont {Sandholzer}, \citenamefont {Minguzzi},
		\citenamefont {Desbuquois}, \citenamefont {Messer},\ and\ \citenamefont
		{Esslinger}}]{Gorg_2019}%
	\BibitemOpen
	\bibfield  {author} {\bibinfo {author} {\bibfnamefont {F.}~\bibnamefont
			{G{\"o}rg}}, \bibinfo {author} {\bibfnamefont {K.}~\bibnamefont
			{Sandholzer}}, \bibinfo {author} {\bibfnamefont {J.}~\bibnamefont
			{Minguzzi}}, \bibinfo {author} {\bibfnamefont {R.}~\bibnamefont
			{Desbuquois}}, \bibinfo {author} {\bibfnamefont {M.}~\bibnamefont {Messer}},\
		and\ \bibinfo {author} {\bibfnamefont {T.}~\bibnamefont {Esslinger}},\
	}\bibfield  {title} {\bibinfo {title} {{Realization of density-dependent
				Peierls phases to engineer quantized gauge fields coupled to ultracold
				matter}},\ }\href {https://doi.org/10.1038/s41567-019-0615-4} {\bibfield
		{journal} {\bibinfo  {journal} {Nature Physics}\ }\textbf {\bibinfo {volume}
			{15}},\ \bibinfo {pages} {1161} (\bibinfo {year} {2019})}\BibitemShut
	{NoStop}%
	\bibitem [{\citenamefont {Schweizer}\ \emph {et~al.}(2019)\citenamefont
		{Schweizer}, \citenamefont {Grusdt}, \citenamefont {Berngruber},
		\citenamefont {Barbiero}, \citenamefont {Demler}, \citenamefont {Goldman},
		\citenamefont {Bloch},\ and\ \citenamefont {Aidelsburger}}]{Schweizer_2019}%
	\BibitemOpen
	\bibfield  {author} {\bibinfo {author} {\bibfnamefont {C.}~\bibnamefont
			{Schweizer}}, \bibinfo {author} {\bibfnamefont {F.}~\bibnamefont {Grusdt}},
		\bibinfo {author} {\bibfnamefont {M.}~\bibnamefont {Berngruber}}, \bibinfo
		{author} {\bibfnamefont {L.}~\bibnamefont {Barbiero}}, \bibinfo {author}
		{\bibfnamefont {E.}~\bibnamefont {Demler}}, \bibinfo {author} {\bibfnamefont
			{N.}~\bibnamefont {Goldman}}, \bibinfo {author} {\bibfnamefont
			{I.}~\bibnamefont {Bloch}},\ and\ \bibinfo {author} {\bibfnamefont
			{M.}~\bibnamefont {Aidelsburger}},\ }\bibfield  {title} {\bibinfo {title}
		{Floquet approach to $\mathbb{Z}$2 lattice gauge theories with ultracold
			atoms in optical lattices},\ }\href
	{https://doi.org/10.1038/s41567-019-0649-7} {\bibfield  {journal} {\bibinfo
			{journal} {Nature Physics}\ }\textbf {\bibinfo {volume} {15}},\ \bibinfo
		{pages} {1168} (\bibinfo {year} {2019})}\BibitemShut {NoStop}%
	\bibitem [{\citenamefont {Baier}\ \emph {et~al.}(2016)\citenamefont {Baier},
		\citenamefont {Mark}, \citenamefont {Petter}, \citenamefont {Aikawa},
		\citenamefont {Chomaz}, \citenamefont {Cai}, \citenamefont {Baranov},
		\citenamefont {Zoller},\ and\ \citenamefont {Ferlaino}}]{Baier_2016}%
	\BibitemOpen
	\bibfield  {author} {\bibinfo {author} {\bibfnamefont {S.}~\bibnamefont
			{Baier}}, \bibinfo {author} {\bibfnamefont {M.~J.}\ \bibnamefont {Mark}},
		\bibinfo {author} {\bibfnamefont {D.}~\bibnamefont {Petter}}, \bibinfo
		{author} {\bibfnamefont {K.}~\bibnamefont {Aikawa}}, \bibinfo {author}
		{\bibfnamefont {L.}~\bibnamefont {Chomaz}}, \bibinfo {author} {\bibfnamefont
			{Z.}~\bibnamefont {Cai}}, \bibinfo {author} {\bibfnamefont {M.}~\bibnamefont
			{Baranov}}, \bibinfo {author} {\bibfnamefont {P.}~\bibnamefont {Zoller}},\
		and\ \bibinfo {author} {\bibfnamefont {F.}~\bibnamefont {Ferlaino}},\
	}\bibfield  {title} {\bibinfo {title} {{Extended Bose-Hubbard models with
				ultracold magnetic atoms}},\ }\href {https://doi.org/10.1126/science.aac9812}
	{\bibfield  {journal} {\bibinfo  {journal} {Science}\ }\textbf {\bibinfo
			{volume} {352}},\ \bibinfo {pages} {201} (\bibinfo {year}
		{2016})}\BibitemShut {NoStop}%
	\bibitem [{\citenamefont {Gadway}(2015)}]{Gadway_2015}%
	\BibitemOpen
	\bibfield  {author} {\bibinfo {author} {\bibfnamefont {B.}~\bibnamefont
			{Gadway}},\ }\bibfield  {title} {\bibinfo {title} {{Atom-optics approach to
				studying transport phenomena}},\ }\href
	{https://doi.org/10.1103/PhysRevA.92.043606} {\bibfield  {journal} {\bibinfo
			{journal} {Phys. Rev. A}\ }\textbf {\bibinfo {volume} {92}},\ \bibinfo
		{pages} {043606} (\bibinfo {year} {2015})}\BibitemShut {NoStop}%
	\bibitem [{\citenamefont {Meier}\ \emph {et~al.}(2016)\citenamefont {Meier},
		\citenamefont {An},\ and\ \citenamefont {Gadway}}]{Gadway_2016}%
	\BibitemOpen
	\bibfield  {author} {\bibinfo {author} {\bibfnamefont {E.~J.}\ \bibnamefont
			{Meier}}, \bibinfo {author} {\bibfnamefont {F.~A.}\ \bibnamefont {An}},\ and\
		\bibinfo {author} {\bibfnamefont {B.}~\bibnamefont {Gadway}},\ }\bibfield
	{title} {\bibinfo {title} {{Atom-optics simulator of lattice transport
				phenomena}},\ }\href {https://doi.org/10.1103/PhysRevA.93.051602} {\bibfield
		{journal} {\bibinfo  {journal} {Phys. Rev. A}\ }\textbf {\bibinfo {volume}
			{93}},\ \bibinfo {pages} {051602} (\bibinfo {year} {2016})}\BibitemShut
	{NoStop}%
	\bibitem [{\citenamefont {An}\ \emph {et~al.}(2018)\citenamefont {An},
		\citenamefont {Meier}, \citenamefont {Ang'ong'a},\ and\ \citenamefont
		{Gadway}}]{An_2018}%
	\BibitemOpen
	\bibfield  {author} {\bibinfo {author} {\bibfnamefont {F.~A.}\ \bibnamefont
			{An}}, \bibinfo {author} {\bibfnamefont {E.~J.}\ \bibnamefont {Meier}},
		\bibinfo {author} {\bibfnamefont {J.}~\bibnamefont {Ang'ong'a}},\ and\
		\bibinfo {author} {\bibfnamefont {B.}~\bibnamefont {Gadway}},\ }\bibfield
	{title} {\bibinfo {title} {{Correlated Dynamics in a Synthetic Lattice of
				Momentum States}},\ }\href {https://doi.org/10.1103/PhysRevLett.120.040407}
	{\bibfield  {journal} {\bibinfo  {journal} {Phys. Rev. Lett.}\ }\textbf
		{\bibinfo {volume} {120}},\ \bibinfo {pages} {040407} (\bibinfo {year}
		{2018})}\BibitemShut {NoStop}%
	\bibitem [{\citenamefont {Gou}\ \emph {et~al.}(2020)\citenamefont {Gou},
		\citenamefont {Chen}, \citenamefont {Xie}, \citenamefont {Xiao},
		\citenamefont {Deng}, \citenamefont {Gadway}, \citenamefont {Yi},\ and\
		\citenamefont {Yan}}]{Yan_2020}%
	\BibitemOpen
	\bibfield  {author} {\bibinfo {author} {\bibfnamefont {W.}~\bibnamefont
			{Gou}}, \bibinfo {author} {\bibfnamefont {T.}~\bibnamefont {Chen}}, \bibinfo
		{author} {\bibfnamefont {D.}~\bibnamefont {Xie}}, \bibinfo {author}
		{\bibfnamefont {T.}~\bibnamefont {Xiao}}, \bibinfo {author} {\bibfnamefont
			{T.-S.}\ \bibnamefont {Deng}}, \bibinfo {author} {\bibfnamefont
			{B.}~\bibnamefont {Gadway}}, \bibinfo {author} {\bibfnamefont
			{W.}~\bibnamefont {Yi}},\ and\ \bibinfo {author} {\bibfnamefont
			{B.}~\bibnamefont {Yan}},\ }\bibfield  {title} {\bibinfo {title} {{Tunable
				Nonreciprocal Quantum Transport through a Dissipative Aharonov-Bohm Ring in
				Ultracold Atoms}},\ }\href {https://doi.org/10.1103/PhysRevLett.124.070402}
	{\bibfield  {journal} {\bibinfo  {journal} {Phys. Rev. Lett.}\ }\textbf
		{\bibinfo {volume} {124}},\ \bibinfo {pages} {070402} (\bibinfo {year}
		{2020})}\BibitemShut {NoStop}%
	\bibitem [{\citenamefont {Vrijsen}\ \emph {et~al.}(2011)\citenamefont
		{Vrijsen}, \citenamefont {Hosten}, \citenamefont {Lee}, \citenamefont
		{Bernon},\ and\ \citenamefont {Kasevich}}]{Kasevich_2011}%
	\BibitemOpen
	\bibfield  {author} {\bibinfo {author} {\bibfnamefont {G.}~\bibnamefont
			{Vrijsen}}, \bibinfo {author} {\bibfnamefont {O.}~\bibnamefont {Hosten}},
		\bibinfo {author} {\bibfnamefont {J.}~\bibnamefont {Lee}}, \bibinfo {author}
		{\bibfnamefont {S.}~\bibnamefont {Bernon}},\ and\ \bibinfo {author}
		{\bibfnamefont {M.~A.}\ \bibnamefont {Kasevich}},\ }\bibfield  {title}
	{\bibinfo {title} {{Raman Lasing with a Cold Atom Gain Medium in a
				High-Finesse Optical Cavity}},\ }\href
	{https://doi.org/10.1103/PhysRevLett.107.063904} {\bibfield  {journal}
		{\bibinfo  {journal} {Phys. Rev. Lett.}\ }\textbf {\bibinfo {volume} {107}},\
		\bibinfo {pages} {063904} (\bibinfo {year} {2011})}\BibitemShut {NoStop}%
	\bibitem [{\citenamefont {Bohnet}\ \emph {et~al.}(2012)\citenamefont {Bohnet},
		\citenamefont {Chen}, \citenamefont {Weiner}, \citenamefont {Meiser},
		\citenamefont {Holland},\ and\ \citenamefont {Thompson}}]{Thompson_2012}%
	\BibitemOpen
	\bibfield  {author} {\bibinfo {author} {\bibfnamefont {J.~G.}\ \bibnamefont
			{Bohnet}}, \bibinfo {author} {\bibfnamefont {Z.}~\bibnamefont {Chen}},
		\bibinfo {author} {\bibfnamefont {J.~M.}\ \bibnamefont {Weiner}}, \bibinfo
		{author} {\bibfnamefont {D.}~\bibnamefont {Meiser}}, \bibinfo {author}
		{\bibfnamefont {M.~J.}\ \bibnamefont {Holland}},\ and\ \bibinfo {author}
		{\bibfnamefont {J.~K.}\ \bibnamefont {Thompson}},\ }\bibfield  {title}
	{\bibinfo {title} {{A steady-state superradiant laser with less than one
				intracavity photon}},\ }\href {https://doi.org/10.1038/nature10920}
	{\bibfield  {journal} {\bibinfo  {journal} {Nature}\ }\textbf {\bibinfo
			{volume} {484}},\ \bibinfo {pages} {78} (\bibinfo {year} {2012})}\BibitemShut
	{NoStop}%
	\bibitem [{\citenamefont {Dicke}(1954)}]{Dicke_1954}%
	\BibitemOpen
	\bibfield  {author} {\bibinfo {author} {\bibfnamefont {R.~H.}\ \bibnamefont
			{Dicke}},\ }\bibfield  {title} {\bibinfo {title} {{Coherence in Spontaneous
				Radiation Processes}},\ }\href {https://doi.org/10.1103/PhysRev.93.99}
	{\bibfield  {journal} {\bibinfo  {journal} {Phys. Rev.}\ }\textbf {\bibinfo
			{volume} {93}},\ \bibinfo {pages} {99} (\bibinfo {year} {1954})}\BibitemShut
	{NoStop}%
	\bibitem [{\citenamefont {Gross}\ and\ \citenamefont
		{Haroche}(1982)}]{Haroche_1982}%
	\BibitemOpen
	\bibfield  {author} {\bibinfo {author} {\bibfnamefont {M.}~\bibnamefont
			{Gross}}\ and\ \bibinfo {author} {\bibfnamefont {S.}~\bibnamefont
			{Haroche}},\ }\bibfield  {title} {\bibinfo {title} {{Superradiance: An essay
				on the theory of collective spontaneous emission}},\ }\href
	{https://doi.org/https://doi.org/10.1016/0370-1573(82)90102-8} {\bibfield
		{journal} {\bibinfo  {journal} {Physics Reports}\ }\textbf {\bibinfo {volume}
			{93}},\ \bibinfo {pages} {301} (\bibinfo {year} {1982})}\BibitemShut
	{NoStop}%
	\bibitem [{\citenamefont {Schneble}\ \emph {et~al.}(2004)\citenamefont
		{Schneble}, \citenamefont {Campbell}, \citenamefont {Streed}, \citenamefont
		{Boyd}, \citenamefont {Pritchard},\ and\ \citenamefont
		{Ketterle}}]{Ketterle_2004}%
	\BibitemOpen
	\bibfield  {author} {\bibinfo {author} {\bibfnamefont {D.}~\bibnamefont
			{Schneble}}, \bibinfo {author} {\bibfnamefont {G.~K.}\ \bibnamefont
			{Campbell}}, \bibinfo {author} {\bibfnamefont {E.~W.}\ \bibnamefont
			{Streed}}, \bibinfo {author} {\bibfnamefont {M.}~\bibnamefont {Boyd}},
		\bibinfo {author} {\bibfnamefont {D.~E.}\ \bibnamefont {Pritchard}},\ and\
		\bibinfo {author} {\bibfnamefont {W.}~\bibnamefont {Ketterle}},\ }\bibfield
	{title} {\bibinfo {title} {{Raman amplification of matter waves}},\ }\href
	{https://doi.org/10.1103/PhysRevA.69.041601} {\bibfield  {journal} {\bibinfo
			{journal} {Phys. Rev. A}\ }\textbf {\bibinfo {volume} {69}},\ \bibinfo
		{pages} {041601} (\bibinfo {year} {2004})}\BibitemShut {NoStop}%
	\bibitem [{\citenamefont {Yoshikawa}\ \emph {et~al.}(2004)\citenamefont
		{Yoshikawa}, \citenamefont {Sugiura}, \citenamefont {Torii},\ and\
		\citenamefont {Kuga}}]{Kuga_2004}%
	\BibitemOpen
	\bibfield  {author} {\bibinfo {author} {\bibfnamefont {Y.}~\bibnamefont
			{Yoshikawa}}, \bibinfo {author} {\bibfnamefont {T.}~\bibnamefont {Sugiura}},
		\bibinfo {author} {\bibfnamefont {Y.}~\bibnamefont {Torii}},\ and\ \bibinfo
		{author} {\bibfnamefont {T.}~\bibnamefont {Kuga}},\ }\bibfield  {title}
	{\bibinfo {title} {{Observation of superradiant Raman scattering in a
				Bose-Einstein condensate}},\ }\href
	{https://doi.org/10.1103/PhysRevA.69.041603} {\bibfield  {journal} {\bibinfo
			{journal} {Phys. Rev. A}\ }\textbf {\bibinfo {volume} {69}},\ \bibinfo
		{pages} {041603} (\bibinfo {year} {2004})}\BibitemShut {NoStop}%
	\bibitem [{\citenamefont {Cola}\ and\ \citenamefont
		{Piovella}(2004)}]{Piovella_2004}%
	\BibitemOpen
	\bibfield  {author} {\bibinfo {author} {\bibfnamefont {M.~M.}\ \bibnamefont
			{Cola}}\ and\ \bibinfo {author} {\bibfnamefont {N.}~\bibnamefont
			{Piovella}},\ }\bibfield  {title} {\bibinfo {title} {{Theory of collective
				Raman scattering from a Bose-Einstein condensate}},\ }\href
	{https://doi.org/10.1103/PhysRevA.70.045601} {\bibfield  {journal} {\bibinfo
			{journal} {Phys. Rev. A}\ }\textbf {\bibinfo {volume} {70}},\ \bibinfo
		{pages} {045601} (\bibinfo {year} {2004})}\BibitemShut {NoStop}%
	\bibitem [{\citenamefont {Wang}\ and\ \citenamefont
		{Yelin}(2005)}]{Yelin_2005}%
	\BibitemOpen
	\bibfield  {author} {\bibinfo {author} {\bibfnamefont {T.}~\bibnamefont
			{Wang}}\ and\ \bibinfo {author} {\bibfnamefont {S.~F.}\ \bibnamefont
			{Yelin}},\ }\bibfield  {title} {\bibinfo {title} {{Theory for Raman
				superradiance in atomic gases}},\ }\href
	{https://doi.org/10.1103/PhysRevA.72.043804} {\bibfield  {journal} {\bibinfo
			{journal} {Phys. Rev. A}\ }\textbf {\bibinfo {volume} {72}},\ \bibinfo
		{pages} {043804} (\bibinfo {year} {2005})}\BibitemShut {NoStop}%
	\bibitem [{\citenamefont {Laflamme}\ \emph {et~al.}(2017)\citenamefont
		{Laflamme}, \citenamefont {Yang},\ and\ \citenamefont
		{Zoller}}]{Zoller_2017}%
	\BibitemOpen
	\bibfield  {author} {\bibinfo {author} {\bibfnamefont {C.}~\bibnamefont
			{Laflamme}}, \bibinfo {author} {\bibfnamefont {D.}~\bibnamefont {Yang}},\
		and\ \bibinfo {author} {\bibfnamefont {P.}~\bibnamefont {Zoller}},\
	}\bibfield  {title} {\bibinfo {title} {Continuous measurement of an atomic
			current},\ }\href {https://doi.org/10.1103/PhysRevA.95.043843} {\bibfield
		{journal} {\bibinfo  {journal} {Phys. Rev. A}\ }\textbf {\bibinfo {volume}
			{95}},\ \bibinfo {pages} {043843} (\bibinfo {year} {2017})}\BibitemShut
	{NoStop}%
	\bibitem [{\citenamefont {Yang}\ \emph {et~al.}(2018)\citenamefont {Yang},
		\citenamefont {Laflamme}, \citenamefont {Vasilyev}, \citenamefont {Baranov},\
		and\ \citenamefont {Zoller}}]{Zoller_2018}%
	\BibitemOpen
	\bibfield  {author} {\bibinfo {author} {\bibfnamefont {D.}~\bibnamefont
			{Yang}}, \bibinfo {author} {\bibfnamefont {C.}~\bibnamefont {Laflamme}},
		\bibinfo {author} {\bibfnamefont {D.~V.}\ \bibnamefont {Vasilyev}}, \bibinfo
		{author} {\bibfnamefont {M.~A.}\ \bibnamefont {Baranov}},\ and\ \bibinfo
		{author} {\bibfnamefont {P.}~\bibnamefont {Zoller}},\ }\bibfield  {title}
	{\bibinfo {title} {{Theory of a Quantum Scanning Microscope for Cold
				Atoms}},\ }\href {https://doi.org/10.1103/PhysRevLett.120.133601} {\bibfield
		{journal} {\bibinfo  {journal} {Phys. Rev. Lett.}\ }\textbf {\bibinfo
			{volume} {120}},\ \bibinfo {pages} {133601} (\bibinfo {year}
		{2018})}\BibitemShut {NoStop}%
	\bibitem [{\citenamefont {Geier}\ \emph {et~al.}(2021)\citenamefont {Geier},
		\citenamefont {Reichstetter},\ and\ \citenamefont {Hauke}}]{Geier_2021}%
	\BibitemOpen
	\bibfield  {author} {\bibinfo {author} {\bibfnamefont {K.~T.}\ \bibnamefont
			{Geier}}, \bibinfo {author} {\bibfnamefont {J.}~\bibnamefont
			{Reichstetter}},\ and\ \bibinfo {author} {\bibfnamefont {P.}~\bibnamefont
			{Hauke}},\ }\href@noop {} {\bibinfo {title} {{Non-invasive measurement of
				currents in analog quantum simulators}}} (\bibinfo {year} {2021}),\ \Eprint
	{https://arxiv.org/abs/2106.12599} {arXiv:2106.12599 [quant-ph]} \BibitemShut
	{NoStop}%
	\bibitem [{SI()}]{SI}%
	\BibitemOpen
	\href@noop {} {}\bibinfo {note} {See Supplemental Material}\BibitemShut
	{NoStop}%
	\bibitem [{\citenamefont {Fan}\ \emph {et~al.}(2020)\citenamefont {Fan},
		\citenamefont {Chen},\ and\ \citenamefont {Jia}}]{Fan_2020}%
	\BibitemOpen
	\bibfield  {author} {\bibinfo {author} {\bibfnamefont {J.}~\bibnamefont
			{Fan}}, \bibinfo {author} {\bibfnamefont {G.}~\bibnamefont {Chen}},\ and\
		\bibinfo {author} {\bibfnamefont {S.}~\bibnamefont {Jia}},\ }\bibfield
	{title} {\bibinfo {title} {{Atomic self-organization emerging from tunable
				quadrature coupling}},\ }\href {https://doi.org/10.1103/PhysRevA.101.063627}
	{\bibfield  {journal} {\bibinfo  {journal} {Phys. Rev. A}\ }\textbf {\bibinfo
			{volume} {101}},\ \bibinfo {pages} {063627} (\bibinfo {year}
		{2020})}\BibitemShut {NoStop}%
	\bibitem [{\citenamefont {Lin}\ \emph {et~al.}(2021)\citenamefont {Lin},
		\citenamefont {Rosa-Medina}, \citenamefont {Ferri}, \citenamefont {Finger},
		\citenamefont {Kroeger}, \citenamefont {Donner}, \citenamefont {Esslinger},\
		and\ \citenamefont {Chitra}}]{Lin_2021}%
	\BibitemOpen
	\bibfield  {author} {\bibinfo {author} {\bibfnamefont {R.}~\bibnamefont
			{Lin}}, \bibinfo {author} {\bibfnamefont {R.}~\bibnamefont {Rosa-Medina}},
		\bibinfo {author} {\bibfnamefont {F.}~\bibnamefont {Ferri}}, \bibinfo
		{author} {\bibfnamefont {F.}~\bibnamefont {Finger}}, \bibinfo {author}
		{\bibfnamefont {K.}~\bibnamefont {Kroeger}}, \bibinfo {author} {\bibfnamefont
			{T.}~\bibnamefont {Donner}}, \bibinfo {author} {\bibfnamefont
			{T.}~\bibnamefont {Esslinger}},\ and\ \bibinfo {author} {\bibfnamefont
			{R.}~\bibnamefont {Chitra}},\ }\href@noop {} {\bibinfo {title}
		{Dissipation-engineered family of nearly dark states in many-body cavity-atom
			systems}} (\bibinfo {year} {2021}),\ \Eprint
	{https://arxiv.org/abs/2109.00422} {arXiv:2109.00422 [cond-mat.quant-gas]}
	\BibitemShut {NoStop}%
	\bibitem [{\citenamefont {Kroeze}\ \emph {et~al.}(2018)\citenamefont {Kroeze},
		\citenamefont {Guo}, \citenamefont {Vaidya}, \citenamefont {Keeling},\ and\
		\citenamefont {Lev}}]{Lev_2018}%
	\BibitemOpen
	\bibfield  {author} {\bibinfo {author} {\bibfnamefont {R.~M.}\ \bibnamefont
			{Kroeze}}, \bibinfo {author} {\bibfnamefont {Y.}~\bibnamefont {Guo}},
		\bibinfo {author} {\bibfnamefont {V.~D.}\ \bibnamefont {Vaidya}}, \bibinfo
		{author} {\bibfnamefont {J.}~\bibnamefont {Keeling}},\ and\ \bibinfo {author}
		{\bibfnamefont {B.~L.}\ \bibnamefont {Lev}},\ }\bibfield  {title} {\bibinfo
		{title} {{Spinor Self-Ordering of a Quantum Gas in a Cavity}},\ }\href
	{https://doi.org/10.1103/PhysRevLett.121.163601} {\bibfield  {journal}
		{\bibinfo  {journal} {Phys. Rev. Lett.}\ }\textbf {\bibinfo {volume} {121}},\
		\bibinfo {pages} {163601} (\bibinfo {year} {2018})}\BibitemShut {NoStop}%
	\bibitem [{\citenamefont {Ferri}\ \emph {et~al.}(2021)\citenamefont {Ferri},
		\citenamefont {Rosa-Medina}, \citenamefont {Finger}, \citenamefont {Dogra},
		\citenamefont {Soriente}, \citenamefont {Zilberberg}, \citenamefont
		{Donner},\ and\ \citenamefont {Esslinger}}]{Ferri_2021}%
	\BibitemOpen
	\bibfield  {author} {\bibinfo {author} {\bibfnamefont {F.}~\bibnamefont
			{Ferri}}, \bibinfo {author} {\bibfnamefont {R.}~\bibnamefont {Rosa-Medina}},
		\bibinfo {author} {\bibfnamefont {F.}~\bibnamefont {Finger}}, \bibinfo
		{author} {\bibfnamefont {N.}~\bibnamefont {Dogra}}, \bibinfo {author}
		{\bibfnamefont {M.}~\bibnamefont {Soriente}}, \bibinfo {author}
		{\bibfnamefont {O.}~\bibnamefont {Zilberberg}}, \bibinfo {author}
		{\bibfnamefont {T.}~\bibnamefont {Donner}},\ and\ \bibinfo {author}
		{\bibfnamefont {T.}~\bibnamefont {Esslinger}},\ }\bibfield  {title} {\bibinfo
		{title} {{Emerging Dissipative Phases in a Superradiant Quantum Gas with
				Tunable Decay}},\ }\href {https://doi.org/10.1103/PhysRevX.11.041046}
	{\bibfield  {journal} {\bibinfo  {journal} {Phys. Rev. X}\ }\textbf {\bibinfo
			{volume} {11}},\ \bibinfo {pages} {041046} (\bibinfo {year}
		{2021})}\BibitemShut {NoStop}%
	\bibitem [{\citenamefont {Mivehvar}\ \emph {et~al.}(2021)\citenamefont
		{Mivehvar}, \citenamefont {Piazza}, \citenamefont {Donner},\ and\
		\citenamefont {Ritsch}}]{Mivehvar_2021}%
	\BibitemOpen
	\bibfield  {author} {\bibinfo {author} {\bibfnamefont {F.}~\bibnamefont
			{Mivehvar}}, \bibinfo {author} {\bibfnamefont {F.}~\bibnamefont {Piazza}},
		\bibinfo {author} {\bibfnamefont {T.}~\bibnamefont {Donner}},\ and\ \bibinfo
		{author} {\bibfnamefont {H.}~\bibnamefont {Ritsch}},\ }\bibfield  {title}
	{\bibinfo {title} {{Cavity QED with quantum gases: new paradigms in many-body
				physics}},\ }\href {https://doi.org/10.1080/00018732.2021.1969727} {\bibfield
		{journal} {\bibinfo  {journal} {Advances in Physics}\ }\textbf {\bibinfo
			{volume} {70}},\ \bibinfo {pages} {1} (\bibinfo {year} {2021})}\BibitemShut
	{NoStop}%
	\bibitem [{\citenamefont {Ke{\ss}ler}\ \emph {et~al.}(2014)\citenamefont
		{Ke{\ss}ler}, \citenamefont {Klinder}, \citenamefont {Wolke},\ and\
		\citenamefont {Hemmerich}}]{Kessler_2014}%
	\BibitemOpen
	\bibfield  {author} {\bibinfo {author} {\bibfnamefont {H.}~\bibnamefont
			{Ke{\ss}ler}}, \bibinfo {author} {\bibfnamefont {J.}~\bibnamefont {Klinder}},
		\bibinfo {author} {\bibfnamefont {M.}~\bibnamefont {Wolke}},\ and\ \bibinfo
		{author} {\bibfnamefont {A.}~\bibnamefont {Hemmerich}},\ }\bibfield  {title}
	{\bibinfo {title} {Optomechanical atom-cavity interaction in the sub-recoil
			regime},\ }\href {https://doi.org/10.1088/1367-2630/16/5/053008} {\ \textbf
		{\bibinfo {volume} {16}},\ \bibinfo {pages} {053008} (\bibinfo {year}
		{2014})}\BibitemShut {NoStop}%
	\bibitem [{\citenamefont {Norcia}\ \emph {et~al.}(2016)\citenamefont {Norcia},
		\citenamefont {Winchester}, \citenamefont {Cline},\ and\ \citenamefont
		{Thompson}}]{Norcia_2016}%
	\BibitemOpen
	\bibfield  {author} {\bibinfo {author} {\bibfnamefont {M.~A.}\ \bibnamefont
			{Norcia}}, \bibinfo {author} {\bibfnamefont {M.~N.}\ \bibnamefont
			{Winchester}}, \bibinfo {author} {\bibfnamefont {J.~R.~K.}\ \bibnamefont
			{Cline}},\ and\ \bibinfo {author} {\bibfnamefont {J.~K.}\ \bibnamefont
			{Thompson}},\ }\bibfield  {title} {\bibinfo {title} {{Superradiance on the
				millihertz linewidth strontium clock transition}},\ }\bibfield  {journal}
	{\bibinfo  {journal} {Science Advances}\ }\textbf {\bibinfo {volume} {2}},\
	\href {https://doi.org/10.1126/sciadv.1601231} {10.1126/sciadv.1601231}
	(\bibinfo {year} {2016})\BibitemShut {NoStop}%
	\bibitem [{\citenamefont {Laske}\ \emph {et~al.}(2019)\citenamefont {Laske},
		\citenamefont {Winter},\ and\ \citenamefont {Hemmerich}}]{Hemmerich_2019}%
	\BibitemOpen
	\bibfield  {author} {\bibinfo {author} {\bibfnamefont {T.}~\bibnamefont
			{Laske}}, \bibinfo {author} {\bibfnamefont {H.}~\bibnamefont {Winter}},\ and\
		\bibinfo {author} {\bibfnamefont {A.}~\bibnamefont {Hemmerich}},\ }\bibfield
	{title} {\bibinfo {title} {{Pulse Delay Time Statistics in a Superradiant
				Laser with Calcium Atoms}},\ }\href
	{https://doi.org/10.1103/PhysRevLett.123.103601} {\bibfield  {journal}
		{\bibinfo  {journal} {Phys. Rev. Lett.}\ }\textbf {\bibinfo {volume} {123}},\
		\bibinfo {pages} {103601} (\bibinfo {year} {2019})}\BibitemShut {NoStop}%
	\bibitem [{\citenamefont {Ferioli}\ \emph {et~al.}(2021)\citenamefont
		{Ferioli}, \citenamefont {Glicenstein}, \citenamefont {Robicheaux},
		\citenamefont {Sutherland}, \citenamefont {Browaeys},\ and\ \citenamefont
		{Ferrier-Barbut}}]{Ferioli_2021}%
	\BibitemOpen
	\bibfield  {author} {\bibinfo {author} {\bibfnamefont {G.}~\bibnamefont
			{Ferioli}}, \bibinfo {author} {\bibfnamefont {A.}~\bibnamefont
			{Glicenstein}}, \bibinfo {author} {\bibfnamefont {F.}~\bibnamefont
			{Robicheaux}}, \bibinfo {author} {\bibfnamefont {R.~T.}\ \bibnamefont
			{Sutherland}}, \bibinfo {author} {\bibfnamefont {A.}~\bibnamefont
			{Browaeys}},\ and\ \bibinfo {author} {\bibfnamefont {I.}~\bibnamefont
			{Ferrier-Barbut}},\ }\bibfield  {title} {\bibinfo {title} {{Laser-Driven
				Superradiant Ensembles of Two-Level Atoms near Dicke Regime}},\ }\href
	{https://doi.org/10.1103/PhysRevLett.127.243602} {\bibfield  {journal}
		{\bibinfo  {journal} {Phys. Rev. Lett.}\ }\textbf {\bibinfo {volume} {127}},\
		\bibinfo {pages} {243602} (\bibinfo {year} {2021})}\BibitemShut {NoStop}%
	\bibitem [{\citenamefont {Mandel}\ \emph {et~al.}(1995)\citenamefont {Mandel},
		\citenamefont {Wolf},\ and\ \citenamefont {Press}}]{Mandel_1995}%
	\BibitemOpen
	\bibfield  {author} {\bibinfo {author} {\bibfnamefont {L.}~\bibnamefont
			{Mandel}}, \bibinfo {author} {\bibfnamefont {E.}~\bibnamefont {Wolf}},\ and\
		\bibinfo {author} {\bibfnamefont {C.~U.}\ \bibnamefont {Press}},\ }\href
	{https://doi.org/https://doi.org/10.1017/CBO9781139644105} {\emph {\bibinfo
			{title} {{Optical Coherence and Quantum Optics}}}},\ EBL-Schweitzer\
	(\bibinfo  {publisher} {Cambridge University Press},\ \bibinfo {year}
	{1995})\BibitemShut {NoStop}%
	\bibitem [{\citenamefont {Sias}\ \emph {et~al.}(2008)\citenamefont {Sias},
		\citenamefont {Lignier}, \citenamefont {Singh}, \citenamefont {Zenesini},
		\citenamefont {Ciampini}, \citenamefont {Morsch},\ and\ \citenamefont
		{Arimondo}}]{Arimondo_2008}%
	\BibitemOpen
	\bibfield  {author} {\bibinfo {author} {\bibfnamefont {C.}~\bibnamefont
			{Sias}}, \bibinfo {author} {\bibfnamefont {H.}~\bibnamefont {Lignier}},
		\bibinfo {author} {\bibfnamefont {Y.~P.}\ \bibnamefont {Singh}}, \bibinfo
		{author} {\bibfnamefont {A.}~\bibnamefont {Zenesini}}, \bibinfo {author}
		{\bibfnamefont {D.}~\bibnamefont {Ciampini}}, \bibinfo {author}
		{\bibfnamefont {O.}~\bibnamefont {Morsch}},\ and\ \bibinfo {author}
		{\bibfnamefont {E.}~\bibnamefont {Arimondo}},\ }\bibfield  {title} {\bibinfo
		{title} {{Observation of Photon-Assisted Tunneling in Optical Lattices}},\
	}\href {https://doi.org/10.1103/PhysRevLett.100.040404} {\bibfield  {journal}
		{\bibinfo  {journal} {Phys. Rev. Lett.}\ }\textbf {\bibinfo {volume} {100}},\
		\bibinfo {pages} {040404} (\bibinfo {year} {2008})}\BibitemShut {NoStop}%
	\bibitem [{\citenamefont {An}(2020)}]{AnPhD_2020}%
	\BibitemOpen
	\bibfield  {author} {\bibinfo {author} {\bibfnamefont {F.~A.}\ \bibnamefont
			{An}},\ }\emph {\bibinfo {title} {{The cold atom toolbox in momentum
				space}}},\ \href@noop {} {Ph.D. thesis},\ \bibinfo  {school} {University of
		Illinois at Urbana-Champaign} (\bibinfo {year} {2020})\BibitemShut {NoStop}%
	\bibitem [{\citenamefont {Kroeger}\ \emph {et~al.}(2020)\citenamefont
		{Kroeger}, \citenamefont {Dogra}, \citenamefont {Rosa-Medina}, \citenamefont
		{Paluch}, \citenamefont {Ferri}, \citenamefont {Donner},\ and\ \citenamefont
		{Esslinger}}]{Kroeger_2020}%
	\BibitemOpen
	\bibfield  {author} {\bibinfo {author} {\bibfnamefont {K.}~\bibnamefont
			{Kroeger}}, \bibinfo {author} {\bibfnamefont {N.}~\bibnamefont {Dogra}},
		\bibinfo {author} {\bibfnamefont {R.}~\bibnamefont {Rosa-Medina}}, \bibinfo
		{author} {\bibfnamefont {M.}~\bibnamefont {Paluch}}, \bibinfo {author}
		{\bibfnamefont {F.}~\bibnamefont {Ferri}}, \bibinfo {author} {\bibfnamefont
			{T.}~\bibnamefont {Donner}},\ and\ \bibinfo {author} {\bibfnamefont
			{T.}~\bibnamefont {Esslinger}},\ }\bibfield  {title} {\bibinfo {title}
		{{Continuous feedback on a quantum gas coupled to an optical cavity}},\
	}\href {https://doi.org/10.1088/1367-2630/ab73cc} {\bibfield  {journal}
		{\bibinfo  {journal} {New Journal of Physics}\ }\textbf {\bibinfo {volume}
			{22}},\ \bibinfo {pages} {033020} (\bibinfo {year} {2020})}\BibitemShut
	{NoStop}%
	\bibitem [{\citenamefont {Gong}\ \emph {et~al.}(2018)\citenamefont {Gong},
		\citenamefont {Ashida}, \citenamefont {Kawabata}, \citenamefont {Takasan},
		\citenamefont {Higashikawa},\ and\ \citenamefont {Ueda}}]{Gong_2018}%
	\BibitemOpen
	\bibfield  {author} {\bibinfo {author} {\bibfnamefont {Z.}~\bibnamefont
			{Gong}}, \bibinfo {author} {\bibfnamefont {Y.}~\bibnamefont {Ashida}},
		\bibinfo {author} {\bibfnamefont {K.}~\bibnamefont {Kawabata}}, \bibinfo
		{author} {\bibfnamefont {K.}~\bibnamefont {Takasan}}, \bibinfo {author}
		{\bibfnamefont {S.}~\bibnamefont {Higashikawa}},\ and\ \bibinfo {author}
		{\bibfnamefont {M.}~\bibnamefont {Ueda}},\ }\bibfield  {title} {\bibinfo
		{title} {{Topological Phases of Non-Hermitian Systems}},\ }\href
	{https://doi.org/10.1103/PhysRevX.8.031079} {\bibfield  {journal} {\bibinfo
			{journal} {Phys. Rev. X}\ }\textbf {\bibinfo {volume} {8}},\ \bibinfo {pages}
		{031079} (\bibinfo {year} {2018})}\BibitemShut {NoStop}%
	\bibitem [{\citenamefont {Ozawa}\ and\ \citenamefont
		{Price}(2019)}]{Ozawa_2019}%
	\BibitemOpen
	\bibfield  {author} {\bibinfo {author} {\bibfnamefont {T.}~\bibnamefont
			{Ozawa}}\ and\ \bibinfo {author} {\bibfnamefont {H.~M.}\ \bibnamefont
			{Price}},\ }\bibfield  {title} {\bibinfo {title} {{Topological quantum matter
				in synthetic dimensions}},\ }\href
	{https://doi.org/10.1038/s42254-019-0045-3} {\bibfield  {journal} {\bibinfo
			{journal} {Nature Reviews Physics}\ }\textbf {\bibinfo {volume} {1}},\
		\bibinfo {pages} {349} (\bibinfo {year} {2019})}\BibitemShut {NoStop}%
	\bibitem [{\citenamefont {Kroeze}\ \emph {et~al.}(2019)\citenamefont {Kroeze},
		\citenamefont {Guo},\ and\ \citenamefont {Lev}}]{Lev_2019}%
	\BibitemOpen
	\bibfield  {author} {\bibinfo {author} {\bibfnamefont {R.~M.}\ \bibnamefont
			{Kroeze}}, \bibinfo {author} {\bibfnamefont {Y.}~\bibnamefont {Guo}},\ and\
		\bibinfo {author} {\bibfnamefont {B.~L.}\ \bibnamefont {Lev}},\ }\bibfield
	{title} {\bibinfo {title} {{Dynamical Spin-Orbit Coupling of a Quantum
				Gas}},\ }\href {https://doi.org/10.1103/PhysRevLett.123.160404} {\bibfield
		{journal} {\bibinfo  {journal} {Phys. Rev. Lett.}\ }\textbf {\bibinfo
			{volume} {123}},\ \bibinfo {pages} {160404} (\bibinfo {year}
		{2019})}\BibitemShut {NoStop}%
	\bibitem [{\citenamefont {Halati}\ \emph {et~al.}(2019)\citenamefont {Halati},
		\citenamefont {Sheikhan},\ and\ \citenamefont {Kollath}}]{Kollath_2019}%
	\BibitemOpen
	\bibfield  {author} {\bibinfo {author} {\bibfnamefont {C.-M.}\ \bibnamefont
			{Halati}}, \bibinfo {author} {\bibfnamefont {A.}~\bibnamefont {Sheikhan}},\
		and\ \bibinfo {author} {\bibfnamefont {C.}~\bibnamefont {Kollath}},\
	}\bibfield  {title} {\bibinfo {title} {Cavity-induced spin-orbit coupling in
			an interacting bosonic wire},\ }\href
	{https://doi.org/10.1103/PhysRevA.99.033604} {\bibfield  {journal} {\bibinfo
			{journal} {Phys. Rev. A}\ }\textbf {\bibinfo {volume} {99}},\ \bibinfo
		{pages} {033604} (\bibinfo {year} {2019})}\BibitemShut {NoStop}%
	\bibitem [{\citenamefont {Ostermann}\ \emph {et~al.}(2021)\citenamefont
		{Ostermann}, \citenamefont {Ritsch},\ and\ \citenamefont
		{Mivehvar}}]{Farokh_2021}%
	\BibitemOpen
	\bibfield  {author} {\bibinfo {author} {\bibfnamefont {S.}~\bibnamefont
			{Ostermann}}, \bibinfo {author} {\bibfnamefont {H.}~\bibnamefont {Ritsch}},\
		and\ \bibinfo {author} {\bibfnamefont {F.}~\bibnamefont {Mivehvar}},\
	}\bibfield  {title} {\bibinfo {title} {Many-body phases of a planar
			bose-einstein condensate with cavity-induced spin-orbit coupling},\ }\href
	{https://doi.org/10.1103/PhysRevA.103.023302} {\bibfield  {journal} {\bibinfo
			{journal} {Phys. Rev. A}\ }\textbf {\bibinfo {volume} {103}},\ \bibinfo
		{pages} {023302} (\bibinfo {year} {2021})}\BibitemShut {NoStop}%
	\bibitem [{\citenamefont {Landig}\ \emph {et~al.}(2016)\citenamefont {Landig},
		\citenamefont {Hruby}, \citenamefont {Dogra}, \citenamefont {Landini},
		\citenamefont {Mottl}, \citenamefont {Donner},\ and\ \citenamefont
		{Esslinger}}]{Landig_2016}%
	\BibitemOpen
	\bibfield  {author} {\bibinfo {author} {\bibfnamefont {R.}~\bibnamefont
			{Landig}}, \bibinfo {author} {\bibfnamefont {L.}~\bibnamefont {Hruby}},
		\bibinfo {author} {\bibfnamefont {N.}~\bibnamefont {Dogra}}, \bibinfo
		{author} {\bibfnamefont {M.}~\bibnamefont {Landini}}, \bibinfo {author}
		{\bibfnamefont {R.}~\bibnamefont {Mottl}}, \bibinfo {author} {\bibfnamefont
			{T.}~\bibnamefont {Donner}},\ and\ \bibinfo {author} {\bibfnamefont
			{T.}~\bibnamefont {Esslinger}},\ }\bibfield  {title} {\bibinfo {title}
		{{Quantum phases from competing short- and long-range interactions in an
				optical lattice}},\ }\href {https://doi.org/10.1038/nature17409} {\bibfield
		{journal} {\bibinfo  {journal} {Nature}\ }\textbf {\bibinfo {volume} {532}},\
		\bibinfo {pages} {476} (\bibinfo {year} {2016})}\BibitemShut {NoStop}%
	\bibitem [{\citenamefont {Halati}\ \emph {et~al.}(2017)\citenamefont {Halati},
		\citenamefont {Sheikhan},\ and\ \citenamefont {Kollath}}]{Halati_2017}%
	\BibitemOpen
	\bibfield  {author} {\bibinfo {author} {\bibfnamefont {C.-M.}\ \bibnamefont
			{Halati}}, \bibinfo {author} {\bibfnamefont {A.}~\bibnamefont {Sheikhan}},\
		and\ \bibinfo {author} {\bibfnamefont {C.}~\bibnamefont {Kollath}},\
	}\bibfield  {title} {\bibinfo {title} {{Cavity-induced artificial gauge field
				in a Bose-Hubbard ladder}},\ }\href
	{https://doi.org/10.1103/PhysRevA.96.063621} {\bibfield  {journal} {\bibinfo
			{journal} {Phys. Rev. A}\ }\textbf {\bibinfo {volume} {96}},\ \bibinfo
		{pages} {063621} (\bibinfo {year} {2017})}\BibitemShut {NoStop}%
	\bibitem [{\citenamefont {Chanda}\ \emph {et~al.}(2021)\citenamefont {Chanda},
		\citenamefont {Kraus}, \citenamefont {Morigi},\ and\ \citenamefont
		{Zakrzewski}}]{Chanda_2021}%
	\BibitemOpen
	\bibfield  {author} {\bibinfo {author} {\bibfnamefont {T.}~\bibnamefont
			{Chanda}}, \bibinfo {author} {\bibfnamefont {R.}~\bibnamefont {Kraus}},
		\bibinfo {author} {\bibfnamefont {G.}~\bibnamefont {Morigi}},\ and\ \bibinfo
		{author} {\bibfnamefont {J.}~\bibnamefont {Zakrzewski}},\ }\bibfield  {title}
	{\bibinfo {title} {{Self-organized topological insulator due to
				cavity-mediated correlated tunneling}},\ }\href
	{https://doi.org/10.22331/q-2021-07-13-501} {\bibfield  {journal} {\bibinfo
			{journal} {Quantum}\ }\textbf {\bibinfo {volume} {5}},\ \bibinfo {pages}
		{501} (\bibinfo {year} {2021})}\BibitemShut {NoStop}%
	\bibitem [{\citenamefont {Colella}\ \emph {et~al.}(2022)\citenamefont
		{Colella}, \citenamefont {Kosior}, \citenamefont {Mivehvar},\ and\
		\citenamefont {Ritsch}}]{Colela_2021}%
	\BibitemOpen
	\bibfield  {author} {\bibinfo {author} {\bibfnamefont {E.}~\bibnamefont
			{Colella}}, \bibinfo {author} {\bibfnamefont {A.}~\bibnamefont {Kosior}},
		\bibinfo {author} {\bibfnamefont {F.}~\bibnamefont {Mivehvar}},\ and\
		\bibinfo {author} {\bibfnamefont {H.}~\bibnamefont {Ritsch}},\ }\bibfield
	{title} {\bibinfo {title} {{Open Quantum System Simulation of Faraday's
				Induction Law via Dynamical Instabilities}},\ }\href
	{https://doi.org/10.1103/PhysRevLett.128.070603} {\bibfield  {journal}
		{\bibinfo  {journal} {Phys. Rev. Lett.}\ }\textbf {\bibinfo {volume} {128}},\
		\bibinfo {pages} {070603} (\bibinfo {year} {2022})}\BibitemShut {NoStop}%
\end{thebibliography}

\begin{thebibliography}{24}%
	\makeatletter
	\providecommand \@ifxundefined [1]{%
		\@ifx{#1\undefined}
	}%
	\providecommand \@ifnum [1]{%
		\ifnum #1\expandafter \@firstoftwo
		\else \expandafter \@secondoftwo
		\fi
	}%
	\providecommand \@ifx [1]{%
		\ifx #1\expandafter \@firstoftwo
		\else \expandafter \@secondoftwo
		\fi
	}%
	\providecommand \natexlab [1]{#1}%
	\providecommand \enquote  [1]{``#1''}%
	\providecommand \bibnamefont  [1]{#1}%
	\providecommand \bibfnamefont [1]{#1}%
	\providecommand \citenamefont [1]{#1}%
	\providecommand \href@noop [0]{\@secondoftwo}%
	\providecommand \href [0]{\begingroup \@sanitize@url \@href}%
	\providecommand \@href[1]{\@@startlink{#1}\@@href}%
	\providecommand \@@href[1]{\endgroup#1\@@endlink}%
	\providecommand \@sanitize@url [0]{\catcode `\\12\catcode `\$12\catcode
		`\&12\catcode `\#12\catcode `\^12\catcode `\_12\catcode `\%12\relax}%
	\providecommand \@@startlink[1]{}%
	\providecommand \@@endlink[0]{}%
	\providecommand \url  [0]{\begingroup\@sanitize@url \@url }%
	\providecommand \@url [1]{\endgroup\@href {#1}{\urlprefix }}%
	\providecommand \urlprefix  [0]{URL }%
	\providecommand \Eprint [0]{\href }%
	\providecommand \doibase [0]{https://doi.org/}%
	\providecommand \selectlanguage [0]{\@gobble}%
	\providecommand \bibinfo  [0]{\@secondoftwo}%
	\providecommand \bibfield  [0]{\@secondoftwo}%
	\providecommand \translation [1]{[#1]}%
	\providecommand \BibitemOpen [0]{}%
	\providecommand \bibitemStop [0]{}%
	\providecommand \bibitemNoStop [0]{.\EOS\space}%
	\providecommand \EOS [0]{\spacefactor3000\relax}%
	\providecommand \BibitemShut  [1]{\csname bibitem#1\endcsname}%
	\let\auto@bib@innerbib\@empty
	%</preamble>
	\bibitem [{\citenamefont {Ferri}\ \emph {et~al.}(2021)\citenamefont {Ferri},
		\citenamefont {Rosa-Medina}, \citenamefont {Finger}, \citenamefont {Dogra},
		\citenamefont {Soriente}, \citenamefont {Zilberberg}, \citenamefont
		{Donner},\ and\ \citenamefont {Esslinger}}]{Ferri_2021}%
	\BibitemOpen
	\bibfield  {author} {\bibinfo {author} {\bibfnamefont {F.}~\bibnamefont
			{Ferri}}, \bibinfo {author} {\bibfnamefont {R.}~\bibnamefont {Rosa-Medina}},
		\bibinfo {author} {\bibfnamefont {F.}~\bibnamefont {Finger}}, \bibinfo
		{author} {\bibfnamefont {N.}~\bibnamefont {Dogra}}, \bibinfo {author}
		{\bibfnamefont {M.}~\bibnamefont {Soriente}}, \bibinfo {author}
		{\bibfnamefont {O.}~\bibnamefont {Zilberberg}}, \bibinfo {author}
		{\bibfnamefont {T.}~\bibnamefont {Donner}},\ and\ \bibinfo {author}
		{\bibfnamefont {T.}~\bibnamefont {Esslinger}},\ }\bibfield  {title} {\bibinfo
		{title} {{Emerging Dissipative Phases in a Superradiant Quantum Gas with
				Tunable Decay}},\ }\href {https://doi.org/10.1103/PhysRevX.11.041046}
	{\bibfield  {journal} {\bibinfo  {journal} {Phys. Rev. X}\ }\textbf {\bibinfo
			{volume} {11}},\ \bibinfo {pages} {041046} (\bibinfo {year}
		{2021})}\BibitemShut {NoStop}%
	\bibitem [{\citenamefont {Gadway}\ \emph {et~al.}(2009)\citenamefont {Gadway},
		\citenamefont {Pertot}, \citenamefont {Reimann}, \citenamefont {Cohen},\ and\
		\citenamefont {Schneble}}]{Gadway_2009}%
	\BibitemOpen
	\bibfield  {author} {\bibinfo {author} {\bibfnamefont {B.}~\bibnamefont
			{Gadway}}, \bibinfo {author} {\bibfnamefont {D.}~\bibnamefont {Pertot}},
		\bibinfo {author} {\bibfnamefont {R.}~\bibnamefont {Reimann}}, \bibinfo
		{author} {\bibfnamefont {M.~G.}\ \bibnamefont {Cohen}},\ and\ \bibinfo
		{author} {\bibfnamefont {D.}~\bibnamefont {Schneble}},\ }\bibfield  {title}
	{\bibinfo {title} {Analysis of {K}apitza-{D}irac diffraction patterns beyond
			the raman-nath regime},\ }\href {https://doi.org/10.1364/OE.17.019173}
	{\bibfield  {journal} {\bibinfo  {journal} {Opt. Express}\ }\textbf {\bibinfo
			{volume} {17}},\ \bibinfo {pages} {19173} (\bibinfo {year}
		{2009})}\BibitemShut {NoStop}%
	\bibitem [{\citenamefont {Le~Kien}\ \emph {et~al.}(2013)\citenamefont
		{Le~Kien}, \citenamefont {Schneeweiss},\ and\ \citenamefont
		{Rauschenbeutel}}]{Le_2013}%
	\BibitemOpen
	\bibfield  {author} {\bibinfo {author} {\bibfnamefont {F.}~\bibnamefont
			{Le~Kien}}, \bibinfo {author} {\bibfnamefont {P.}~\bibnamefont
			{Schneeweiss}},\ and\ \bibinfo {author} {\bibfnamefont {A.}~\bibnamefont
			{Rauschenbeutel}},\ }\bibfield  {title} {\bibinfo {title} {{Dynamical
				polarizability of atoms in arbitrary light fields: general theory and
				application to cesium}},\ }\href {https://doi.org/10.1140/epjd/e2013-30729-x}
	{\bibfield  {journal} {\bibinfo  {journal} {The European Physical Journal D}\
		}\textbf {\bibinfo {volume} {67}},\ \bibinfo {pages} {92} (\bibinfo {year}
		{2013})}\BibitemShut {NoStop}%
	\bibitem [{\citenamefont {Landini}\ \emph {et~al.}(2018)\citenamefont
		{Landini}, \citenamefont {Dogra}, \citenamefont {Kroeger}, \citenamefont
		{Hruby}, \citenamefont {Donner},\ and\ \citenamefont
		{Esslinger}}]{Landini_2018}%
	\BibitemOpen
	\bibfield  {author} {\bibinfo {author} {\bibfnamefont {M.}~\bibnamefont
			{Landini}}, \bibinfo {author} {\bibfnamefont {N.}~\bibnamefont {Dogra}},
		\bibinfo {author} {\bibfnamefont {K.}~\bibnamefont {Kroeger}}, \bibinfo
		{author} {\bibfnamefont {L.}~\bibnamefont {Hruby}}, \bibinfo {author}
		{\bibfnamefont {T.}~\bibnamefont {Donner}},\ and\ \bibinfo {author}
		{\bibfnamefont {T.}~\bibnamefont {Esslinger}},\ }\bibfield  {title} {\bibinfo
		{title} {{Formation of a Spin Texture in a Quantum Gas Coupled to a
				Cavity}},\ }\href {https://doi.org/10.1103/PhysRevLett.120.223602} {\bibfield
		{journal} {\bibinfo  {journal} {Phys. Rev. Lett.}\ }\textbf {\bibinfo
			{volume} {120}},\ \bibinfo {pages} {223602} (\bibinfo {year}
		{2018})}\BibitemShut {NoStop}%
	\bibitem [{\citenamefont {Dogra}\ \emph {et~al.}(2019)\citenamefont {Dogra},
		\citenamefont {Landini}, \citenamefont {Kroeger}, \citenamefont {Hruby},
		\citenamefont {Donner},\ and\ \citenamefont {Esslinger}}]{Dogra_2019}%
	\BibitemOpen
	\bibfield  {author} {\bibinfo {author} {\bibfnamefont {N.}~\bibnamefont
			{Dogra}}, \bibinfo {author} {\bibfnamefont {M.}~\bibnamefont {Landini}},
		\bibinfo {author} {\bibfnamefont {K.}~\bibnamefont {Kroeger}}, \bibinfo
		{author} {\bibfnamefont {L.}~\bibnamefont {Hruby}}, \bibinfo {author}
		{\bibfnamefont {T.}~\bibnamefont {Donner}},\ and\ \bibinfo {author}
		{\bibfnamefont {T.}~\bibnamefont {Esslinger}},\ }\bibfield  {title} {\bibinfo
		{title} {{Dissipation-induced structural instability and chiral dynamics in a
				quantum gas}},\ }\href {https://doi.org/10.1126/science.aaw4465} {\bibfield
		{journal} {\bibinfo  {journal} {Science}\ }\textbf {\bibinfo {volume}
			{366}},\ \bibinfo {pages} {1496} (\bibinfo {year} {2019})}\BibitemShut
	{NoStop}%
	\bibitem [{\citenamefont {Reinaudi}\ \emph {et~al.}(2007)\citenamefont
		{Reinaudi}, \citenamefont {Lahaye}, \citenamefont {Wang},\ and\ \citenamefont
		{Gu\'{e}ry-Odelin}}]{Reinaudi_2007}%
	\BibitemOpen
	\bibfield  {author} {\bibinfo {author} {\bibfnamefont {G.}~\bibnamefont
			{Reinaudi}}, \bibinfo {author} {\bibfnamefont {T.}~\bibnamefont {Lahaye}},
		\bibinfo {author} {\bibfnamefont {Z.}~\bibnamefont {Wang}},\ and\ \bibinfo
		{author} {\bibfnamefont {D.}~\bibnamefont {Gu\'{e}ry-Odelin}},\ }\bibfield
	{title} {\bibinfo {title} {{Strong saturation absorption imaging of dense
				clouds of ultracold atoms}},\ }\href {https://doi.org/10.1364/OL.32.003143}
	{\bibfield  {journal} {\bibinfo  {journal} {Opt. Lett.}\ }\textbf {\bibinfo
			{volume} {32}},\ \bibinfo {pages} {3143} (\bibinfo {year}
		{2007})}\BibitemShut {NoStop}%
	\bibitem [{\citenamefont {Goldman}\ \emph {et~al.}(2014)\citenamefont
		{Goldman}, \citenamefont {Juzeli{\={u}}nas}, \citenamefont {Öhberg},\ and\
		\citenamefont {Spielman}}]{Goldman_2014}%
	\BibitemOpen
	\bibfield  {author} {\bibinfo {author} {\bibfnamefont {N.}~\bibnamefont
			{Goldman}}, \bibinfo {author} {\bibfnamefont {G.}~\bibnamefont
			{Juzeli{\={u}}nas}}, \bibinfo {author} {\bibfnamefont {P.}~\bibnamefont
			{Öhberg}},\ and\ \bibinfo {author} {\bibfnamefont {I.~B.}\ \bibnamefont
			{Spielman}},\ }\bibfield  {title} {\bibinfo {title} {{Light-induced gauge
				fields for ultracold atoms}},\ }\href
	{https://doi.org/10.1088/0034-4885/77/12/126401} {\bibfield  {journal}
		{\bibinfo  {journal} {Reports on Progress in Physics}\ }\textbf {\bibinfo
			{volume} {77}},\ \bibinfo {pages} {126401} (\bibinfo {year}
		{2014})}\BibitemShut {NoStop}%
	\bibitem [{\citenamefont {Stamper-Kurn}\ and\ \citenamefont
		{Ueda}(2013)}]{Stamper-Kurn_2013}%
	\BibitemOpen
	\bibfield  {author} {\bibinfo {author} {\bibfnamefont {D.~M.}\ \bibnamefont
			{Stamper-Kurn}}\ and\ \bibinfo {author} {\bibfnamefont {M.}~\bibnamefont
			{Ueda}},\ }\bibfield  {title} {\bibinfo {title} {{Spinor Bose gases:
				Symmetries, magnetism, and quantum dynamics}},\ }\href
	{https://doi.org/10.1103/RevModPhys.85.1191} {\bibfield  {journal} {\bibinfo
			{journal} {Rev. Mod. Phys.}\ }\textbf {\bibinfo {volume} {85}},\ \bibinfo
		{pages} {1191} (\bibinfo {year} {2013})}\BibitemShut {NoStop}%
	\bibitem [{\citenamefont {An}\ \emph {et~al.}(2018)\citenamefont {An},
		\citenamefont {Meier}, \citenamefont {Ang'ong'a},\ and\ \citenamefont
		{Gadway}}]{An_2018}%
	\BibitemOpen
	\bibfield  {author} {\bibinfo {author} {\bibfnamefont {F.~A.}\ \bibnamefont
			{An}}, \bibinfo {author} {\bibfnamefont {E.~J.}\ \bibnamefont {Meier}},
		\bibinfo {author} {\bibfnamefont {J.}~\bibnamefont {Ang'ong'a}},\ and\
		\bibinfo {author} {\bibfnamefont {B.}~\bibnamefont {Gadway}},\ }\bibfield
	{title} {\bibinfo {title} {{Correlated Dynamics in a Synthetic Lattice of
				Momentum States}},\ }\href {https://doi.org/10.1103/PhysRevLett.120.040407}
	{\bibfield  {journal} {\bibinfo  {journal} {Phys. Rev. Lett.}\ }\textbf
		{\bibinfo {volume} {120}},\ \bibinfo {pages} {040407} (\bibinfo {year}
		{2018})}\BibitemShut {NoStop}%
	\bibitem [{\citenamefont {Chen}\ \emph {et~al.}(2021)\citenamefont {Chen},
		\citenamefont {Xie}, \citenamefont {Gadway},\ and\ \citenamefont
		{Yan}}]{Chen_2021}%
	\BibitemOpen
	\bibfield  {author} {\bibinfo {author} {\bibfnamefont {T.}~\bibnamefont
			{Chen}}, \bibinfo {author} {\bibfnamefont {D.}~\bibnamefont {Xie}}, \bibinfo
		{author} {\bibfnamefont {B.}~\bibnamefont {Gadway}},\ and\ \bibinfo {author}
		{\bibfnamefont {B.}~\bibnamefont {Yan}},\ }\href@noop {} {\bibinfo {title} {A
			{G}ross-{P}itaevskii-equation description of the momentum-state lattice:
			roles of the trap and many-body interactions}} (\bibinfo {year} {2021}),\
	\Eprint {https://arxiv.org/abs/2103.14205} {arXiv:2103.14205
		[cond-mat.quant-gas]} \BibitemShut {NoStop}%
	\bibitem [{\citenamefont {An}\ \emph {et~al.}(2021)\citenamefont {An},
		\citenamefont {Sundar}, \citenamefont {Hou}, \citenamefont {Luo},
		\citenamefont {Meier}, \citenamefont {Zhang}, \citenamefont {Hazzard},\ and\
		\citenamefont {Gadway}}]{An_2021}%
	\BibitemOpen
	\bibfield  {author} {\bibinfo {author} {\bibfnamefont {F.~A.}\ \bibnamefont
			{An}}, \bibinfo {author} {\bibfnamefont {B.}~\bibnamefont {Sundar}}, \bibinfo
		{author} {\bibfnamefont {J.}~\bibnamefont {Hou}}, \bibinfo {author}
		{\bibfnamefont {X.-W.}\ \bibnamefont {Luo}}, \bibinfo {author} {\bibfnamefont
			{E.~J.}\ \bibnamefont {Meier}}, \bibinfo {author} {\bibfnamefont
			{C.}~\bibnamefont {Zhang}}, \bibinfo {author} {\bibfnamefont {K.~R.~A.}\
			\bibnamefont {Hazzard}},\ and\ \bibinfo {author} {\bibfnamefont
			{B.}~\bibnamefont {Gadway}},\ }\bibfield  {title} {\bibinfo {title}
		{Nonlinear dynamics in a synthetic momentum-state lattice},\ }\href
	{https://doi.org/10.1103/PhysRevLett.127.130401} {\bibfield  {journal}
		{\bibinfo  {journal} {Phys. Rev. Lett.}\ }\textbf {\bibinfo {volume} {127}},\
		\bibinfo {pages} {130401} (\bibinfo {year} {2021})}\BibitemShut {NoStop}%
	\bibitem [{\citenamefont {Deng}\ \emph {et~al.}(1999)\citenamefont {Deng},
		\citenamefont {Hagley}, \citenamefont {Wen}, \citenamefont {Trippenbach},
		\citenamefont {Band}, \citenamefont {Julienne}, \citenamefont {Simsarian},
		\citenamefont {Helmerson}, \citenamefont {Rolston},\ and\ \citenamefont
		{Phillips}}]{Deng_1999}%
	\BibitemOpen
	\bibfield  {author} {\bibinfo {author} {\bibfnamefont {L.}~\bibnamefont
			{Deng}}, \bibinfo {author} {\bibfnamefont {E.~W.}\ \bibnamefont {Hagley}},
		\bibinfo {author} {\bibfnamefont {J.}~\bibnamefont {Wen}}, \bibinfo {author}
		{\bibfnamefont {M.}~\bibnamefont {Trippenbach}}, \bibinfo {author}
		{\bibfnamefont {Y.}~\bibnamefont {Band}}, \bibinfo {author} {\bibfnamefont
			{P.~S.}\ \bibnamefont {Julienne}}, \bibinfo {author} {\bibfnamefont {J.~E.}\
			\bibnamefont {Simsarian}}, \bibinfo {author} {\bibfnamefont {K.}~\bibnamefont
			{Helmerson}}, \bibinfo {author} {\bibfnamefont {S.~L.}\ \bibnamefont
			{Rolston}},\ and\ \bibinfo {author} {\bibfnamefont {W.~D.}\ \bibnamefont
			{Phillips}},\ }\bibfield  {title} {\bibinfo {title} {{Four-wave mixing with
				matter waves}},\ }\href {https://doi.org/10.1038/18395} {\bibfield  {journal}
		{\bibinfo  {journal} {Nature}\ }\textbf {\bibinfo {volume} {398}},\ \bibinfo
		{pages} {218} (\bibinfo {year} {1999})}\BibitemShut {NoStop}%
	\bibitem [{\citenamefont {Gross}\ and\ \citenamefont
		{Haroche}(1982)}]{Haroche_1982}%
	\BibitemOpen
	\bibfield  {author} {\bibinfo {author} {\bibfnamefont {M.}~\bibnamefont
			{Gross}}\ and\ \bibinfo {author} {\bibfnamefont {S.}~\bibnamefont
			{Haroche}},\ }\bibfield  {title} {\bibinfo {title} {{Superradiance: An essay
				on the theory of collective spontaneous emission}},\ }\href
	{https://doi.org/https://doi.org/10.1016/0370-1573(82)90102-8} {\bibfield
		{journal} {\bibinfo  {journal} {Physics Reports}\ }\textbf {\bibinfo {volume}
			{93}},\ \bibinfo {pages} {301} (\bibinfo {year} {1982})}\BibitemShut
	{NoStop}%
	\bibitem [{\citenamefont {Mandel}\ \emph {et~al.}(1995)\citenamefont {Mandel},
		\citenamefont {Wolf},\ and\ \citenamefont {Press}}]{Mandel_1995}%
	\BibitemOpen
	\bibfield  {author} {\bibinfo {author} {\bibfnamefont {L.}~\bibnamefont
			{Mandel}}, \bibinfo {author} {\bibfnamefont {E.}~\bibnamefont {Wolf}},\ and\
		\bibinfo {author} {\bibfnamefont {C.~U.}\ \bibnamefont {Press}},\ }\href
	{https://doi.org/https://doi.org/10.1017/CBO9781139644105} {\emph {\bibinfo
			{title} {{Optical Coherence and Quantum Optics}}}},\ EBL-Schweitzer\
	(\bibinfo  {publisher} {Cambridge University Press},\ \bibinfo {year}
	{1995})\BibitemShut {NoStop}%
	\bibitem [{\citenamefont {Shampine}\ and\ \citenamefont
		{Reichelt}(1997)}]{Shampine_1997}%
	\BibitemOpen
	\bibfield  {author} {\bibinfo {author} {\bibfnamefont {L.~F.}\ \bibnamefont
			{Shampine}}\ and\ \bibinfo {author} {\bibfnamefont {M.~W.}\ \bibnamefont
			{Reichelt}},\ }\bibfield  {title} {\bibinfo {title} {{The MATLAB ODE
				Suite}},\ }\href {https://doi.org/https://doi.org/10.1137/S1064827594276424}
	{\bibfield  {journal} {\bibinfo  {journal} {SIAM Journal on Scientific
				Computing}\ }\textbf {\bibinfo {volume} {18}},\ \bibinfo {pages} {1}
		(\bibinfo {year} {1997})}\BibitemShut {NoStop}%
	\bibitem [{\citenamefont {Lode}(2016)}]{axel16}%
	\BibitemOpen
	\bibfield  {author} {\bibinfo {author} {\bibfnamefont {A.~U.~J.}\
			\bibnamefont {Lode}},\ }\bibfield  {title} {\bibinfo {title}
		{Multiconfigurational time-dependent {H}artree method for bosons with
			internal degrees of freedom: Theory and composite fragmentation of
			multicomponent bose-einstein condensates},\ }\href
	{https://doi.org/10.1103/PhysRevA.93.063601} {\bibfield  {journal} {\bibinfo
			{journal} {Phys. Rev. A}\ }\textbf {\bibinfo {volume} {93}},\ \bibinfo
		{pages} {063601} (\bibinfo {year} {2016})}\BibitemShut {NoStop}%
	\bibitem [{\citenamefont {Alon}\ \emph {et~al.}(2008)\citenamefont {Alon},
		\citenamefont {Streltsov},\ and\ \citenamefont {Cederbaum}}]{alon08}%
	\BibitemOpen
	\bibfield  {author} {\bibinfo {author} {\bibfnamefont {O.~E.}\ \bibnamefont
			{Alon}}, \bibinfo {author} {\bibfnamefont {A.~I.}\ \bibnamefont
			{Streltsov}},\ and\ \bibinfo {author} {\bibfnamefont {L.~S.}\ \bibnamefont
			{Cederbaum}},\ }\bibfield  {title} {\bibinfo {title} {{Multiconfigurational
				time-dependent Hartree method for bosons: Many-body dynamics of bosonic
				systems}},\ }\href {https://doi.org/10.1103/PhysRevA.77.033613} {\bibfield
		{journal} {\bibinfo  {journal} {Phys. Rev. A}\ }\textbf {\bibinfo {volume}
			{77}},\ \bibinfo {pages} {033613} (\bibinfo {year} {2008})}\BibitemShut
	{NoStop}%
	\bibitem [{\citenamefont {Fasshauer}\ and\ \citenamefont
		{Lode}(2016)}]{fasshauer16}%
	\BibitemOpen
	\bibfield  {author} {\bibinfo {author} {\bibfnamefont {E.}~\bibnamefont
			{Fasshauer}}\ and\ \bibinfo {author} {\bibfnamefont {A.~U.~J.}\ \bibnamefont
			{Lode}},\ }\bibfield  {title} {\bibinfo {title} {Multiconfigurational
			time-dependent {H}artree method for fermions: {I}mplementation, exactness,
			and few-fermion tunneling to open space},\ }\href
	{https://doi.org/10.1103/PhysRevA.93.033635} {\bibfield  {journal} {\bibinfo
			{journal} {Phys. Rev. A}\ }\textbf {\bibinfo {volume} {93}},\ \bibinfo
		{pages} {033635} (\bibinfo {year} {2016})}\BibitemShut {NoStop}%
	\bibitem [{\citenamefont {Lin}\ \emph {et~al.}(2020)\citenamefont {Lin},
		\citenamefont {Molignini}, \citenamefont {Papariello}, \citenamefont
		{Tsatsos}, \citenamefont {L{\'{e}}v{\^{e}}que}, \citenamefont {Weiner},
		\citenamefont {Fasshauer}, \citenamefont {Chitra},\ and\ \citenamefont
		{Lode}}]{lin20}%
	\BibitemOpen
	\bibfield  {author} {\bibinfo {author} {\bibfnamefont {R.}~\bibnamefont
			{Lin}}, \bibinfo {author} {\bibfnamefont {P.}~\bibnamefont {Molignini}},
		\bibinfo {author} {\bibfnamefont {L.}~\bibnamefont {Papariello}}, \bibinfo
		{author} {\bibfnamefont {M.~C.}\ \bibnamefont {Tsatsos}}, \bibinfo {author}
		{\bibfnamefont {C.}~\bibnamefont {L{\'{e}}v{\^{e}}que}}, \bibinfo {author}
		{\bibfnamefont {S.~E.}\ \bibnamefont {Weiner}}, \bibinfo {author}
		{\bibfnamefont {E.}~\bibnamefont {Fasshauer}}, \bibinfo {author}
		{\bibfnamefont {R.}~\bibnamefont {Chitra}},\ and\ \bibinfo {author}
		{\bibfnamefont {A.~U.~J.}\ \bibnamefont {Lode}},\ }\bibfield  {title}
	{\bibinfo {title} {{MCTDH-X}: {T}he multiconfigurational time-dependent
			{H}artree method for indistinguishable particles software},\ }\href
	{https://doi.org/10.1088/2058-9565/ab788b} {\bibfield  {journal} {\bibinfo
			{journal} {Quantum Science and Technology}\ }\textbf {\bibinfo {volume}
			{5}},\ \bibinfo {pages} {024004} (\bibinfo {year} {2020})}\BibitemShut
	{NoStop}%
	\bibitem [{\citenamefont {Lode}\ \emph {et~al.}(2020)\citenamefont {Lode},
		\citenamefont {L\'ev\^eque}, \citenamefont {Madsen}, \citenamefont
		{Streltsov},\ and\ \citenamefont {Alon}}]{axel20}%
	\BibitemOpen
	\bibfield  {author} {\bibinfo {author} {\bibfnamefont {A.~U.~J.}\
			\bibnamefont {Lode}}, \bibinfo {author} {\bibfnamefont {C.}~\bibnamefont
			{L\'ev\^eque}}, \bibinfo {author} {\bibfnamefont {L.~B.}\ \bibnamefont
			{Madsen}}, \bibinfo {author} {\bibfnamefont {A.~I.}\ \bibnamefont
			{Streltsov}},\ and\ \bibinfo {author} {\bibfnamefont {O.~E.}\ \bibnamefont
			{Alon}},\ }\bibfield  {title} {\bibinfo {title} {Colloquium:
			{M}ulticonfigurational time-dependent {H}artree approaches for
			indistinguishable particles},\ }\href
	{https://doi.org/10.1103/RevModPhys.92.011001} {\bibfield  {journal}
		{\bibinfo  {journal} {Rev. Mod. Phys.}\ }\textbf {\bibinfo {volume} {92}},\
		\bibinfo {pages} {011001} (\bibinfo {year} {2020})}\BibitemShut {NoStop}%
	\bibitem [{\citenamefont {Lode}\ \emph {et~al.}(2021)\citenamefont {Lode},
		\citenamefont {Tsatsos}, \citenamefont {Fasshauer}, \citenamefont {Lin},
		\citenamefont {Papariello}, \citenamefont {Molignini}, \citenamefont
		{L\'{e}v\^{e}que},\ and\ \citenamefont {Weiner}}]{ultracold}%
	\BibitemOpen
	\bibfield  {author} {\bibinfo {author} {\bibfnamefont {A.~U.~J.}\
			\bibnamefont {Lode}}, \bibinfo {author} {\bibfnamefont {M.~C.}\ \bibnamefont
			{Tsatsos}}, \bibinfo {author} {\bibfnamefont {E.}~\bibnamefont {Fasshauer}},
		\bibinfo {author} {\bibfnamefont {R.}~\bibnamefont {Lin}}, \bibinfo {author}
		{\bibfnamefont {L.}~\bibnamefont {Papariello}}, \bibinfo {author}
		{\bibfnamefont {P.}~\bibnamefont {Molignini}}, \bibinfo {author}
		{\bibfnamefont {C.}~\bibnamefont {L\'{e}v\^{e}que}},\ and\ \bibinfo {author}
		{\bibfnamefont {S.~E.}\ \bibnamefont {Weiner}},\ }\href
	{http://ultracold.org} {\bibinfo {title} {{MCTDH-X:} the time-dependent
			multiconfigurational {H}artree for indistinguishable particles software}}
	(\bibinfo {year} {2021})\BibitemShut {NoStop}%
	\bibitem [{\citenamefont {Mewes}\ \emph {et~al.}(1996)\citenamefont {Mewes},
		\citenamefont {Andrews}, \citenamefont {van Druten}, \citenamefont {Kurn},
		\citenamefont {Durfee}, \citenamefont {Townsend},\ and\ \citenamefont
		{Ketterle}}]{Ketterle_1996}%
	\BibitemOpen
	\bibfield  {author} {\bibinfo {author} {\bibfnamefont {M.-O.}\ \bibnamefont
			{Mewes}}, \bibinfo {author} {\bibfnamefont {M.~R.}\ \bibnamefont {Andrews}},
		\bibinfo {author} {\bibfnamefont {N.~J.}\ \bibnamefont {van Druten}},
		\bibinfo {author} {\bibfnamefont {D.~M.}\ \bibnamefont {Kurn}}, \bibinfo
		{author} {\bibfnamefont {D.~S.}\ \bibnamefont {Durfee}}, \bibinfo {author}
		{\bibfnamefont {C.~G.}\ \bibnamefont {Townsend}},\ and\ \bibinfo {author}
		{\bibfnamefont {W.}~\bibnamefont {Ketterle}},\ }\bibfield  {title} {\bibinfo
		{title} {{Collective Excitations of a Bose-Einstein Condensate in a Magnetic
				Trap}},\ }\href {https://doi.org/10.1103/PhysRevLett.77.988} {\bibfield
		{journal} {\bibinfo  {journal} {Phys. Rev. Lett.}\ }\textbf {\bibinfo
			{volume} {77}},\ \bibinfo {pages} {988} (\bibinfo {year} {1996})}\BibitemShut
	{NoStop}%
	\bibitem [{\citenamefont {Dalfovo}\ \emph {et~al.}(1999)\citenamefont
		{Dalfovo}, \citenamefont {Giorgini}, \citenamefont {Pitaevskii},\ and\
		\citenamefont {Stringari}}]{Dalfovo_1998}%
	\BibitemOpen
	\bibfield  {author} {\bibinfo {author} {\bibfnamefont {F.}~\bibnamefont
			{Dalfovo}}, \bibinfo {author} {\bibfnamefont {S.}~\bibnamefont {Giorgini}},
		\bibinfo {author} {\bibfnamefont {L.~P.}\ \bibnamefont {Pitaevskii}},\ and\
		\bibinfo {author} {\bibfnamefont {S.}~\bibnamefont {Stringari}},\ }\bibfield
	{title} {\bibinfo {title} {{Theory of Bose-Einstein condensation in trapped
				gases}},\ }\href {https://doi.org/10.1103/RevModPhys.71.463} {\bibfield
		{journal} {\bibinfo  {journal} {Rev. Mod. Phys.}\ }\textbf {\bibinfo {volume}
			{71}},\ \bibinfo {pages} {463} (\bibinfo {year} {1999})}\BibitemShut
	{NoStop}%
	\bibitem [{\citenamefont {Chevy}\ \emph {et~al.}(2002)\citenamefont {Chevy},
		\citenamefont {Bretin}, \citenamefont {Rosenbusch}, \citenamefont {Madison},\
		and\ \citenamefont {Dalibard}}]{Chevy_2002}%
	\BibitemOpen
	\bibfield  {author} {\bibinfo {author} {\bibfnamefont {F.}~\bibnamefont
			{Chevy}}, \bibinfo {author} {\bibfnamefont {V.}~\bibnamefont {Bretin}},
		\bibinfo {author} {\bibfnamefont {P.}~\bibnamefont {Rosenbusch}}, \bibinfo
		{author} {\bibfnamefont {K.~W.}\ \bibnamefont {Madison}},\ and\ \bibinfo
		{author} {\bibfnamefont {J.}~\bibnamefont {Dalibard}},\ }\bibfield  {title}
	{\bibinfo {title} {{Transverse Breathing Mode of an Elongated Bose-Einstein
				Condensate}},\ }\href {https://doi.org/10.1103/PhysRevLett.88.250402}
	{\bibfield  {journal} {\bibinfo  {journal} {Phys. Rev. Lett.}\ }\textbf
		{\bibinfo {volume} {88}},\ \bibinfo {pages} {250402} (\bibinfo {year}
		{2002})}\BibitemShut {NoStop}%
\end{thebibliography}

\end{document}